\documentclass[longauth]{aa}
\usepackage{graphicx,txfonts}
\usepackage{natbib}

\newdimen\captwidth
\newdimen\figwidth

\newcommand{\zR}{\zeta^2 ~\mathrm{Reticuli}}
\newcommand{\mjup}{M_\mathrm{Jup}}
\newcommand{\caii}{\ion{Ca}{ii}}

\newcommand{\rd}{\mathrm{d}}

\begin{document}

\title{Can eccentric debris disks be long-lived?} 
\subtitle{A first numerical investigation \& application to $\zR$}
\author{V. Faramaz$^1$ \and H. Beust$^1$ \and P. Th\'ebault$^2$ \and J.-C. Augereau$^1$ \and A. Bonsor$^1$ \and C. del Burgo$^3$ \and \\ S. Ertel$^1$ \and J.P. Marshall$^4$ \and J. Milli $^{1,5}$ \and B. Montesinos$^6$ \and A. Mora$^7$ \and G. Bryden$^8$ \and W. Danchi$^9$ \and C. Eiroa$^4$ \and G.J. White$^{10,11}$ \and S. Wolf$^{12}$}
\institute{UJF-Grenoble 1 / CNRS-INSU, Institut de
Plan\'etologie et d'Astrophysique de Grenoble (IPAG) UMR 5274,
Grenoble, F-38041, France \and LESIA, Observatoire de Paris, 92195, Meudon, France \and Instituto Nacional de Astrof\'\i sica, \'Optica y Electr\'onica, Luis Enrique Erro 1, Sta. Ma. Tonantzintla, Puebla, Mexico \and Universidad Aut\'onoma de Madrid, Dpto. F\'isica Te\'orica, M\'odulo 15, Facultad de Ciencias, Campus de Cantoblanco, E-28049 Madrid, Spain \and European Southern Observatory, Casilla 19001, Santiago 19, Chile \and Dpt de Astrof\'isica, Centro de Astrobiolog\'ia (INTA-CSIC), ESAC Campus, P.O.Box 78, E-28691, Villanueva de la Ca\~nada, Madrid, Spain \and Aurora Technology B.V., ESA-ESAC, P.O. Box 78, 28691, Villanueva de la Ca\~nada, Madrid, Spain  \and Jet Propulsion Laboratory, California Institute of Technology, 4800 Oak Grove Drive, Pasadena, CA 91109, USA \and NASA Goddard Space Flight Center, Exoplanets and Stellar Astrophysics, Code 667, Greenbelt, MD 20771, USA \and Rutherford Appleton Laboratory, Chilton OX11 0QX, UK \and Department of Physics and Astrophysics, Open University, Walton Hall, Milton Keynes MK7 6AA, UK \and Christian-Albrechts-Universit\"at zu Kiel, Institut f\"ur Theoretische Physik und Astrophysik, Leibnizstr. 15, 24098 Kiel, Germany}
\date{Received 12 August 2013; Accepted 13 December 2013}
\offprints{V. Faramaz}
\mail{Virginie.Faramaz@obs.ujf-grenoble.fr}
\titlerunning{Can eccentric debris disks be long-lived?}
\authorrunning{Faramaz et al.}
\abstract
{Imaging of debris disks has found evidence for both eccentric and offset disks. One hypothesis 
is that these provide evidence for massive perturbers, for example planets or binary companions, that sculpt the observed structures. One such disk was recently observed in the far-IR by the \textit{Herschel}\thanks{\textit{Herschel} Space Observatory is an ESA space observatory with science instruments provided by European-led Principal Investigator consortia and with important participation from NASA.} Space Observatory around $\zR$. In contrast with previously reported systems, the disk is significantly eccentric, and the system is Gyr-old.}
{We aim to investigate the long-term evolution of eccentric structures in debris disks caused by a perturber on an eccentric orbit around the star. We hypothesise that the observed eccentric disk around $\zR$ might be evidence of such a scenario. If so we are able to constrain the mass and orbit of a potential perturber, either a giant planet or binary companion.}
{Analytical techniques are used to predict the effects of a perturber on a debris disk. 
Numerical \textit{N}-body simulations are used to verify these results and further investigate the observable structures that could be produced by eccentric perturbers. The long-term evolution of the disk geometry is examined, with particular application to the $\zR$ system. In addition, synthetic images of the disk are produced for direct comparison with \textit{Herschel} observations.}
{We show that an eccentric companion can produce both the observed offsets and eccentric disks. Such effects are not immediate and we characterise the timescale required for the disk to develop to an eccentric state (and any spirals to vanish).
For the case of $\zR$, we place limits on the mass and orbit of the companion required to produce the observations. Synthetic images show that the pattern observed around $\zR$ can be produced by an eccentric disk seen close to edge-on, and allow us to bring additional constraints on the disk parameters of our model (disk flux, extent).}
{We determine that eccentric planets or stellar companions can induce long-lived eccentric structures in debris disks. Observations of such eccentric structures, thus, provide potential evidence of the presence of such a companion in a planetary system. We consider the specific example of $\zR$, whose observed eccentric disk can be explained by a distant companion (at tens of AU), on an eccentric orbit ($e_\mathrm{p}\gtrsim 0.3$).}  
\maketitle

\section{Introduction}
The first debris disk was discovered in 1984, when the Infrared Astronomical Satellite (IRAS) found a strong IR excess around Vega, revealing the presence of micron-sized dust grains \citep{1984ApJ...278L..23A}. For most debris disks, these grains have a limited lifetime, which, due to Poynting Robertson drag and collisions, is shorter than the system's age. Therefore, this dust is assumed to be replenished by collisional grinding of much larger parent bodies, which are at least kilometre-sized in order for this collisional cascade to be sustained over the system's age  \citep{1993prpl.conf.1253B,2008ApJ...673.1123L}. Consequently, debris disks provide evidence for the existence of solid bodies having reached km-size, and potentially the planetary-size level. 

Spatially resolved structures in debris disks can provide clues to the invisible planetary component of those systems. Such planets may be responsible for sculpting these disks, and may leave their signature through various asymmetries such as wing asymmetries, resonant clumpy structures, warps, spirals, gaps or eccentric ring structures \citep[see, e.g.,][]{1999PhDT........12W}. This diversity is to be compared with the variety of exoplanetary systems\footnote{see www.exoplanets.org} discovered around main sequence stars since 1995 \citep[51 Peg b,][]{1995Natur.378..355M}. 
Dynamical modelling of such asymmetries is the only method to place constraints on the masses and orbital parameters of planets in systems where direct observations are not possible \citep[see, e.g.,][]{1997MNRAS.292..896M,1999ApJ...527..918W,2001A&A...370..447A,2002AJ....124.2305M,2004AIPC..713...93W,2005Natur.435.1067K,
2006MNRAS.372L..14Q,stark08,2009ApJ...693..734C,2011epsc.conf..678E,2012ApJ...750L..21B,
2012A&A...544A..61E,2012A&A...547A..92T}. 

We focus here on cases of eccentric patterns in debris disks. The modelling of this type of asymmetry and its possible link with the dynamical influence of eccentric companions has been investigated in several earlier studies: authoritative work was carried out by \citet{1999ApJ...527..918W,2000ASPC..219..289W} for the case of HR 4796. Another case of interest is the debris disk of Fomalhaut \citep[][ Beust et al, in revision]{2004stapelfeldt,2005Natur.435.1067K,2006MNRAS.372L..14Q,2009ApJ...693..734C,2012ApJ...750L..21B,2013arXiv1305.2222K}.

This pioneering work showed that these large-scale structures arise in systems where debris disks are perturbed by an eccentric companion, on a low inclination orbit relative to the disk \citep{1999ApJ...527..918W}. 
The disk centre of symmetry is offset from the star which may be measured explicitly in high resolution images (e.g., HST scattered light). Furthermore, its periastron is closer to the star and thus hotter and brighter, which results in a two-sided brightness asymmetry.

However, it is important to emphasize that previous studies considered low eccentricity rings ($e\gtrsim 0.02$ for HR4796 and $e=0.11 \pm 0.01$ for Fomalhaut), and were limited to timescales smaller than the typical ages of mature disks ($\leq 10\,$Myr for HR4796 simulations and $\leq 100\,$Myr for Fomalhaut). The issue of whether highly eccentric ring structures could be sustained over very long timescales has not been addressed thus far in the literature. This issue has become very topical because of the discovery of at least two Gyr-old and significantly eccentric debris disks: one around $\zR$ \citep[$e\gtrsim 0.3$][]{2010A&A...518L.131E}, which is used as a reference case in this paper, and another one around HD 202628 \citep[$e\sim 0.18$][]{2012AAS...22050603S,2012AJ....144...45K}.
These systems are both older than Fomalhaut or HR 4796, with disks that are also much more eccentric.

In the present work, we investigate the long-term evolution of highly eccentric structures in debris disks, and their relation to planetary or stellar perturbers, by investigating their evolution over Gyr timescales. One of the questions we address is whether these structures are really Gyr-old, or might have originated from a more recent event, be it a flyby or the late excitement of a sherpherding planet's eccentricity. We also summarize a general modelling method, based on complementary analytical and numerical tools, which we apply to the specific case of $\zR$.

This paper is structured as follows: Sect. ~\ref{sec:tools} presents how a perturber can generate an eccentric ring structure. Useful analytical expressions are derived, to study under which conditions such a pattern can be created. We also show that these predictions can be complemented by numerical studies. Sect. ~\ref{sec:numerical} describes the debris disk of $\zR$, along with newly reduced \textit{Herschel}/PACS images. This debris disk is used as a proxy to determine a numerical set-up. Then, in Sect. ~\ref{sec:results}, the numerical investigation is carried out. From \textit{N}-body simulations, we examine both the onset and suvival of an eccentric pattern and explore their dependencies to the perturber's characteristics.
This modelling approach allows one to put constraints on a perturber at work in shaping a debris disk into an eccentric ring over Gyr timescales, and it is applied to the case of the debris disk of $\zR$. Sect. ~\ref{sec:synthetic} shows synthetic images to perform a full comparison with observations of $\zR$, and retrieve additional constraints on this disk. 
Finally, Sect. ~\ref{sec:conclusion} is devoted to conclusions, discussions and proposing future work.

\section{Footprints of eccentric companions on debris disks}\label{sec:tools}
We have developed a dynamical model to investigate the shaping of a debris disk into an eccentric ring, and the timescales associated with its onset and survival. More specifically, we seek to determine if any perturbers are able to shape and maintain a disk into a significantly eccentric ring structure on Gyr timescales, and whether the asymmetry relaxes or not.

Before presenting our model and our results in detail for this as yet unexplored case, we present the current understanding on how eccentric ring structures arise as a result of the dynamical effect of an eccentric perturber.

\subsection{Basic principle}

For a disk to be shaped into an eccentric ring, it must be perturbed in such a way that its components have both their eccentricities forced to higher values, and their orbits more or less oriented in a common direction. These conditions are both fulfilled if the disk is under the gravitational influence of a perturber, namely a planetary or a stellar companion, (nearly) coplanar to the disk, and on an eccentric orbit.
The eccentricity of the ring causes the disk center of symmetry to be offset from the star, and the disk pericenter to be brighter than the apocenter, since it is closer to the star and thus hotter. This feature was studied by \citet{1999ApJ...527..918W} in the case of a planetary companion, and called the pericenter glow phenomenon.

Spatially resolved imaging is required to determine the structure of debris disks and therefore renewed efforts have been made to image as many debris disks as possible\footnote{see, e.g., www.circumstellardisks.org}.
Most images of resolved debris disks have been obtained so far in the visible or near-IR. At these wavelengths, the emission is dominated by small grains close to the blow-out limit imposed by stellar radiation (sub-micron to micron depending on stellar luminosity). Such grains exhibit complex evolution due to the coupled effects of collisions and radiation pressure \citep[see, e.g.,][]{2007A&A...472..169T}. This may strongly alter, or even mask the dynamical structures imparted by a massive perturber \citep{2012arXiv1209.3969T}. 
Observations at longer wavelengths detect bigger grains less affected by stellar radiation (few tens to few hundreds of micron in size, depending on observing wavelength). These should directly trace the distribution of larger parent bodies and thus more directly reflect the dynamical effect of a perturber \citep[see, e.g.,][for exhaustive reviews]{2010RAA....10..383K,2012arXiv1203.0005M}.

Since the origin of an eccentric pattern is gravitational, we can reasonably assume that large scale asymmetries among an observed dust population already exist amongst the parent planetesimal population that produces it and result from pure gravitational perturbations. This assumption allows us to study the influence of different eccentric perturbers in a simplified way, neglecting the effect of radiation pressure and considering parent planetesimals as massless and collisionless particles in orbit around their host star and perturbed by a companion, be it stellar or planetary.

We assume that at the end of the protoplanetary phase, the planetesimals start from almost circular orbits because of orbital eccentricity damping by primordial gas. We further assume that any perturbing planet among the system is fully formed by the time the gas disappears, and evolves on a significantly eccentric orbit, due to a major perturbing event such as planet-planet scattering. Thus, we can consider the disappearance of the gas as time zero for the onset of planetesimal perturbations by an eccentric companion. From this moment the planetesimal eccentricities start to increase and their lines of apsides tend to align with the planet's.
Under these assumptions, the forced elliptic ring structure takes some time to settle in, and it is preceded by the appearance and disappearance of transient spiral features. These are due to differential precession within the disk: all the planetesimals in the disk have different precession rates (due to their different orbital distances), such that these spiral structures are expected to wind up and finally generate an eccentric ring, as shown by \citet{2004A&A...414.1153A} and \citet{2005A&A...440..937W}.
The characteristic time for reaching this state is of the order of a few precession timescales at the considered distance \citep{2005A&A...440..937W}.
Consequently, the onset of an eccentric ring structure is a matter of timescales, while the value of the disk global eccentricity is to be linked with the planetesimals' forced eccentricity, and thus to the companion's eccentricity.

\subsection{Analytical approach}

We show here how the onset of these structures can be understood from analytical considerations.
Planetesimals are considered to be massless particles. We focus on the secular response of a debris disk to a coplanar perturbing body, either a planet or star. More specifically, both the forced secular eccentricity $e_\mathrm{f}$ and apsidal precession rate $\rd\varpi/\rd t$ of test planetesimals are examined, where $\varpi$ is the longitude of periastron with respect to the direction of the perturber's periastron, i.e., the planet and planetesimal have their periastra aligned when $\varpi=0$ and anti-aligned when $\varpi=\pi$.

When secularly perturbed, the eccentricity of a planetesimal evolves cyclically, the period of which is related to the rate of orbital precession. In particular, if we consider a dynamically cold disk of planetesimals as an initial condition, which, considering the damping effect of the gas during the protoplanetary phase, is a reasonable and classical assumption. In that case, the secular behaviour of a planetesimal can
be understood considering the analytical solution for its eccentricity. In this case, in the Laplace-Lagrange theory, the complex eccentricity of a planetesimal, $z(t)$, can be written:
\begin{equation}
z(t)= e_\mathrm{f} \left\lbrace 1-\exp(IAt) \right\rbrace \qquad,
\end{equation}
where $I^2=-1$, and $A=\frac{\rd\varpi}{\rd t}$ is the secular precession rate.

\begin{figure}[htbp]
\resizebox{\hsize}{!}{
\includegraphics[width=0.45\textwidth]{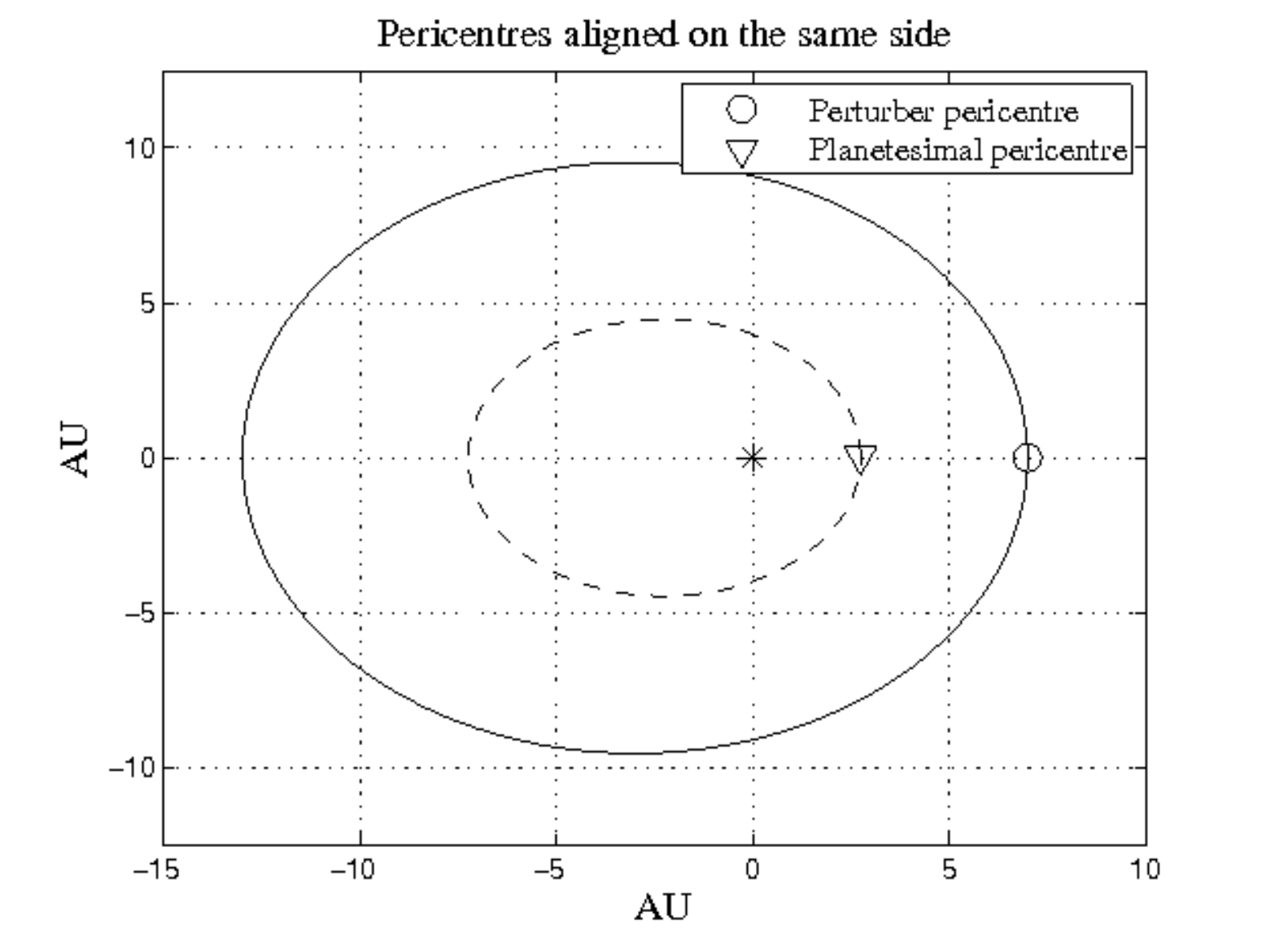}}
\caption{Co-evolution of a planetesimal eccentricity and orbital precession, when acted upon by an eccentric perturber. As the planetesimal orbit precesses, when the longitudes of periastra of the planetesimal and the perturber are equal ($\varpi=0$), i.e., when  the planetesimal and the perturber have their pericentres aligned on the same side of the massive central body, the planetesimal eccentricity is maximum.}
\label{forced_ecc}
\end{figure}

One can see from this expression that the maximum induced eccentricity for a planetesimal is twice the forced eccentricity, i.e., $e_{f,max}=2e_\mathrm{f}$. This occurs when $At=\pi [2\pi]$, i.e., when the longitudes of periastra of the planetesimal and the perturber are equal \citep[$\varpi=0$, see Fig.~\ref{forced_ecc} and, for further details, see e.g.,][]{2005A&A...440..937W,beust2013}.

There are several ways to analytically derive $e_\mathrm{f}$ and $\rd\varpi/\rd t$. The most classical one is to apply linear Laplace-Lagrange theory, i.e. an expansion of the interaction Hamiltonian to second order in ascending powers of the eccentricities of both bodies and an averaging over both orbits \citep[see, e.g., Eq.~6 of][]{2009MNRAS.399.1403M}. However, this approach is valid only for small eccentricities, whereas the perturber's orbital eccentricity $e_\mathrm{p}$ is not necessarily small. Therefore, restricting our analytical study to small $e_\mathrm{p}$ may not be appropriate. 

Another way to proceed is to expand the interaction Hamiltonian in spherical harmonics, to truncate at some order in $\alpha$, where $\alpha$ is the ratio\footnote{$\alpha$ is such that $\alpha<1$, always, and thus $\alpha=a_\mathrm{p}/a$ if $a_\mathrm{p}<a$, and inversely, $\alpha=a/a_\mathrm{p}$ if $a_\mathrm{p}>a$.} between $a$ and $a_\mathrm{p}$, the planetesimal and the perturber's semi-major axis respectively, and to average after over both orbits. 
This permits us to perform an analysis without any restriction on the eccentricities. The resulting Hamiltonian is given by \citet{1999MNRAS.304..720K,2000ApJ...535..385F} or \citet{2006A&A...446..137B}. 

To the lowest order in $\alpha$ (2$^{\mathrm{nd}}$ order, quadrupolar), it yields a forced eccentricity $e_\mathrm{f}$:

\begin{equation}\label{eq:forced2}
e_\mathrm{f} \simeq \frac{5}{4} \frac{\alpha e_\mathrm{p}}{1-e_\mathrm{p}^2}\qquad.
\end{equation}

This expression is given by \citet{2004A&A...414.1153A} and \citet[Eq.~8 of][]{2009MNRAS.399.1403M}. Note that this approach is only valid for small enough values of $\alpha$ to ensure a fast convergence of the expansion, i.e. for orbits with significantly different sizes. It is also valid only far from mean-motion resonances. However, these resonances' spatial extension in semi-major axis (of the order of $\sim 0.1~$AU) is typically two orders of magnitude smaller than the extent of the observed structures (of the order of $\sim 10~$AU), although in the case where particles are on eccentric orbits, these resonances may span much larger ranges in terms of radial distance to the star than their span in semi-major axis would have let suppose. But in any case, the amount of material trapped in resonance can reasonably be assumed to be much smaller than the amount of non-resonant material, all the more since we do not suppose here that the planet has migrated, and thus exclude resonant capture during migration. Therefore, these structures are assumed to result from non-resonant material, and our approach is appropriate. Moreover, resonances will be treated in our N-body integrations, and will be confirmed important only for limited parameter combinations (see Sect.~\ref{sec:results}).

To derive the precession rate in the spherical harmonic expansion case, we follow the method given by \citet{2002ApJ...573..829M}. The variation rate for the Runge-Lenz vector of the orbit is computed, expanded to any given order, and integrated over one orbital period. 

Since after numerical tests, one notices that there is less than two orders of magnitude between terms of the 2$^\mathrm{nd}$ and the 4$^\mathrm{th}$ order (the third order terms cancel out), we retain terms up to 4$^\mathrm{th}$ order in the spherical harmonic expansion, and average the resulting precession rate over the longitude of periastron. The precession rate is:

\begin{eqnarray}\label{eq:timescale}
\frac{\rd\varpi}{\rd t} & = & \frac{3n}{4}\frac{m_\mathrm{p}}{M_\star}\alpha^3 
\frac{\sqrt{1-e^2}}{(1-e_\mathrm{p}^2)^{3/2}}\nonumber\\
&&+\frac{45n}{256}
\frac{m_\mathrm{p}}{M_\star}\alpha^5
\frac{(4+3e^2)(2+3e_\mathrm{p}^2)\sqrt{1-e^2}}{(1-e_\mathrm{p}^2)^{7/2}}\qquad.
\end{eqnarray}

We now follow \citet{1999ApJ...527..918W} and assume that the precession timescale, $t_\mathrm{prec}$, corresponds to the lower limit of the typical dynamical timescale for setting a dynamical steady state:

\begin{equation}
t_\mathrm{prec}=\frac{2\pi}{(\rd\varpi/\rd t)_{a_\mathrm{c}}}\qquad,
\label{tglow}
\end{equation}
where $a_\mathrm{c}$ is the typical semi-major axis of a particle in the ring.

\begin{figure}
\resizebox{\hsize}{!}{\includegraphics[width=0.7\textwidth,height=0.5\textwidth]{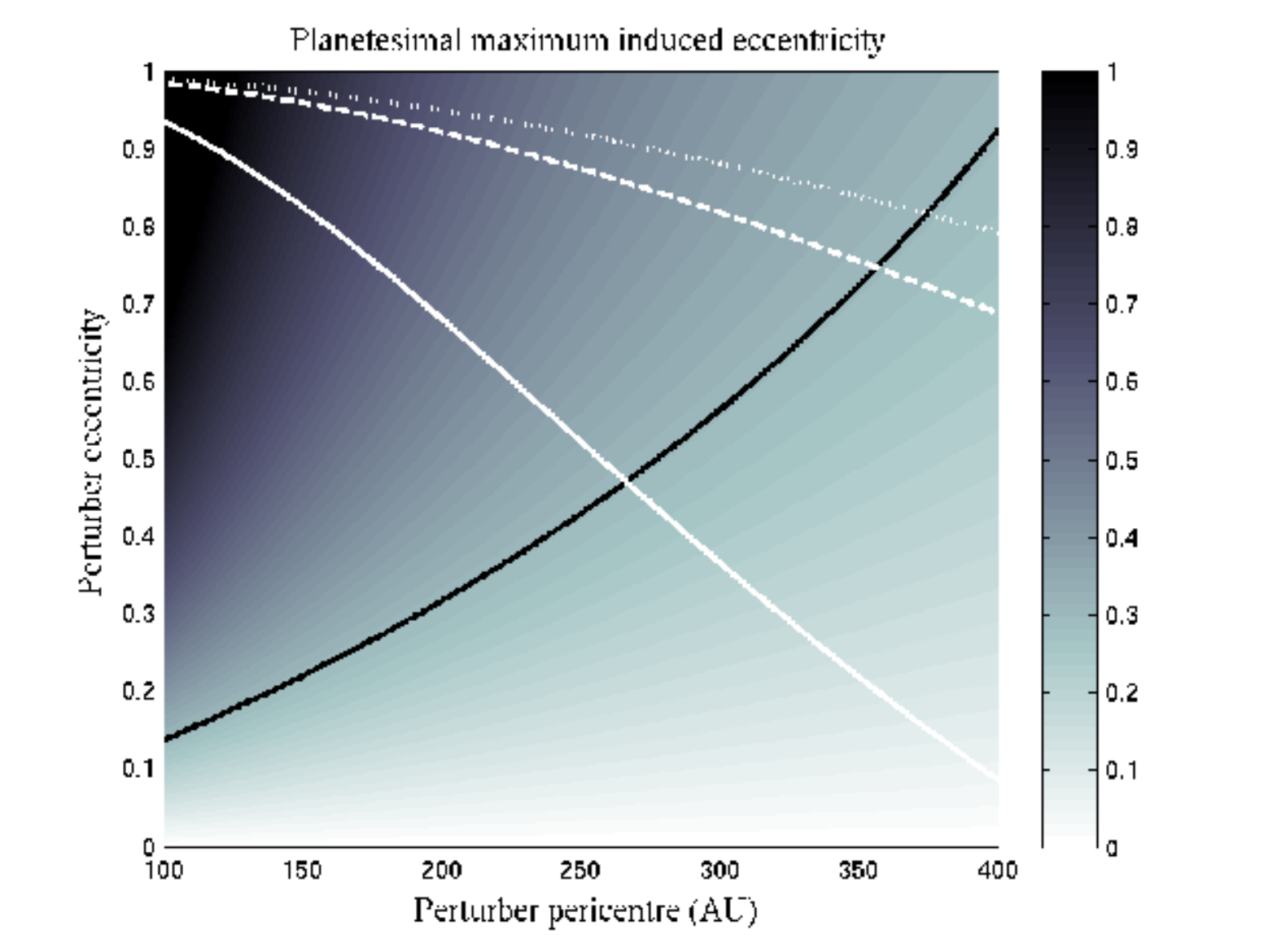}}
\caption[]{Example color map of the maximum induced eccentricity $2 e_\mathrm{f}$ imposed by a planetary perturber on a particle with semi-major axis 100 AU and eccentricity $e=0$, 
as a function of its periastron 
and eccentricity, as estimated from Eq.~(\ref{eq:forced2}). The black line corresponds to a $2 e_\mathrm{f}=0.3$ condition, which is set to mimic the condition for the disk of $\zR$. Note that it does not depend on the mass of the planet. The white lines show the parameters for which the typical timescale to reach a steady state at 100 AU is $t_\mathrm{prec}=1\,$Gyr, using Eq.~(\ref{tglow}). This timescales depends on the mass: $m_\mathrm{p}=0.1\,\mathrm{M}_\mathrm{Jup}$ \textit{(solid line)}, $1\,\mathrm{M}_\mathrm{Jup}$ \textit{(dashed line)} and $2\,\mathrm{M}_\mathrm{Jup}$ \textit{(dotted line)}. For example, a perturber of mass $0.1\,\mathrm{M}_\mathrm{Jup}$, periastron $q_\mathrm{p}=150\,$AU and eccentricity $e_\mathrm{p}=0.4$ should produce a significantly eccentric ring in less than 1 Gyr, although spiral patterns may remain since, as was shown by \citet{2005A&A...440..937W}, it can take several precession timescales for them to vanish.}
\label{fig:precess}
\end{figure}

We can now make analytical predictions of the effect of a perturber on a debris disk, i.e. for any given set of values of $e_\mathrm{p}$ and of the planet periastron $q_\mathrm{p}$, one can derive the precession rate $(\rd\varpi/\rd t)_{a_\mathrm{c}}$ corresponding to planetesimals orbiting at this distance. 
Conversely, one can set this dynamical timescale and the forced eccentricity for a particle with semi-major axis $a_\mathrm{c}$ to correspond to those derived from observations of an eccentric debris disk, and thus make an initial estimate of the perturber's characteristics (see Fig.~\ref{fig:precess}). 

However, the problem is more complex for real disks which have a finite spatial extension, since these estimates depend on radial locations. To first order, it can be seen from Eq.~\ref{eq:forced2}, Eq.~\ref{eq:timescale}, and  Eq.~\ref{tglow} that the forced eccentricity and the secular timescale scale as:

\begin{equation}
e_\mathrm{f} \propto \alpha \qquad,
\label{prop_forced}
\end{equation}
and
\begin{equation}
t_\mathrm{prec} \propto \frac{1}{m_p \alpha^3}\qquad.
\label{prop_secular}
\end{equation}

We now define $a_\mathrm{in}$ and $a_\mathrm{out}$ as the inner and outer limits of the disk in semi-major axis, and define $e_\mathrm{f,min}$ and $e_\mathrm{f,max}$ as the minimum and maximum eccentricities induced across the disk.\footnote{$e_\mathrm{f,min}=e_\mathrm{f,in}$ and $e_\mathrm{f,max}=e_\mathrm{f,out}$ in the case of an inner perturber, and conversely in the case of an outer one.} The minimum and maximum precession timescales, $t_\mathrm{prec,min}$ and $t_\mathrm{prec,max}$, are defined in the same manner. Then, using Eq.~\ref{prop_forced} and  Eq.~\ref{prop_secular}, one obtains:

\begin{equation}
\frac{e_\mathrm{f,max}}{e_\mathrm{f,min}} = \frac{a_\mathrm{out}}{a_\mathrm{in}} \qquad,
\label{ratio_ecc}
\end{equation}
and
\begin{equation}
\frac{t_\mathrm{prec,max}}{t_\mathrm{prec,min}} = \left(\frac{a_\mathrm{out}}{a_\mathrm{in}}\right)^3 \qquad.
\label{ratio_temp}
\end{equation}

It is easy to see from these equations that the secular precession timescale spans a large range of values across the disk. This means that making analytical predictions by setting the wanted values for the forced eccentricity and the secular precession timescale for a particle with semi-major axis at the center of the distribution suffers limitations when applied to an extended disk, especially concerning the timescale.

Eq.~\ref{ratio_ecc} and  Eq.~\ref{ratio_temp} can be rewritten using $\Delta a$, the half width of the disk extent, along with $e_\mathrm{f,c}$ and $t_\mathrm{prec,c}$, the forced eccentricity and secular precession timescale at $a_\mathrm{c}$, respectively:

\begin{equation}
e_\mathrm{f,max/min} = \left(\frac{a_\mathrm{c} \pm \Delta a}{a_\mathrm{c}}\right) e_\mathrm{f,c} \qquad,
\label{central_ecc}
\end{equation}
and
\begin{equation}
t_\mathrm{prec,max/min} = \left(\frac{a_\mathrm{c}\pm \Delta a}{a_\mathrm{c}}\right)^3 t_\mathrm{prec,c}\qquad.
\label{central_temp}
\end{equation}

As an example, we set $a_\mathrm{c}=100$ AU, $\Delta a=25$ AU, $2e_\mathrm{c}=0.3$ and $t_\mathrm{prec,c}=1$ Gyr. These values are close to those derived for the disk of $\zR$ \citep[$e\gtrsim 0.3$ and extent 70-120 AU:][ and Sect.~\ref{sec:numerical} of the present work]{2010A&A...518L.131E}. One obtains:

\begin{equation}
\left\{
\begin{array}{l}
2e_\mathrm{f,min/max} = 0.225-0.375\\
t_\mathrm{prec,min/max} = 0.42-1.95\,\mathrm{Gyr}\\
\end{array}
\right.\qquad.
\end{equation}

In these conditions, the extent of the disk is not expected to affect too much the global eccentricity of the disk, i.e., we should recover in average a global eccentricity corresponding to the forced eccentricity at $a_\mathrm{c}$, once the steady state is reached. 
But the problem is that the extent of the disk affects a lot the timescale to reach this steady state.
This is a limitation of the analytic approach that can be overcome by the use of numerical simulations.

\section{Numerical Investigation: A typical set-up, the highly eccentric, old disk of $\zR$}\label{sec:numerical}
To go beyond the simplified analytical approach and explore the high eccentricity case on Gyr timescales, we resort to numerical tools. We place ourselves in the frame of the restricted 3-body problem, i.e. one central star, a planet and a massless planetesimal. The symplectic N-body code SWIFT-RMVS of \citet{1994Icar..108...18L} is used to integrate the evolution of a ring of planetesimals around a solar mass star, over 1 Gyr. We use a typical timestep of $\sim 1/20$ of the smallest orbital period, and ensure a conservation of energy with a typical error of $\sim 10^{-6}$ on relative energy.
This code has a crucial advantage over an analytical approach: it is able to handle close encounters and scattering processes, along with the short-term variations of the planetesimals orbital elements, whereas these effects are ignored in the analytical approach, for which orbits are averaged, short-term variations are lost and the approach is valid only for $\alpha << 1$, i.e. far from close encounters. 
As will be shown in Sect. ~\ref{sec:results}, the scattering events play a crucial role in the system's evolution.

For the sake of clarity, the $\zR$ system is considered as a proxy for a typical mature and significantly eccentric debris ring. We shall explore different planet-disk configurations, and produce synthetic images for comparison with \textit{Herschel}/PACS observations. As will be shown, the hypothesis of an eccentric debris disk around $\zR$ is fully consistent with the observations.

\subsection{The $\zR$ system}

$\zR$ (HR 1010, HIP 15371) is a G1V  solar-type star \citep{2013A&A...555A..11E} located at 12 pc \citep{2007A&A...474..653V}, of luminosity $L_\star=0.97 L_\odot$, $\log g = 4.51$, and age $\sim 2-3~$ Gyrs \citep{2013A&A...555A..11E}. It has a binary companion $\zeta^1 \mathrm{Reticuli}$, a G2-4V \citep{2006AJ....132..161G,2006A&A...460..695T} star located at a projected distance of 3713 AU from $\zR$ \citep{2001AJ....122.3466M}. Bayesian analysis by \citet{2011ApJS..192....2S} of the proper motions of these stars indicates a very high (near 100\%) probability that the pair are physically connected.

\begin{figure*}
\makebox[\textwidth]{
\includegraphics[width=0.325\textwidth]{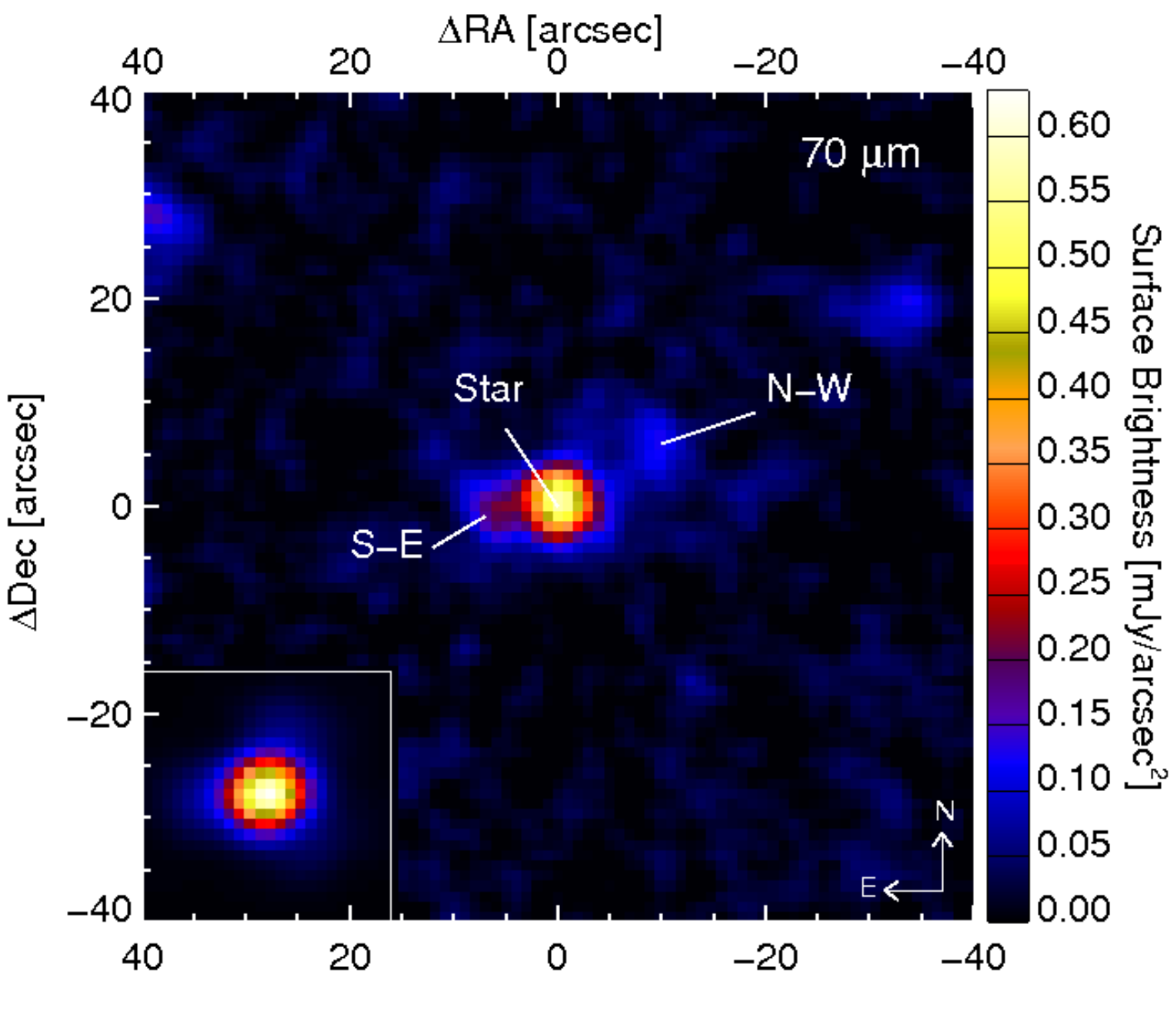}
\includegraphics[width=0.325\textwidth]{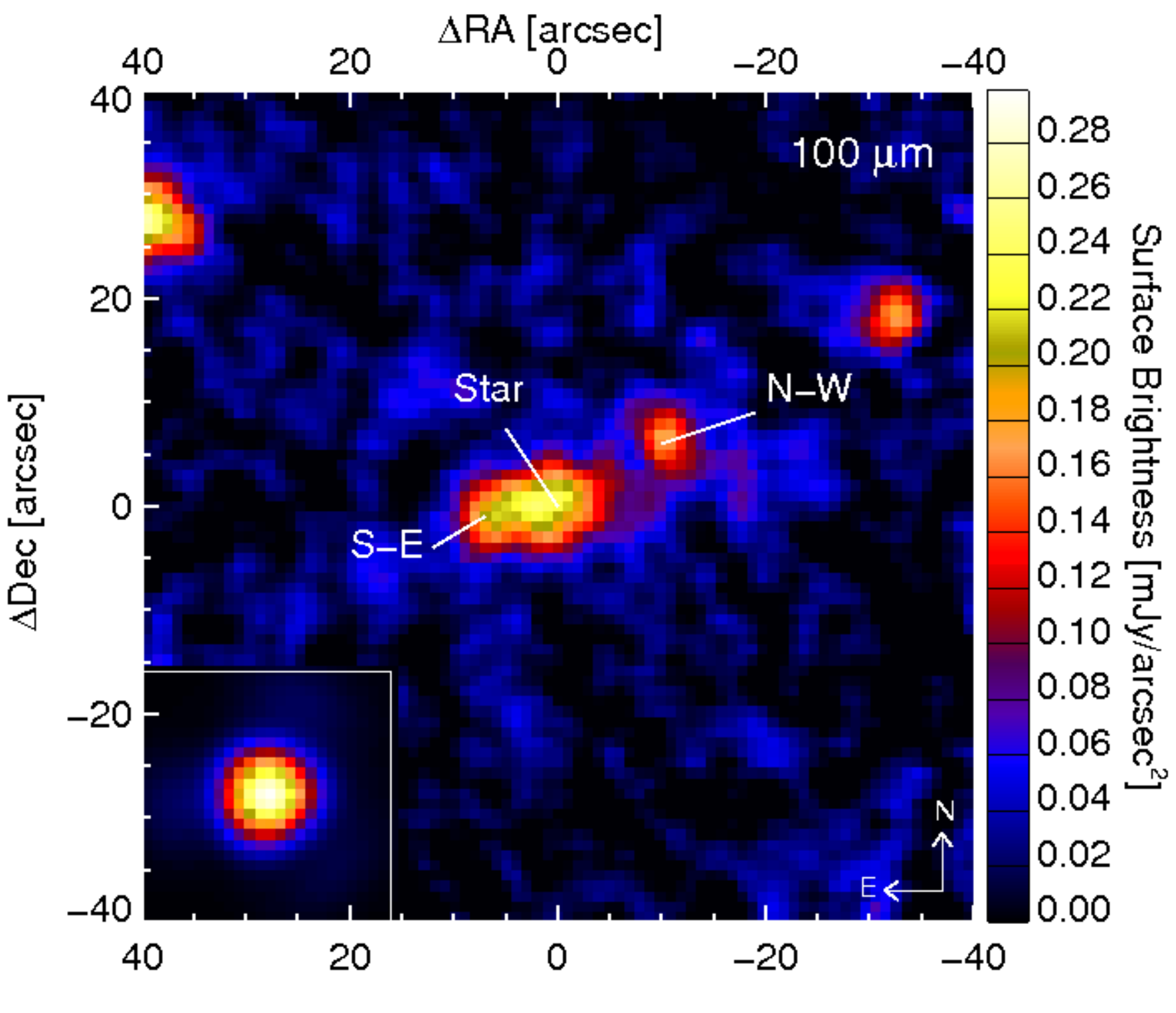}
\includegraphics[width=0.325\textwidth]{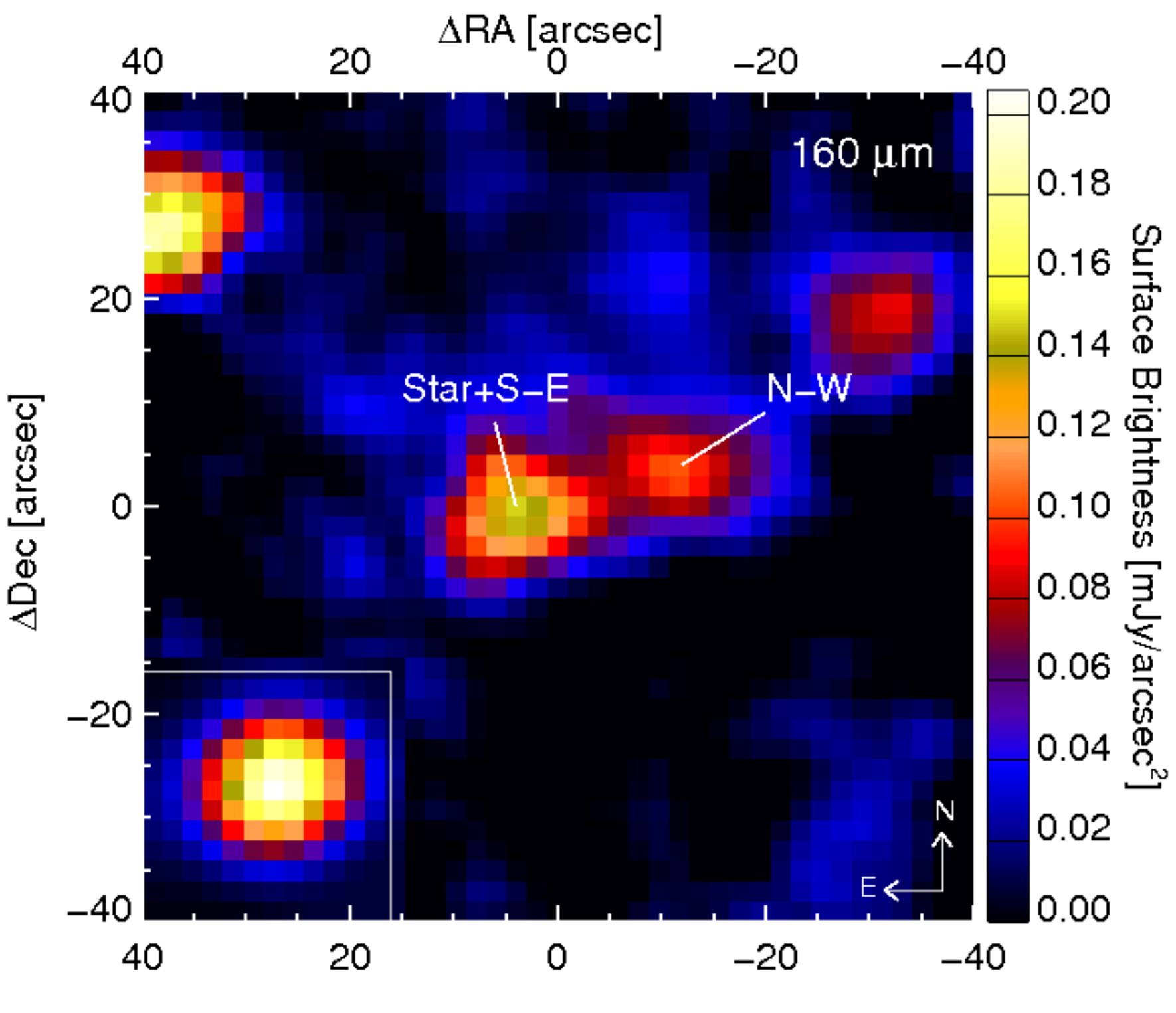}}
\caption[]{\textit{Herschel}/PACS images of $\zR$ at 70, 100 and 160 microns from left to right. North is up and East is left. The inset in the bottom-left corner shows the PSF. The North-West lobe is noted N-W, while the South-East lobe is noted S-E.} 
\label{fig:PACS}
\end{figure*}

The presence of dust around $\zR$ has already been probed with \textit{Spitzer} \citep{2008ApJ...674.1086T}, which suggests a $\sim 150~$K emission at $\sim 4.3~$AU. However, the angular resolution of \textit{Spitzer} is limited, and the dust spatial distribution remained unconstrained.
New observations with \textit{Herschel}/PACS completed the SED, providing the suggestion of an optically thin, $\sim 40~$K, emission at $\sim 100~$AU, with fractional luminosity  $L_{\mathrm{dust}}/L_{\star}\approx 10^{-5}$ \citep{2010A&A...518L.131E}. Moreover, \textit{Herschel}/PACS provided spatially resolved images of the dust thermal emission surrounding $\zR$ at $70~\mathrm{\mu m}$ and $100~\mathrm{\mu m}$ \citep{2010A&A...518L.131E}. 
We present here newly reduced \textit{Herschel}/PACS images (see Fig. ~\ref{fig:PACS}). The images show a double-lobe feature, asymmetric both in position and brightness. Note that at $70~\mathrm{\mu m}$, the probability for alignment with a background source within 10$^{\prime \prime}$ is extremely low, namely $10^{-3}$ \citep{2010A&A...518L.131E}. The disk is not resolved at \textit{Herschel}/SPIRE wavelengths: newly reduced images and star-disk fluxes measurements are presented in Appendix ~\ref{sec:spire}.

As suggested by \citet{2010A&A...518L.131E}, the asymmetry revealed by \textit{Herschel}/PACS in the disk of $\zR$ can be interpreted as a ring-like elliptical structure with $e \gtrsim 0.3$ seen close to edge-on, and extending from $\sim 70$ to $\sim 120~$AU, which is fully consistent with the information derived from the SED \citep{2010A&A...518L.131E}. Alternatively, it could also be interpreted as two clumps from an over-density of dust and planetesimals. In Appendix~\ref{sec:inclination}, we investigate the system inclination on the line of sight, a crucial parameter required to interpret correctly the observed structures. More precisely, we determine the star inclination and assume that the disk and the star are coplanar.
The 50\% probability value is $i=65.5^{\circ}$, i.e. the system is very inclined on the line of sight, which tends to support the eccentric ring scenario.

Without a doubt, this asymmetric structure provides evidence that "something" is dynamically sculpting the disk.
It could be the stellar companion $\zeta^1 \mathrm{Reticuli}$ or a (yet undetected) planet. The latter hypothesis is fully compatible with radial velocity measurements of $\zR$, which suggest there is no Jupiter-mass (or larger) planet interior to $\sim 5-10~$AU \citep{2003Msngr.114...20M} but put no contraints on small planet, or Jupiter-like planet at larger radii. It is also compatible with growing observational evidence for planets at large orbital separation, i.e., a several tens to few hundreds of AU from their host star \citep[see, e.g.,][]{2007ApJ...654..570L,2008Sci...322.1345K,2008Sci...322.1348M,2010Natur.468.1080M}. 

\begin{figure}
   \centering
   \includegraphics[width=\columnwidth]{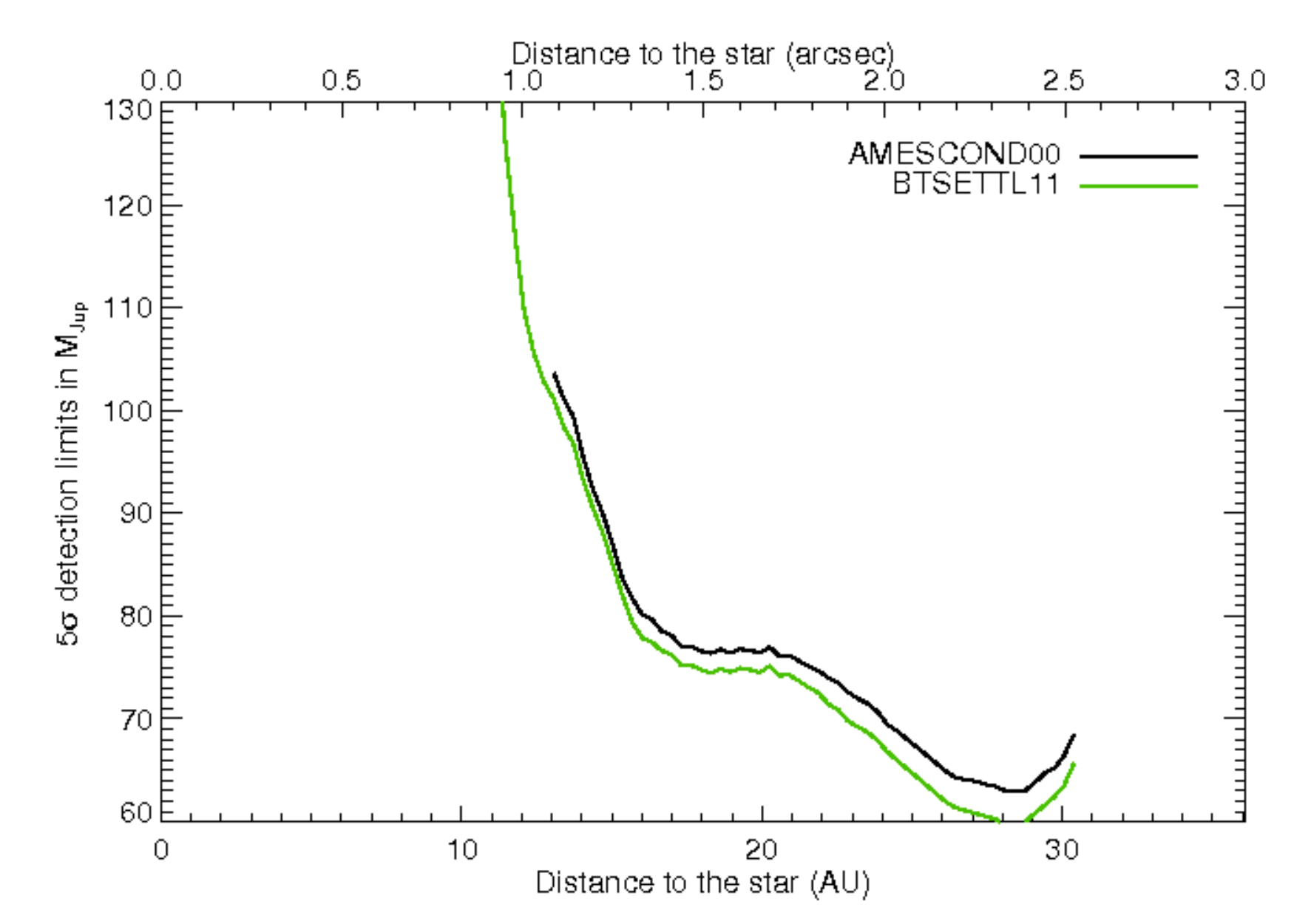}
   \caption{Detection limits set by direct imaging on the presence of brown dwarf / close binary between $1^{\prime \prime}$ and $2.5^{\prime \prime}$ in projected separation.}
    \label{Fig_det_curve}%
    \end{figure}

Constraints from direct imaging are presented in Fig. \ref{Fig_det_curve} using the two evolutionary models COND 2003 \citep{Baraffe2003} or BT-settl 2011 \citep{Allard2011}. Details of the reduction procedure is presented in Appendix \ref{App_det_lim}. These constraints were obtained from VLT/NaCo archival data taken in November 2010 in the Ks band. These data do not provide constraints on companions beyond a projected distance of $\sim 30\,$AU.
The presence of a brown dwarf within $\sim 20\,$AU is still compatible with observations.

\subsection{Numerical set-up}

We consider a ring of 150,000 massless planetesimals uniformly distributed between 70 and 140 AU (except otherwise specified) around a solar mass host star, with initial eccentricities randomly distributed between 0 and 0.05, and initial inclinations between $\mathrm{\pm3^\circ}$, while the remaining angles (longitudes of nodes and periastra) are randomly distributed between 0 and $2\pi$. These values are summarized in Table \ref{tab:init}. This reasonably well mimics the low eccentricities and inclinations expected at the end of the protoplanetary phase. The radial extent of the model disk has been configured to match closely the observed properties of the disk around $\zR$.

Using massless test particles removes both any self-gravity in the disk, and any back-reaction of the disk on the planet. In general, both of these phenomena are significant when the planet mass is comparable to the disk mass. We have no mass estimate for the debris disk of $\zR$. However, a well-studied case is the debris disk of Fomalhaut, which mass was estimated to be $\sim 3-20 \mathrm{M}_{\oplus}$ \citep{2002MNRAS.334..589W,2009ApJ...693..734C}. Since a debris disk loses material over time due to the combined effects of collisional evolution, Poynting-Robertson drag and radiation pressure, and since $\zR$ is much older than Fomalhaut \citep[440 Myr;][]{2012ApJ...754L..20M}, it is reasonable to assume that the debris disk surrounding $\zR$ should not contain more than a few Earth masses. In this case, it is also reasonable to assume that the disk self-gravity and back-reaction are negligible, and the planet will still be able to put its imprint on the disk structure, if its mass is at least $0.1 \mjup \sim 32 \mathrm{M}_{\oplus}$.

\begin{table}[htbp]
\begin{center}
\begin{tabular}{l c}
\hline
\hline
 Parameters &  Values \\
\hline
 Number of particles & 150,000 \\
 Semi-major Axis (AU) & \hspace{0.25cm} $a_{\mathrm{min}}=70$ ; $a_{\mathrm{max}}=140$ \\
 Eccentricity & \hspace{0.25cm} $e_{\mathrm{min}}=0$ ; $e_{\mathrm{max}}=0.05$ \\
 Inclination $(^\circ)$ & \hspace{0.25cm} $i_{\mathrm{min}}=-3$ ; $i_{\mathrm{max}}=3$ \\
\hline  
\end{tabular}
\end{center}
\caption{Initial parameters for the planetesimals used in our N-body simulations}
\label{tab:init}
\end{table}

We consider two planet-disk configurations, both inside and outside the planetesimal belt, performing parametric explorations of their influent orbital elements.  

\subsubsection{Inner perturber}

The first case is that of a planet interior to the initial ring.
In this case, we consider that the inner edge of the disk is located at 70 AU, and make the classical assumption that it is truncated by the chaotic zone generated by coplanar planet \citep[see, e.g., the approach used by][]{2009ApJ...693..734C}. The chaotic zone is defined as where mean-motion resonances overlap, the width of this zone depending on the mass ratio between the central star and the perturber:
\begin{equation}
\frac{\Delta a}{a_{\mathrm{planet}}} = \frac{a_{\mathrm{edge}}-a_{\mathrm{planet}}}{a_{\mathrm{planet}}} = 1.5 \mu^{2/7}\qquad,
\end{equation}
where $\mu = m_{\mathrm{planet}}/m_*$ \citep{1980AJ.....85.1122W,1989Icar...82..402D}.
Consequently, one can deduce the semi-major axis of a planet of a given mass which generates a disk inner edge at 70 AU: 
\begin{equation}\label{eq:semi_maj}
a_{\mathrm{planet}} = \frac{a_{\mathrm{edge}}}{1+1.5\mu^{2/7}}\qquad.
\end{equation}

We choose to perform a parametric exploration of the mass and eccentricity of the perturber, fixing its semi-major axis to the value deduced from the formula above. The values explored are 0.1 $\mjup$, 0.5$\mjup$ and 1 $\mjup$ for the mass and 0.2, 0.4 and 0.6 for the eccentricity (see Table ~\ref{tab:results}). Here the disk initial inner edge is fixed halfway between 70 AU and the planet semi-major axis.

In addition, we also considered the case of more massive perturber such as a brown dwarf, located further inside the system. This was motivated by the observation of a suspicious point source on observations in Ks-band data from ESO archive (ID 086.C-0732(A); PI: L\"ohne,71574), which disappeared after re-reduction of the data, and consequently, since its existence is controversial, we chose not to show these images. However, if it were true, this point source was showing a $42\,\mathrm{M}_\mathrm{Jup}$ brown dwarf located at a projected distance of 17.5 AU from the star. Therefore, we investigated the possibility for the presence of a brown dwarf in the inner parts of the system.
For an inner edge to be produced at 70 AU, the planet's semi-major axis should be 43.8 AU. Since this constraint is less relevant for very eccentric inner perturbers (see Sect.~\ref{sec:results}), we choose here to set this value to the perturber's apastron rather than its semi-major axis. We fix its periastron to 17.5 AU, which leads to an orbital eccentricity of 0.43 and a semi-major axis of 30.6 AU. With such an orbit, analytical predictions indicate that the perturber should excite planetesimals eccentricities up to 0.4. Therefore, this orbit is chosen to test numerically (see Table~\ref{tab:results}).

\subsubsection{Outer perturber}

The second case considered is that of a planet exterior to the ring.
There is indeed growing evidence for planets at large orbital separations, i.e several tens to few hundreds of AU away from their host star \citep[see e.g.][]{2007ApJ...654..570L,2008Sci...322.1345K,2008Sci...322.1348M,2010Natur.468.1080M} and the mass constraints set by direct imaging are loose given the age of the system (see Sect.~\ref{sec:BD}). Therefore, we also investigate the ability of an \emph{external} perturber to shape a disk. 

We consider coplanar outer planetary companions, and explore the impact of the eccentricity, mass and periastron on the disk asymmetry.
While an inner edge is in general considered as a clue for the presence of inner perturbers, it is obviously more delicate to assume that an outer edge is formed in the same manner, since a disk instrinsically has an outer limit.
Therefore, the outer edge is not assumed here to be formed due to resonance overlap, and the planet periastron is fixed instead, to ensure that it does not cross the disk.
In order to explore a great variety of situations despite the CPU-time limitations, we consider a rough parameter space consisting in all the possible combinations between masses $m_\mathrm{p}=0.1-1-2\,\mathrm{M}_\mathrm{Jup}$, periastrons $q_\mathrm{p}=150-200-250\,$AU and eccentricities $e_\mathrm{p}=0.2-0.4-0.6$ (see Table ~\ref{tab:results}).

\begin{table}[htbp]
\caption{Summary of numerical experiments with an inner and an outer perturber, as well as for a brown dwarf and the stellar binary companion $\zeta^1\,$Reticuli. Description of the disk at 1 Gyr: I) Steady state, $e<0.2$, II) Steady state, $e>0.2$, III) Scattered disk, IV) Resonant pattern V) Spiral pattern. The example cases highlighted further in Sect.~\ref{sec:results} are labelled from A to G.}
\begin{center}
\begin{tabular}{l c c c c}
\hline
\hline
 \multicolumn{5}{c}{Inner perturbers} \\
\hline
$m_\mathrm{p}$ $(\mathrm{M}_\mathrm{Jup})$  & \hspace{0.1cm} $a_\mathrm{p}$ (AU) \hspace{0.1cm} & \hspace{0.1cm} $e_\mathrm{p}=0.2$ \hspace{0.1cm} & \hspace{0.1cm} $e_\mathrm{p}=0.4$  \hspace{0.1cm}& \hspace{0.1cm} $e_\mathrm{p}=0.6$ \hspace{0.1cm}\\
\hline
$0.1$ & $63.2$ & I & III,A & III \\
\hline
$0.5$ & $59.8$ & I & II,B & IV,C \\
\hline
$1$ & $57.9$ & I & II & IV \\
\hline
\hline  
\end{tabular}

\vspace{0.5cm}

\begin{tabular}{l c c c c}
\hline
\hline
 \multicolumn{5}{c}{Outer perturbers} \\
\hline
$m_\mathrm{p}$ $(\mathrm{M}_\mathrm{Jup})$  & \hspace{0.1cm} $q_\mathrm{p}$ (AU) \hspace{0.1cm}&\hspace{0.1cm} $e_\mathrm{p}=0.2$ \hspace{0.1cm} & \hspace{0.1cm}$e_\mathrm{p}=0.4$ \hspace{0.1cm}& \hspace{0.1cm} $e_\mathrm{p}=0.6$ \hspace{0.1cm} \\
\hline
 & $150$ & I & II,D & V \\
$0.1$ & $200$ & I & V & V,E \\
 & $250$ & V & V & V \\
\hline
 & $150$ & I & II & III,F \\
$1$ & $200$ & I & II & II \\
 & $250$ & I & I & II \\
\hline
 & $150$ & I & III,G & III \\
$2$ & $200$ & I & I & I \\
 & $250$ & I & I & II \\
\hline
\hline  
\end{tabular}

\vspace{0.5cm}

\begin{tabular}{l c c c }
\hline
\hline
 \multicolumn{4}{c}{Other perturbers} \\
\hline
Perturber & \hspace{0.5cm} $m_\mathrm{p}$ \hspace{0.5cm} & \hspace{0.2cm} Orbital Parameters \hspace{0.2cm} & \hspace{0.1cm} Result \hspace{0.1cm} \\
\hline
Brown Dwarf & $42 \mathrm{M}_\mathrm{Jup}$ & $a_\mathrm{p} = 30.6\,$AU ; $e_\mathrm{p}=0.43$ & II \\
$\zeta^1\,$Reticuli & $1 \mathrm{M}_{\odot}$ & $a_\mathrm{p} = 2046\,$AU ; $e_\mathrm{p}=0.815$ & I \\
\hline
\hline  
\end{tabular}
\end{center}
\label{tab:results}
\end{table}

Note that the perturbers are being put on an initially eccentric orbit, which requires some discussion, since we assumed that the disk is initally symmetric. Indeed, this pictures a situation where the process putting the perturber on its eccentric orbit leaves the disk unperturbed. This may seem rather unrealistic, even though some scenarios can be envisaged. For instance, an inner perturber could acquire its eccentricity via a planet-planet scattering event. However, this event should be such that a single perturber remains in the system. Otherwise, additional perturbations of a second planet would generate an orbital precession of the eccentric perturber, which in turn, could not sculpt the disk into an eccentric shape. This is compatible with observational constraints, since as previously mentionned, radial velocity measurements rule out any massive perturber in the inner system. In the case of an outer perturber, an eccentric outer binary companion may be able to generate such initial conditions (see Sect.~\ref{sec:conclusion}). 
In any case, retrieving realistic initial conditions relies on a complete study of the perturbations induced on the disk for multiple scenarios and, most probably, an extensive parameter space exploration. This study goes beyond the scope of the present paper, and shall be the subject of future work, which motivates our choice of simple initial conditions.

\section{Numerical Investigation: Results}\label{sec:results}
We present here results obtained for both disk-planet set-ups we have considered: an inner and an outer planet, as well as for the case the perturber is the stellar companion $\zeta^1\,$ Ret.
For each case, we will try to find the one that gives an eccentric disk compatible with observational constraints.

\citet{2010A&A...518L.131E} gives a lower limit for the eccentricity of the disk in $\zR$ of 0.3. Therefore, given the uncertainties in the estimation of the disk global eccentricity we compute from our simulations,  a disk global eccentricity lower than 0.2 is discarded in our analysis.
This global eccentricity will be evaluated considering the geometry of an ellipse: the offset $\delta$ of the centre of symmetry from one of the focii of an ellipse is simply the product of its semi-major axis $a$ by its eccentricity $e$, i.e., $\delta=ae$. For a disk from our simulations, $\delta$ can be obtained by calculating the centre of symmetry of the disk, using the positions of the test particles in the heliocentric frame: $\delta$ is the distance of this centre of symmetry to the star. The disk semi-major axis $a$ is determined as follows: the disk is divided into superimposed angular sectors of $3^{\circ}$. For each of these sectors, the radial distribution of the particles is fitted to a Gaussian. This gives us the radial position of the maximum density for each angular sector, and thus a set of points defining the global shape of the disk. It is then straightforward to retrieve $a$ from this set of points by seeking for the major axis, i.e., the maximum distance between two opposite points. Finally, the disk eccentricity is simply $e=\delta/a$.

\subsection{Inner perturber}

We choose 4 illustrative results (see Table ~\ref{tab:results}). For each, we show pole-on projections of the system at 1 Gyr on Fig.~\ref{fig:inner_upperviews}. We also summarize the outputs of our simulations in Table ~\ref{tab:results}.

Best candidates should have a significant orbital eccentricity of $\sim 0.4$. The example of Case B is shown on Fig.~\ref{fig:inner_upperviews} (upper right panel). However, scattering processes may be important, and their study allows us to place constraints on the mass of the perturber.

\subsubsection{Scattered disks}

Inner perturbers may lead to very significant scattering processes. Namely, they can fill the inner parts of the disk instead of producing a well-defined ring. 
Such effects are presented with Case A on Fig ~\ref{fig:inner_upperviews} (upper left panel).

This is an effect appearing in the presence of rather low-mass perturbers. As a matter of fact, such perturbers do not scatter material efficiently enough. This material tends then to populate the inner parts of the system.

As a consequence, there is a lower mass limit for inner perturbers, and in the specific case of $\zR$  system, this lower limit is between 0.1 and 0.5\,$\mjup$.
However, one cannot exclude that another more massive planet produces scattering of the material, blowing it out and leaving an inner hole, so a more correct way to express this constraint would be that perturbers with masses as low as 0.1\,$\mjup$ should be accompanied by another more massive planet to create such a pattern. But this scenario presents difficulties: while this second planet must be massive enough to clear the inner parts of the system of its material, it also must have a limited dynamical effect on the first planet that sculpts the disk: this second planet must be distant and not too massive for the orbit of the first planet to remained unperturbed. Otherwise, this orbit would precess and no longer lead to an eccentric pattern. This is not the purpose of this paper to investigate this scenario, but based on the previous arguments, it would most probably work in a very limited parameter space.

\subsubsection{Resonant patterns}

\begin{figure}
\resizebox{\hsize}{!}{\includegraphics[width=0.7\textwidth,angle=-90]{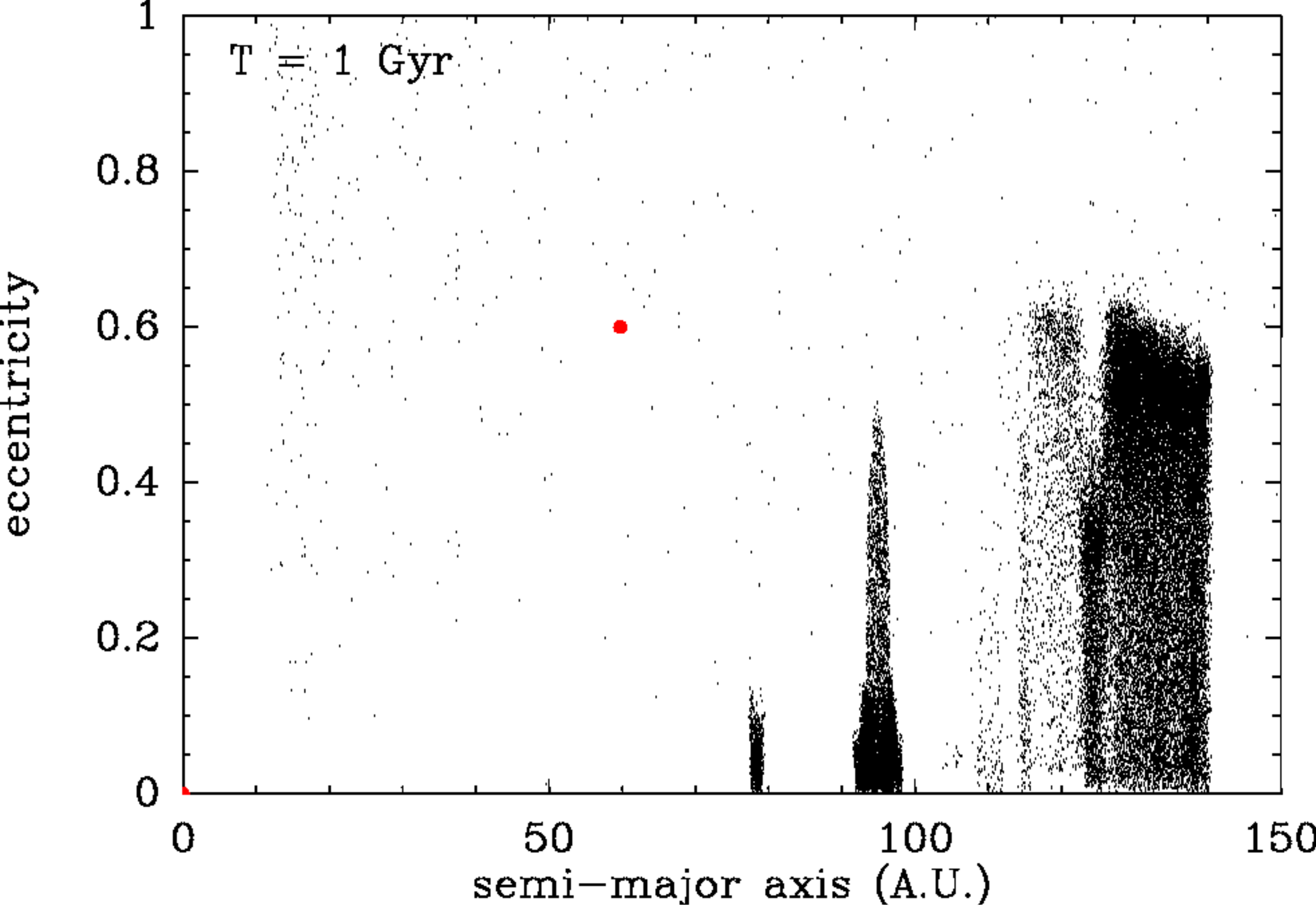}}
\caption[]{Semi-major axis versus eccentricity diagram of the disk at 1 Gyr for the Case C perturber. Overdensities of planetesimals at $\sim 79$ and $\sim 95$ AU correspond to planetesimals respectively in 3:2 and 2:1 mean-motion resonance with the perturber.}
\label{fig:res}
\end{figure}

\begin{figure*}
\makebox[\textwidth]{\includegraphics[width=0.325\textwidth,angle=-90]{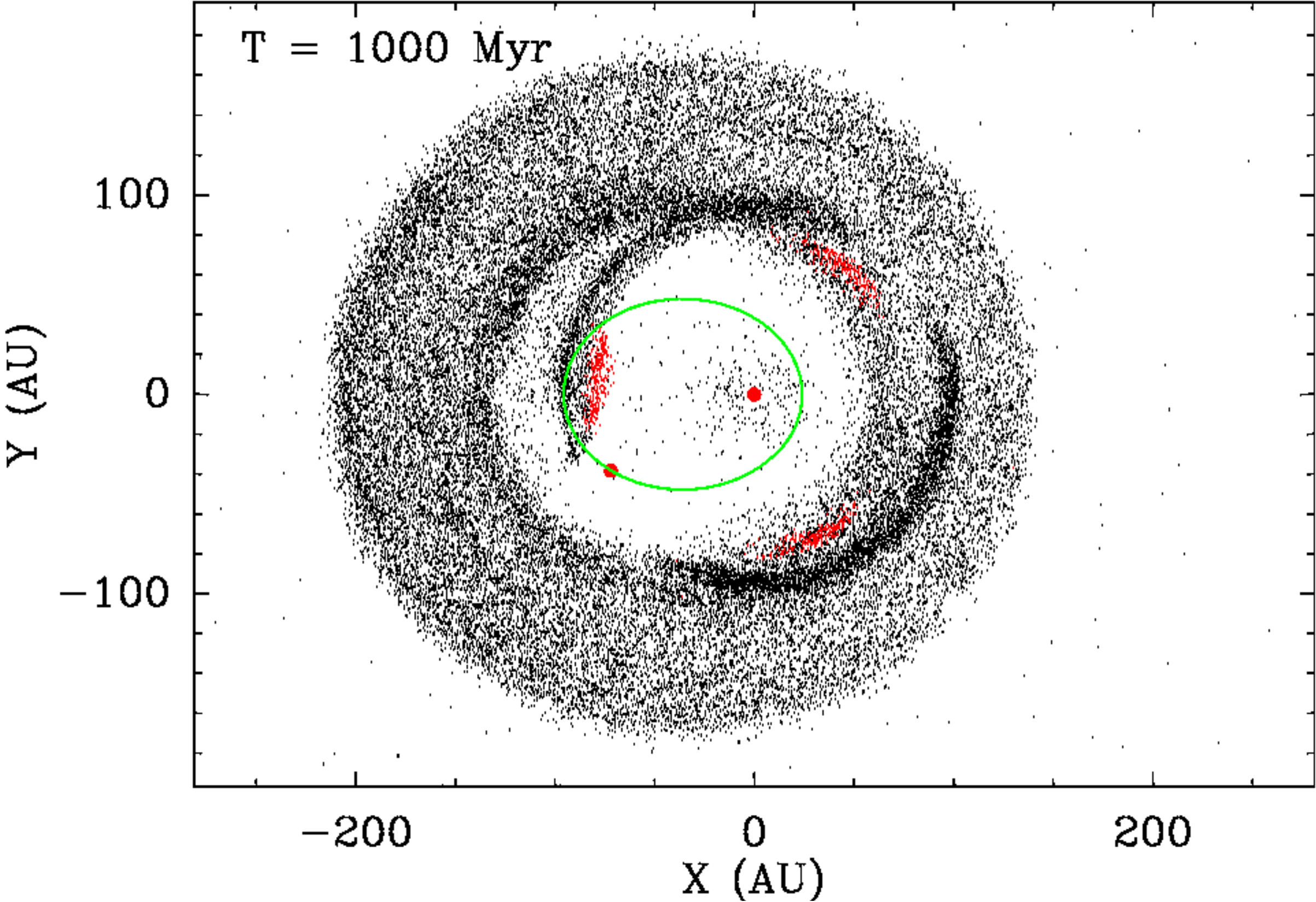}
\hspace{0.75cm}
\includegraphics[width=0.325\textwidth,angle=-90]{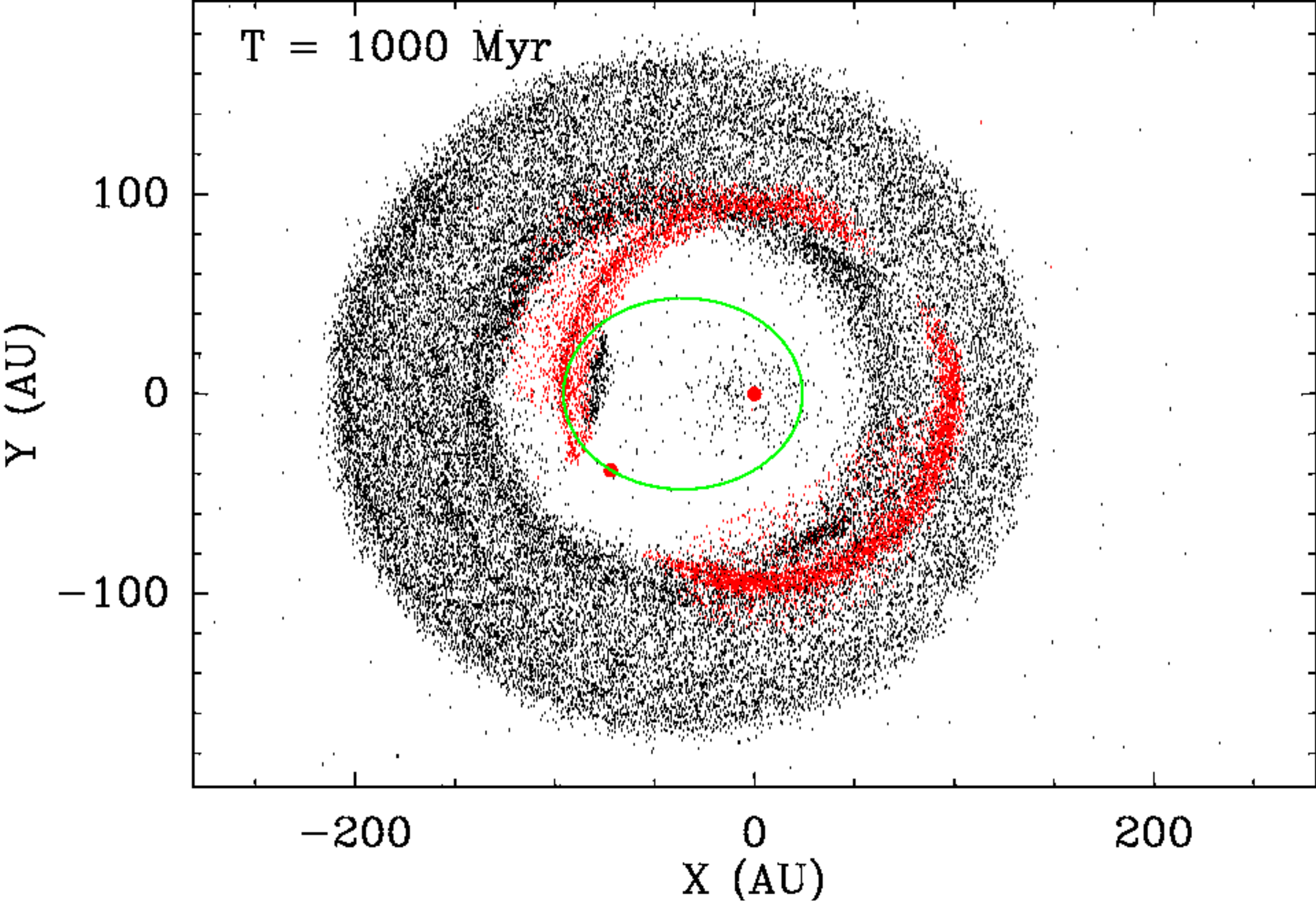}}
\caption[]{Views from above the plane of disks at 1 Gyr in Case C, i.e., where resonant patterns appear. Planetesimals in 3:2 \emph{(left)} and 2:1 \emph{(right)} mean-motion resonance with the perturber are put in evidence in red.}
\label{fig:res1}
\end{figure*}

It is notable that in the case of very eccentric ($e_\mathrm{p}=0.6$) inner perturbers of mass between  0.5 and $1\,\mjup$, resonant clumpy structures may arise. An example is Case C, shown on Fig ~\ref{fig:inner_upperviews} (bottom left panel). In Fig.~\ref{fig:res}, we show a semi-major axis vs eccentricity diagram of the disk: it reveals two populations in resonance with the planet, namely the 3:2 and 2:1 mean-motion resonances, at $\sim 79\,$AU and $\sim 95\,$AU, respectively. These resonant populations are put in evidence in Fig.~\ref{fig:res1}.
Usually, the presence of a population in 3:2 resonance is associated with capture during outward planetary migration, as it is the case in our own Solar System \citep{1993Natur.365..819M,1995AJ....110..420M}. But here, the appearance of these structures is most probably due to the fact that we used the chaotic zone formula (Eq.~\ref{eq:semi_maj}) to determine the perturber semi-major axis, which was derived for perturbers on circular orbits. Therefore, we should expect that this formula works less efficiently with increasing orbital eccentricity of the perturber: the result is that the planet digs into the disk, and consequently, planetesimals unprotected against close encounters by mean-motion resonance are scattered out, leaving the resonant structures apparent. This is supported by the fact that these resonant structures disappear if the constraint given by the chaotic zone formula is applied to the perturber's apastron instead of its semi-major axis, as was done in the case of an inner brown dwarf-type companion. 
Interestingly, the observation of such resonant structures in a system may provide other clues on the dynamical history of a perturber than an outward migration: it could mean that the planet was originally shaping the inner edge of the disk before being put on an eccentric orbit.
However, these are thin structures, and if the disk is seen close to edge-on, as is the case for $\zR$, these would most probably be hidden by the non resonant population.

\subsubsection{Brown dwarf}\label{sec:BD}

\begin{figure*}
\makebox[\textwidth]{\includegraphics[width=0.45\textwidth,height=0.4\textwidth]{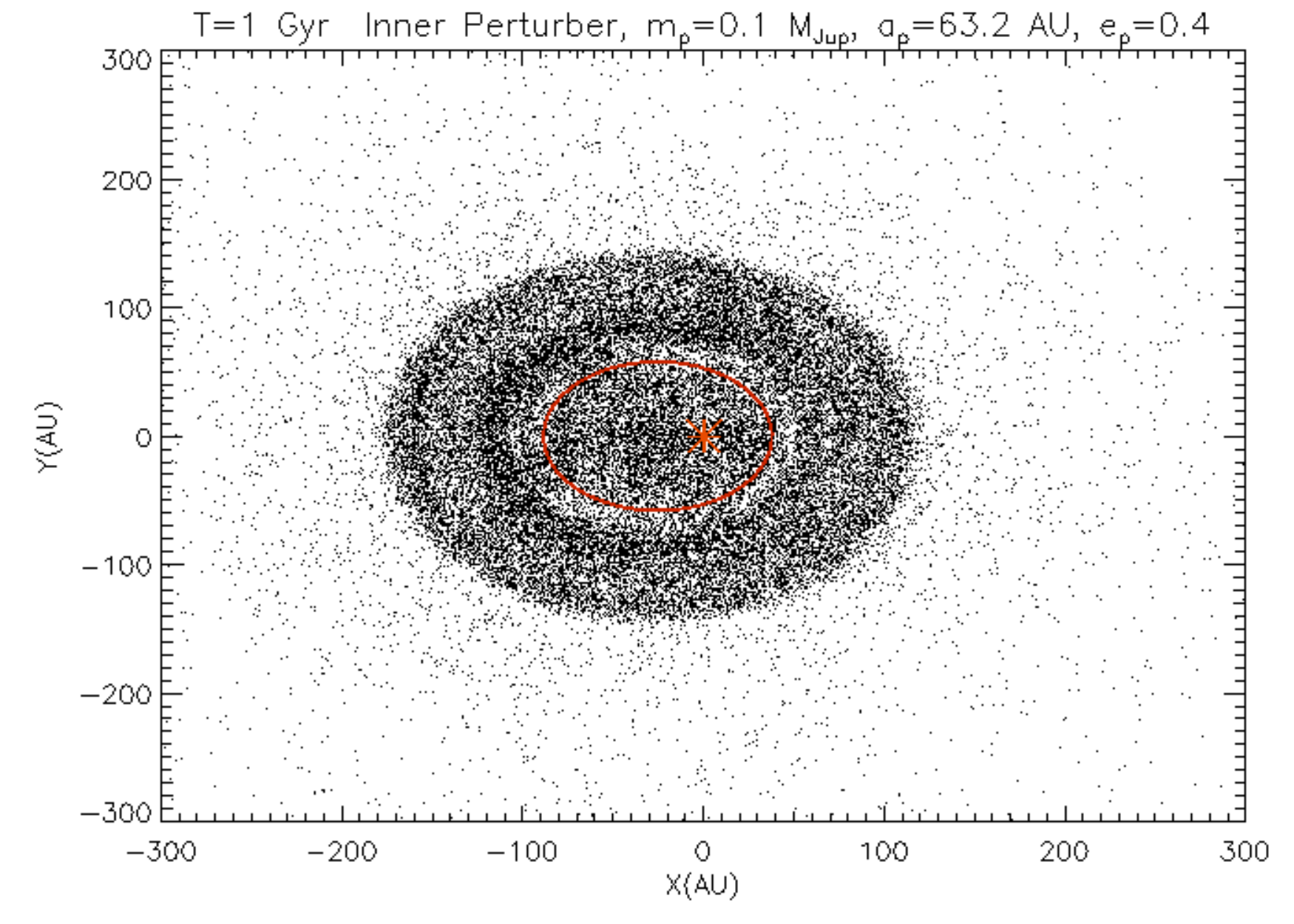}
\hspace{2cm}
\includegraphics[width=0.45\textwidth,height=0.4\textwidth]{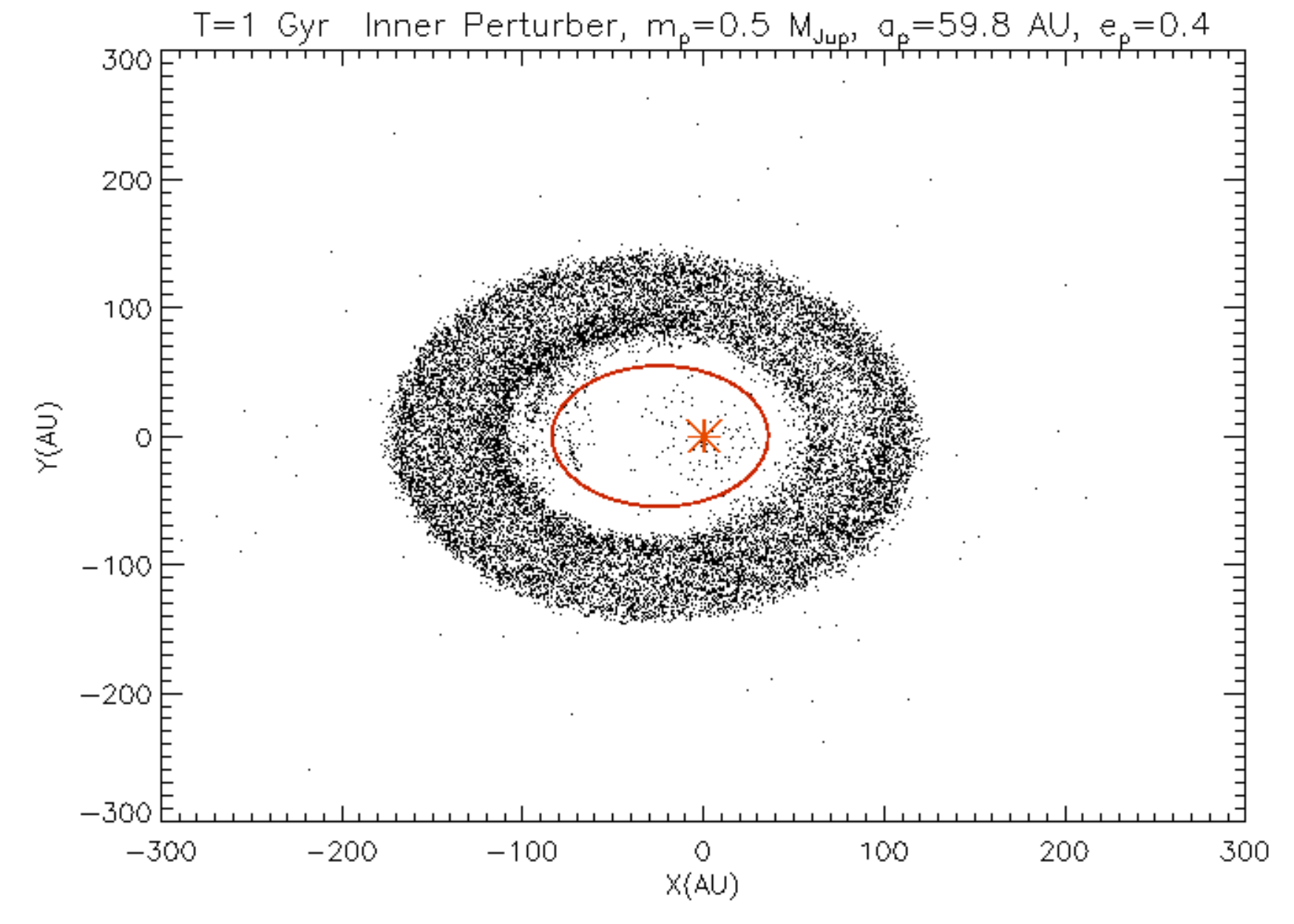}}
\makebox[\textwidth]{\includegraphics[width=0.45\textwidth,height=0.4\textwidth]{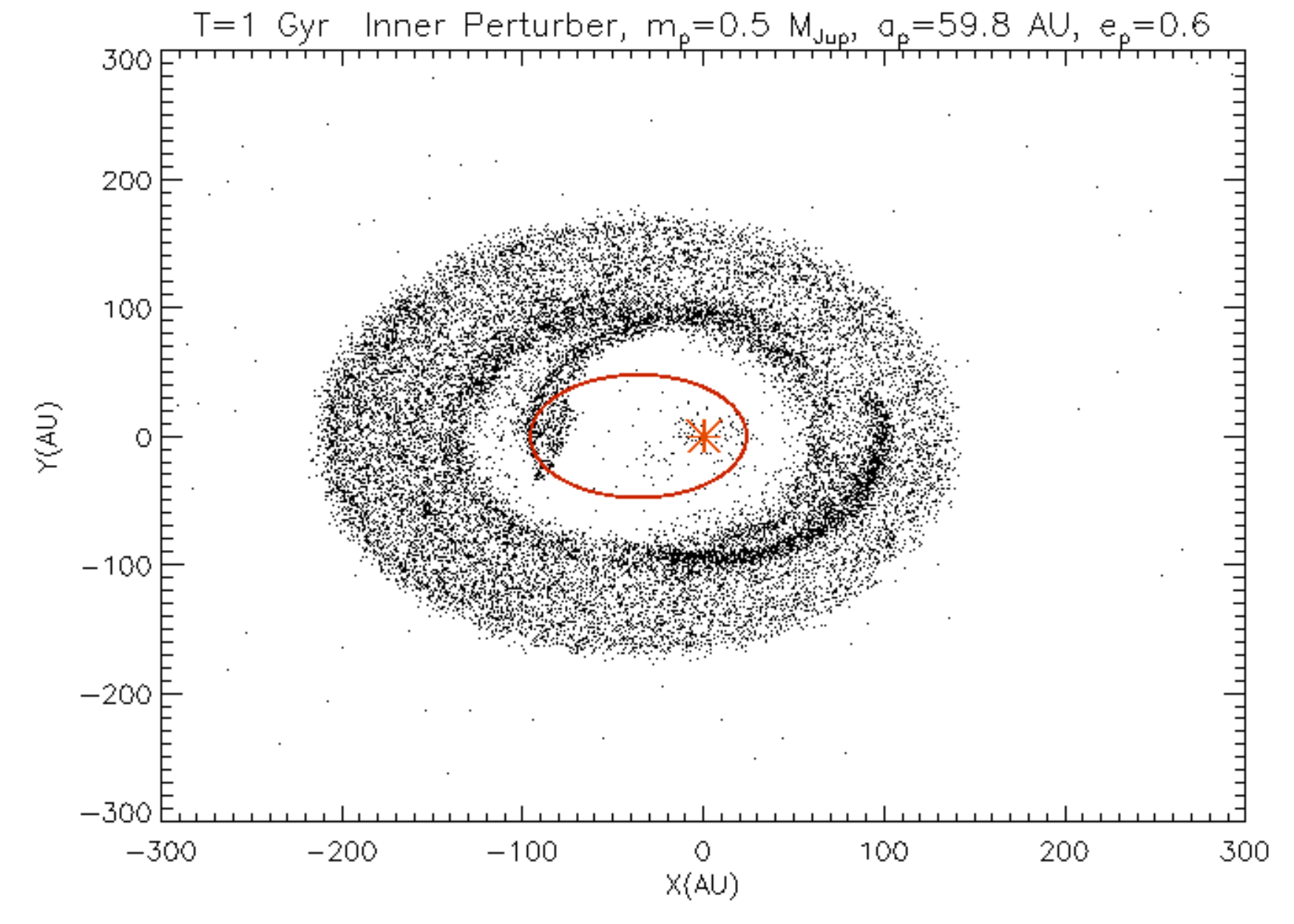}
\hspace{2cm}
\includegraphics[width=0.45\textwidth,height=0.4\textwidth]{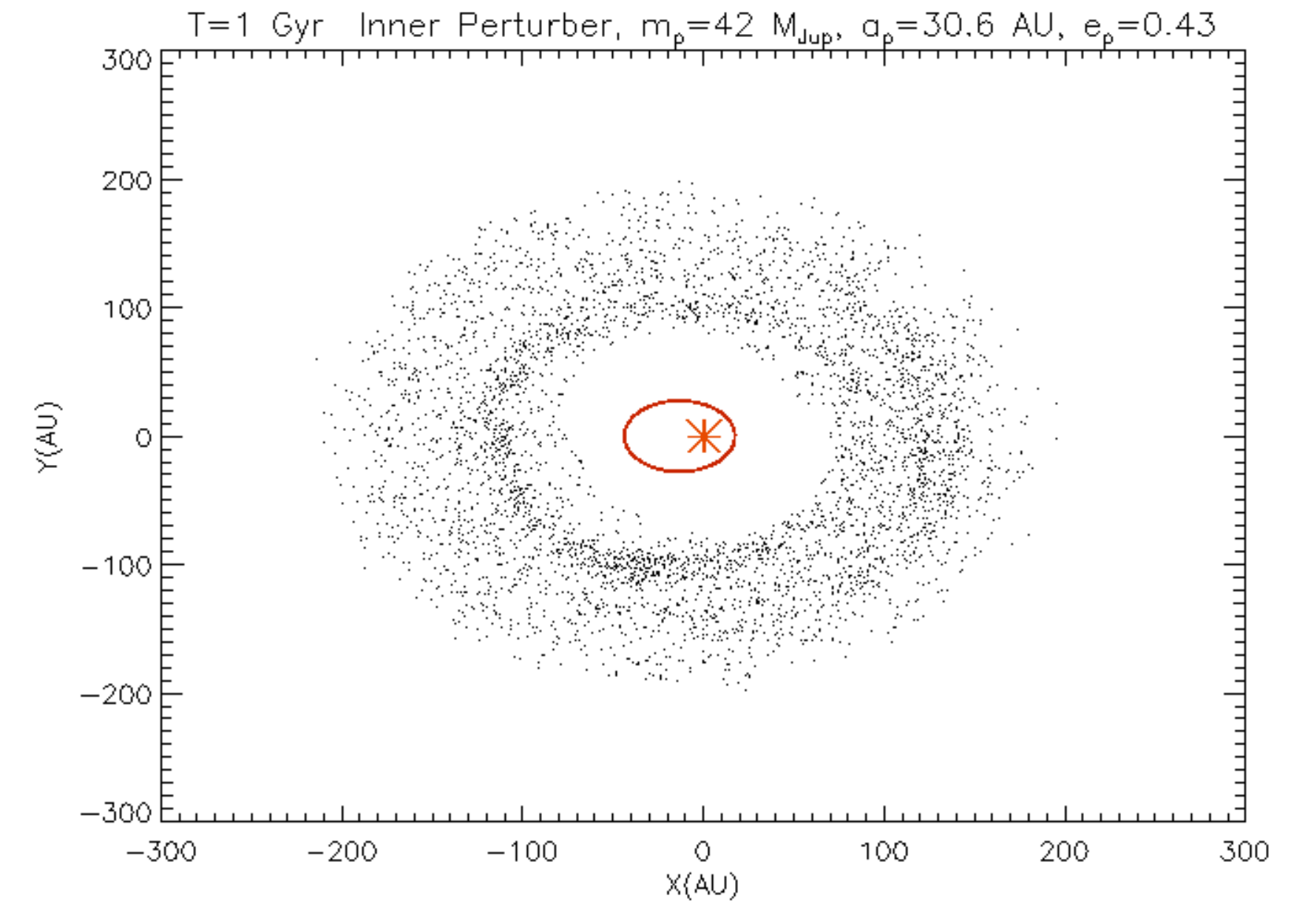}}
\caption[]{Example views from above the plane of disks at 1 Gyr, under the influence of inner perturbers.\textbf{TOP} \emph{Left:} the perturber scatters material in the inner parts of the system (Case A). \emph{Right:} one of the best candidates (Case B).  \textbf{BOTTOM} \emph{Left:} resonant patterns appear (Case C). \emph{Right:} a brown-dwarf perturber can be a good candidate.}
\label{fig:inner_upperviews}
\end{figure*}

Additionally, we investigated the possibility for the presence of a brown dwarf in the inner parts of the system, on an orbit such that the perturber should excite planetesimals eccentricities up to 0.4. 
The disk at 1 Gyr is shown on Fig ~\ref{fig:inner_upperviews} (bottom right panel). 
Its global eccentricity is $\sim 0.2-0.25$, which shows that very massive perturbers in the inner parts of the system can create the wanted pattern.

\subsection{Outer perturber}

We choose 4 illustrative results (see Table ~\ref{tab:results}), and for each, show pole-on projections of the system at 1 Gyr on Fig.~\ref{fig:outer_upperviews}, along with the evolution of the disk offset on 1 Gyr on Fig.~\ref{fig:outer_offsets}. The outputs of our simulations are summarized in Table ~\ref{tab:results}.

\subsubsection{Good candidates}

Case D is shown on Fig.~\ref{fig:outer_upperviews} (upper left panel). The corresponding evolution of the offset clearly shows that the disk is at steady state and significantly eccentric (see Fig.~\ref{fig:outer_offsets}).
The best candidates must have significant eccentricities: none of our 0.2 eccentric perturbers manages to create the desired eccentric structure, even in the limit case predicted analytically where a $0.1\, \mjup$ perturber has a pericentre $q_\mathrm{p}=150\,$AU and eccentricity $e_\mathrm{p}=0.2$ (see Fig. ~\ref{fig:precess}).
As a consequence, the best outer candidates have eccentricities $\sim 0.4-0.6$.

\subsubsection{Spiral patterns}

Case E is shown on Fig.~\ref{fig:outer_upperviews} (upper right panel). The system shows a spiral pattern at 1 Gyr, which, according to analytical predictions, corresponds to one precession timescale (see Fig.~\ref{fig:precess}).
Thus, our \textit{N}-body integrations confirm what was noted by  \citet{2005A&A...440..937W}, i.e., the analytical formula appears to be a lower limit and several precession timescales are sometimes necessary for spiral patterns to vanish. The effect of spirals can also be seen on the evolution of the disk offset, which undergoes oscillations (see Fig. ~\ref{fig:outer_offsets}).

If we had observational proof that the disk of $\zR$ has reached a steady-state and contains no spiral pattern, then the results of our numerical investigation would allow us to place a lower mass limit of 0.1\,$\mjup$ on an outer perturber in a range of periastrons from 150 to 250 AU, based on dynamical timescales criterion, otherwise it takes longer than 1 Gyr to generate a steady-state eccentric disk.
This limit still holds for larger periastrons than the range explored, since dynamical timescales increase with distance.
But here, since the slightly edge-on orientation of the disk and the resolution of the \textit{Herschel}/PACS images, it is extremely difficult to discard  the presence of any spiral pattern in this disk, and no lower mass limit can be clearly put on an outer perturber.

\subsubsection{Scattered disks}

\begin{figure*}
\makebox[\textwidth]{\includegraphics[width=0.45\textwidth,height=0.4\textwidth]{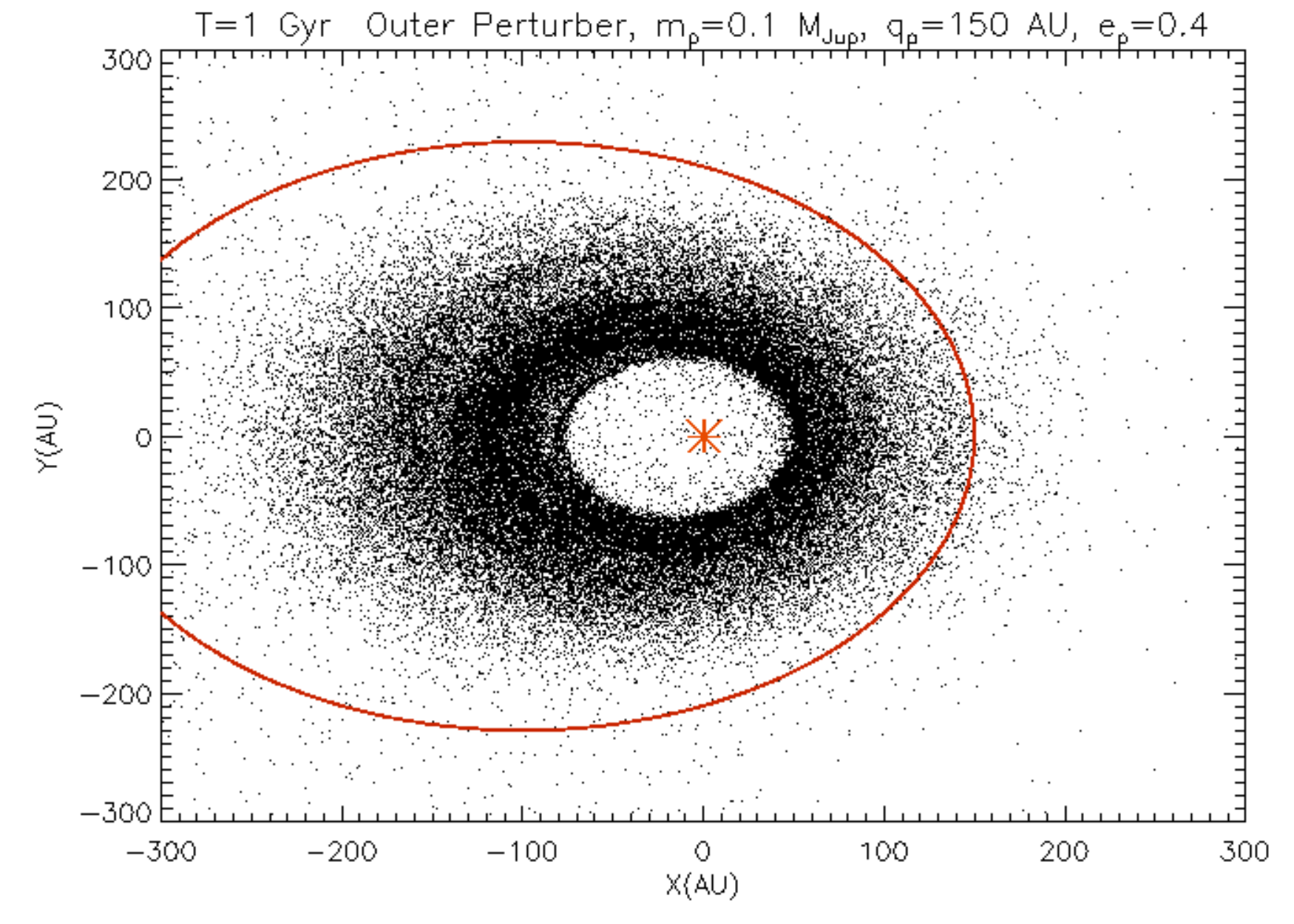}
\hspace{2cm}
\includegraphics[width=0.45\textwidth,height=0.4\textwidth]{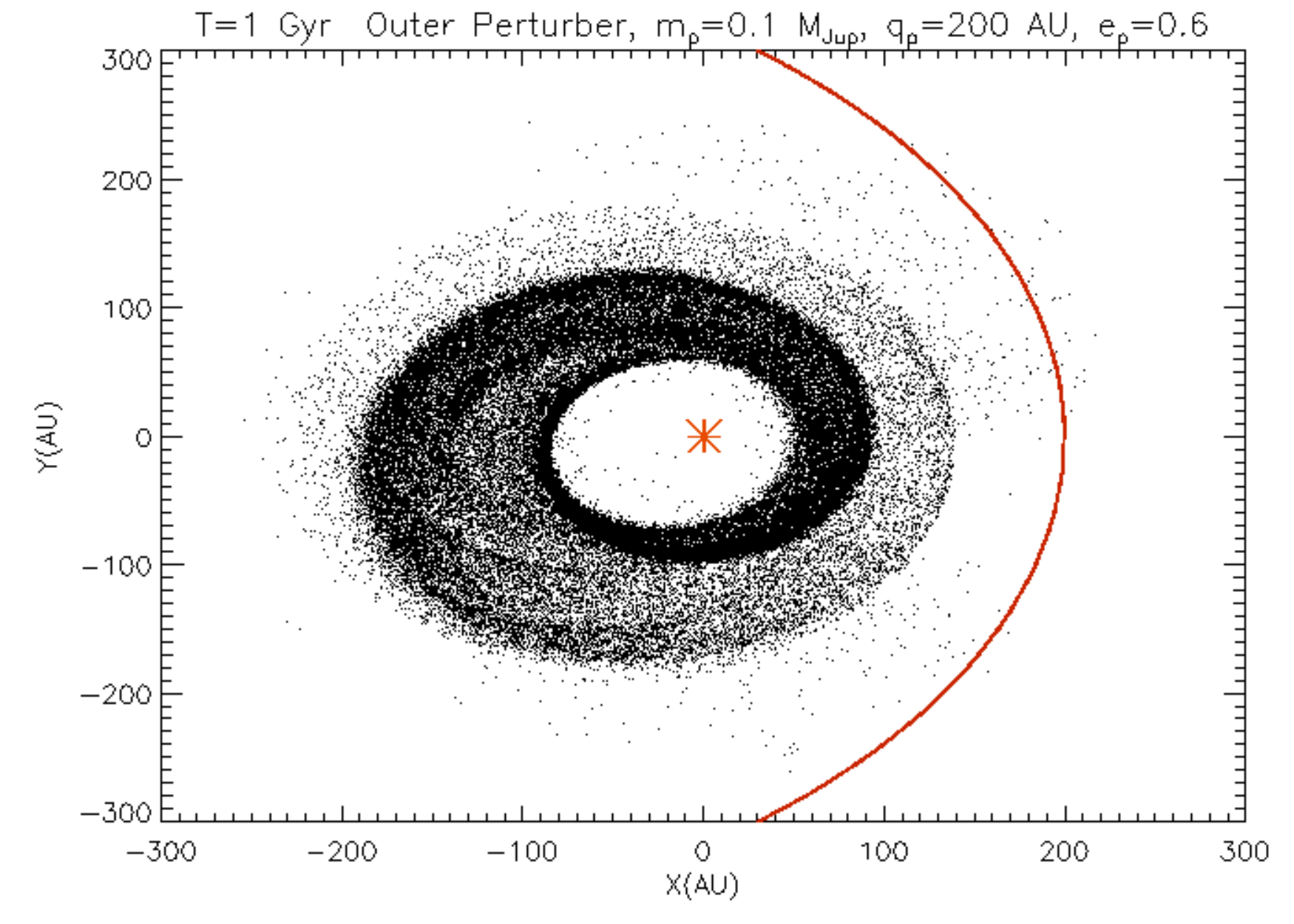}}
\makebox[\textwidth]{\includegraphics[width=0.45\textwidth,height=0.4\textwidth]{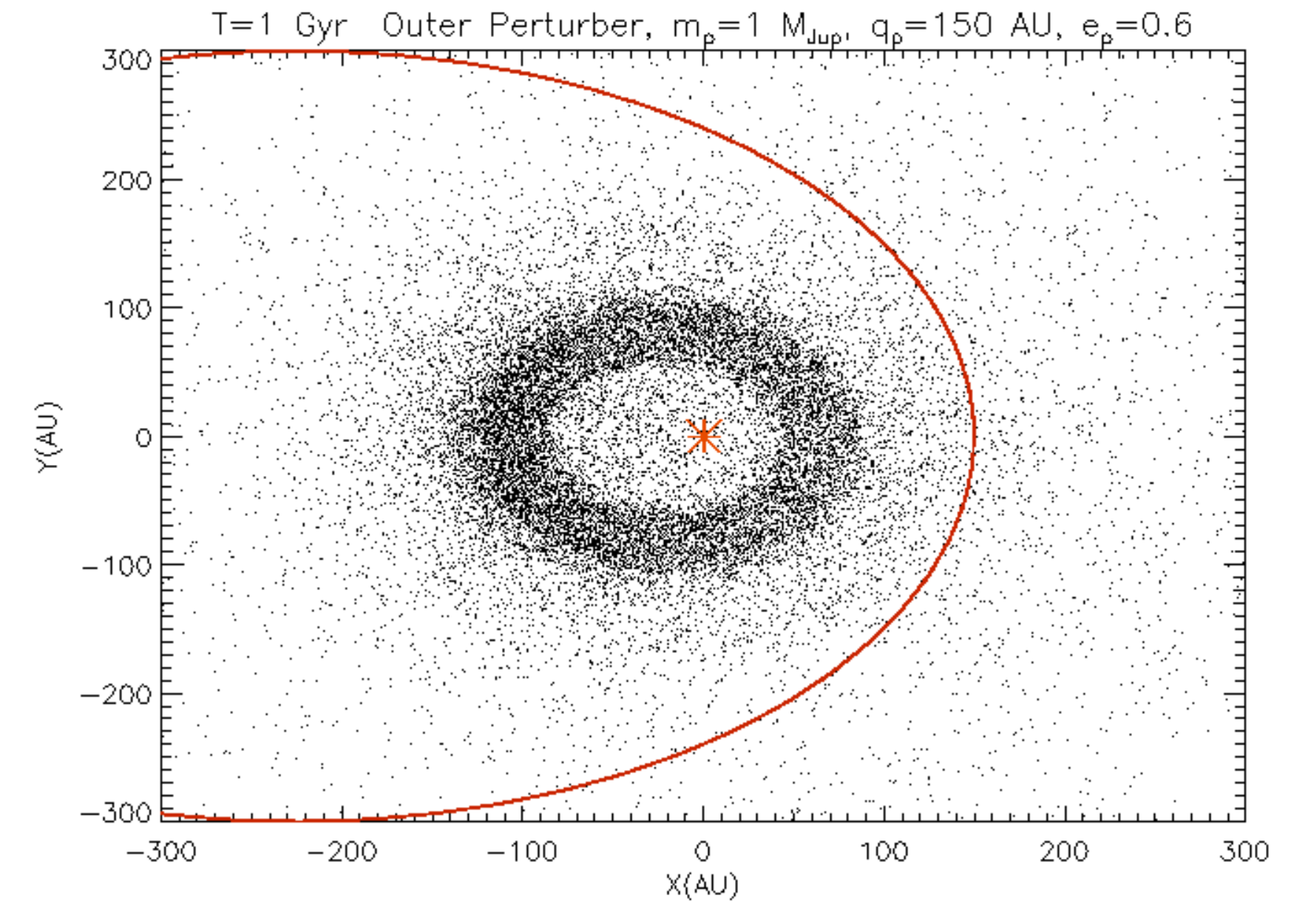}
\hspace{2cm}
\includegraphics[width=0.45\textwidth,height=0.4\textwidth]{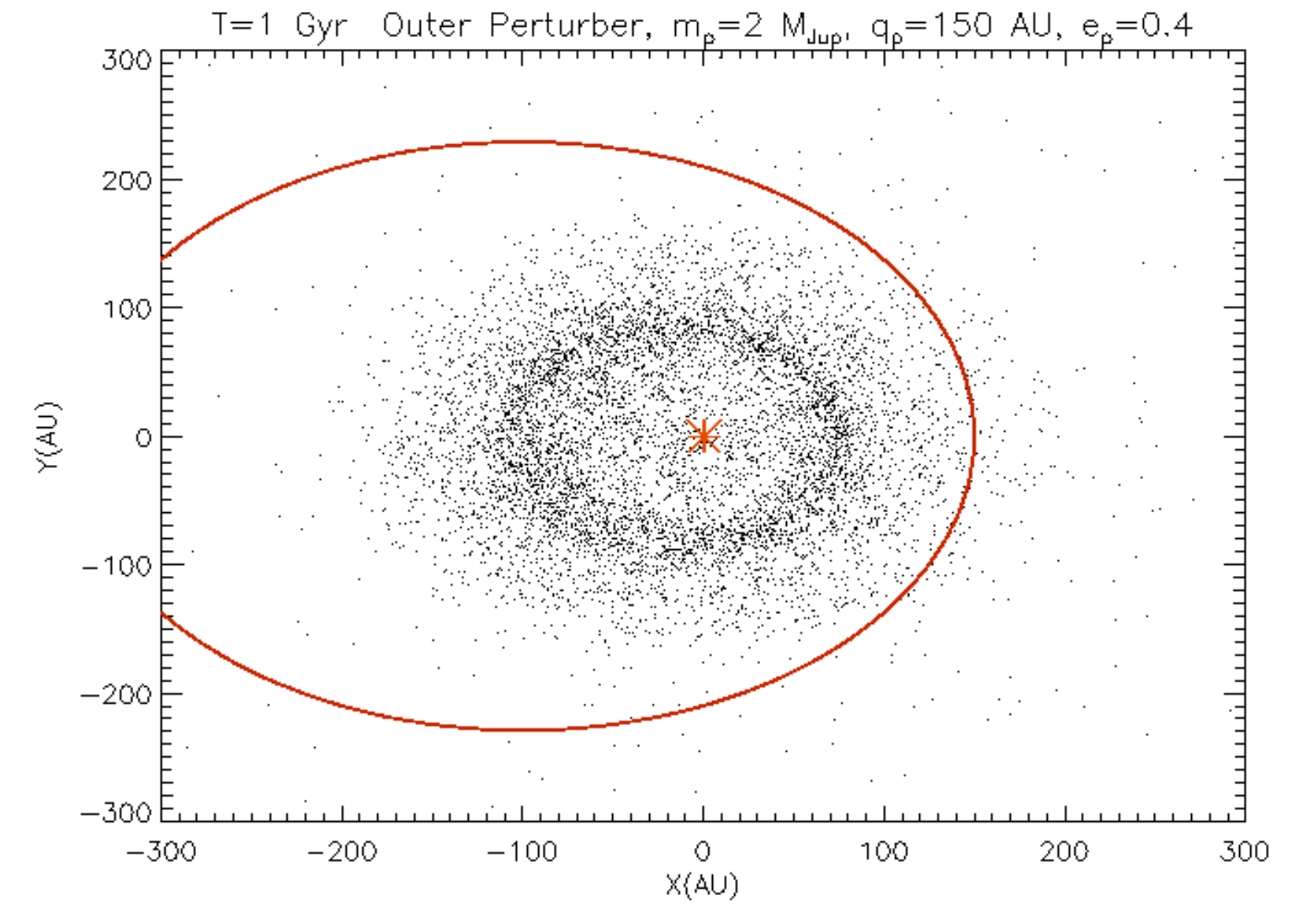}}
\caption[]{Example views from above the disk plane at 1 Gyr, under the influence of outer perturbers. \textbf{TOP} \emph{Left:} one of the best candidates (Case D). \emph{Right:} spirals are still apparent (Case E). \textbf{BOTTOM} \emph{Left:} one might question the contribution of the scattered material to the disk emission (Case F). \emph{Right:} scattering processes tend to destroy the structure (Case G).}
\label{fig:outer_upperviews}
\end{figure*}

\begin{figure*}
\makebox[\textwidth]{\includegraphics[width=0.45\textwidth,height=0.3\textwidth]{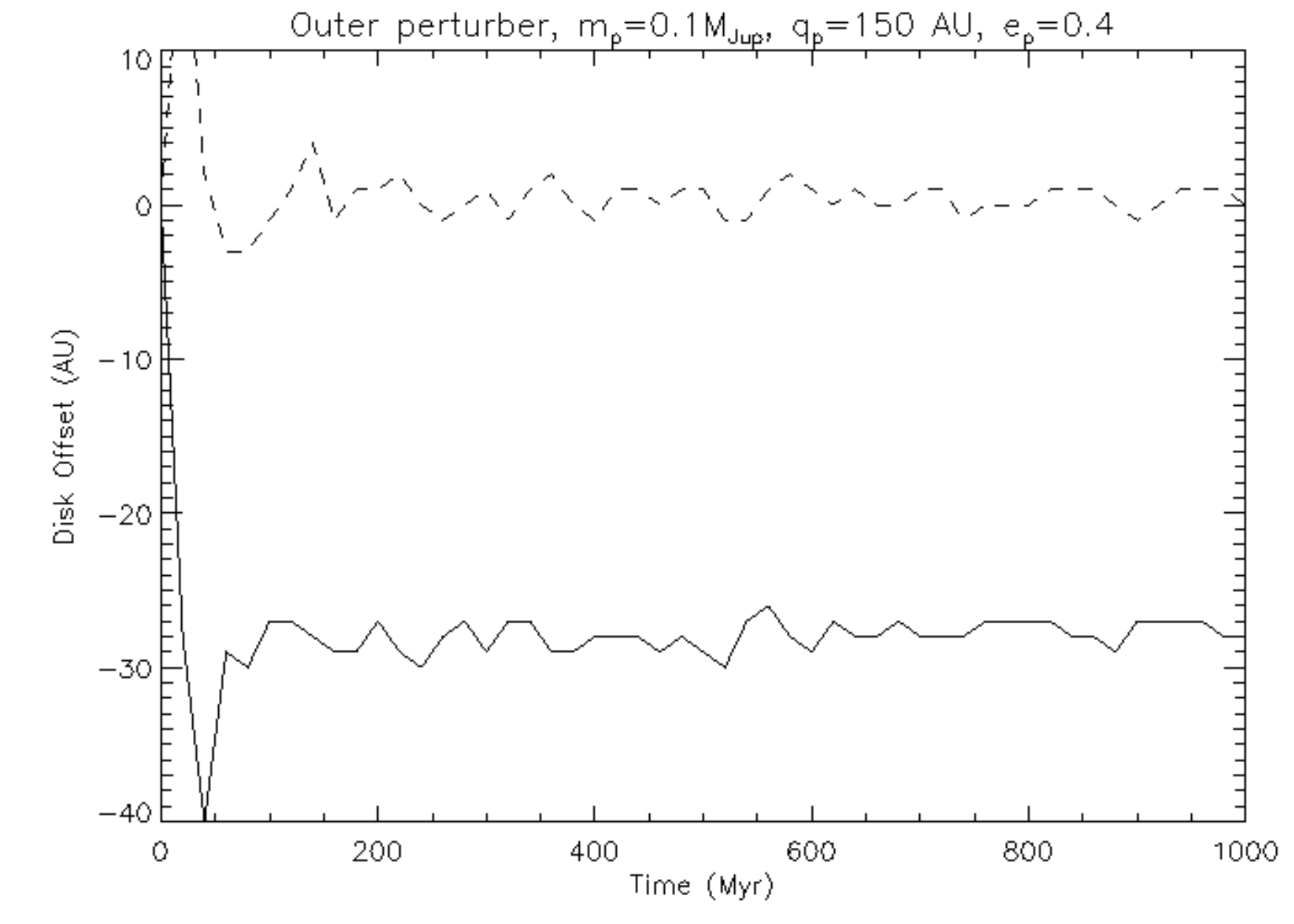}
\hspace{2cm}
\includegraphics[width=0.45\textwidth,height=0.3\textwidth]{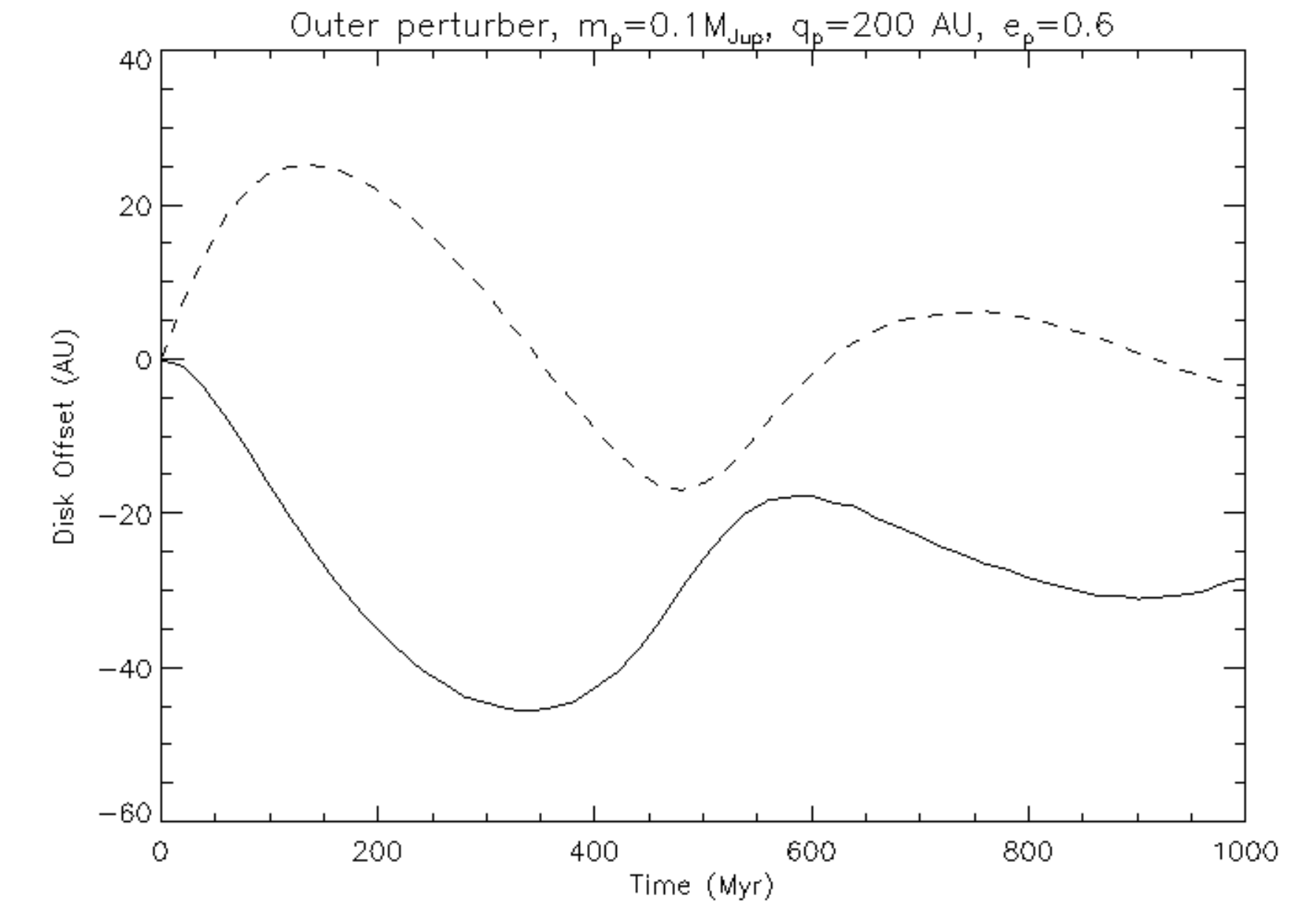}}
\makebox[\textwidth]{\includegraphics[width=0.45\textwidth,height=0.3\textwidth]{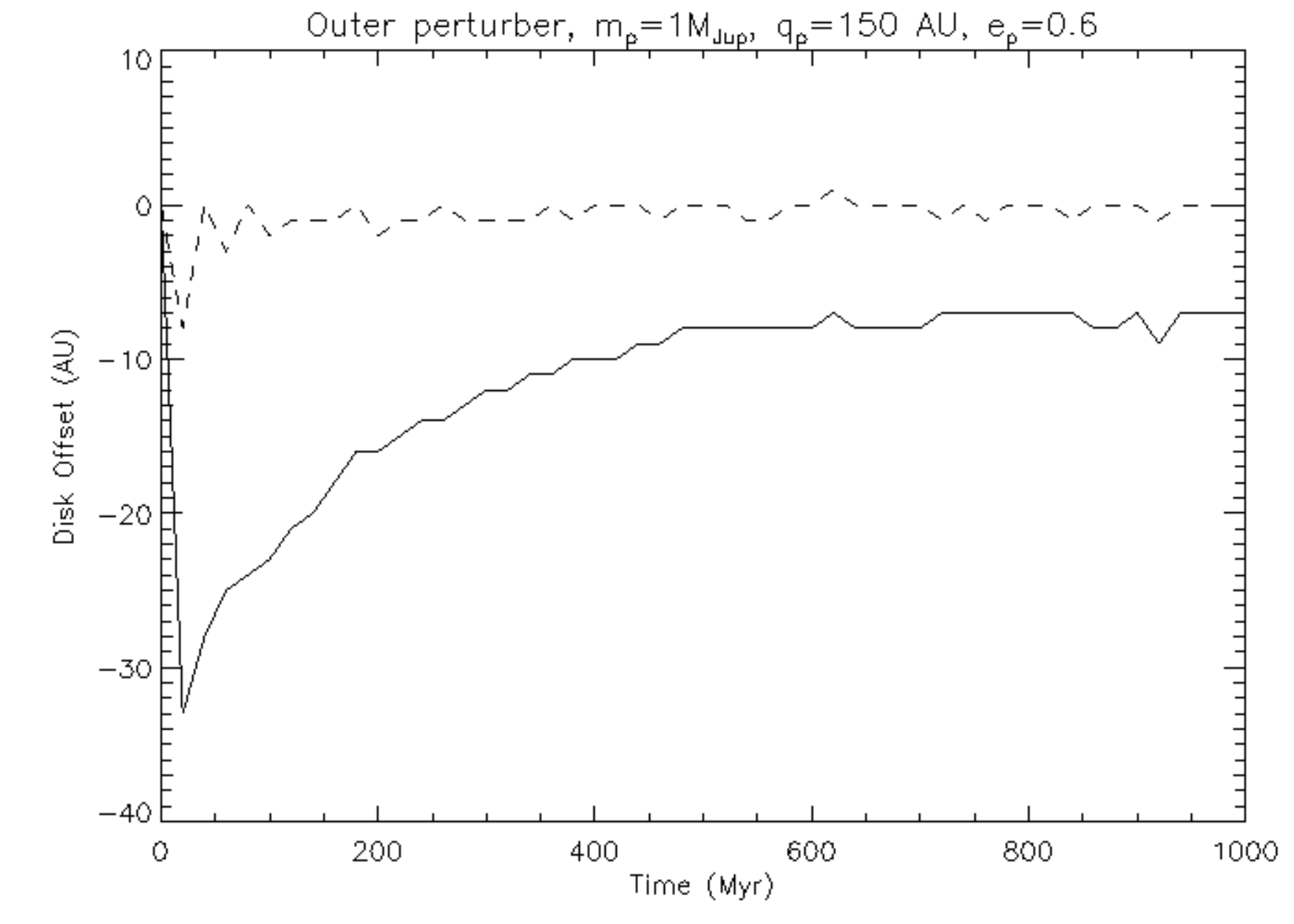}
\hspace{2cm}
\includegraphics[width=0.45\textwidth,height=0.3\textwidth]{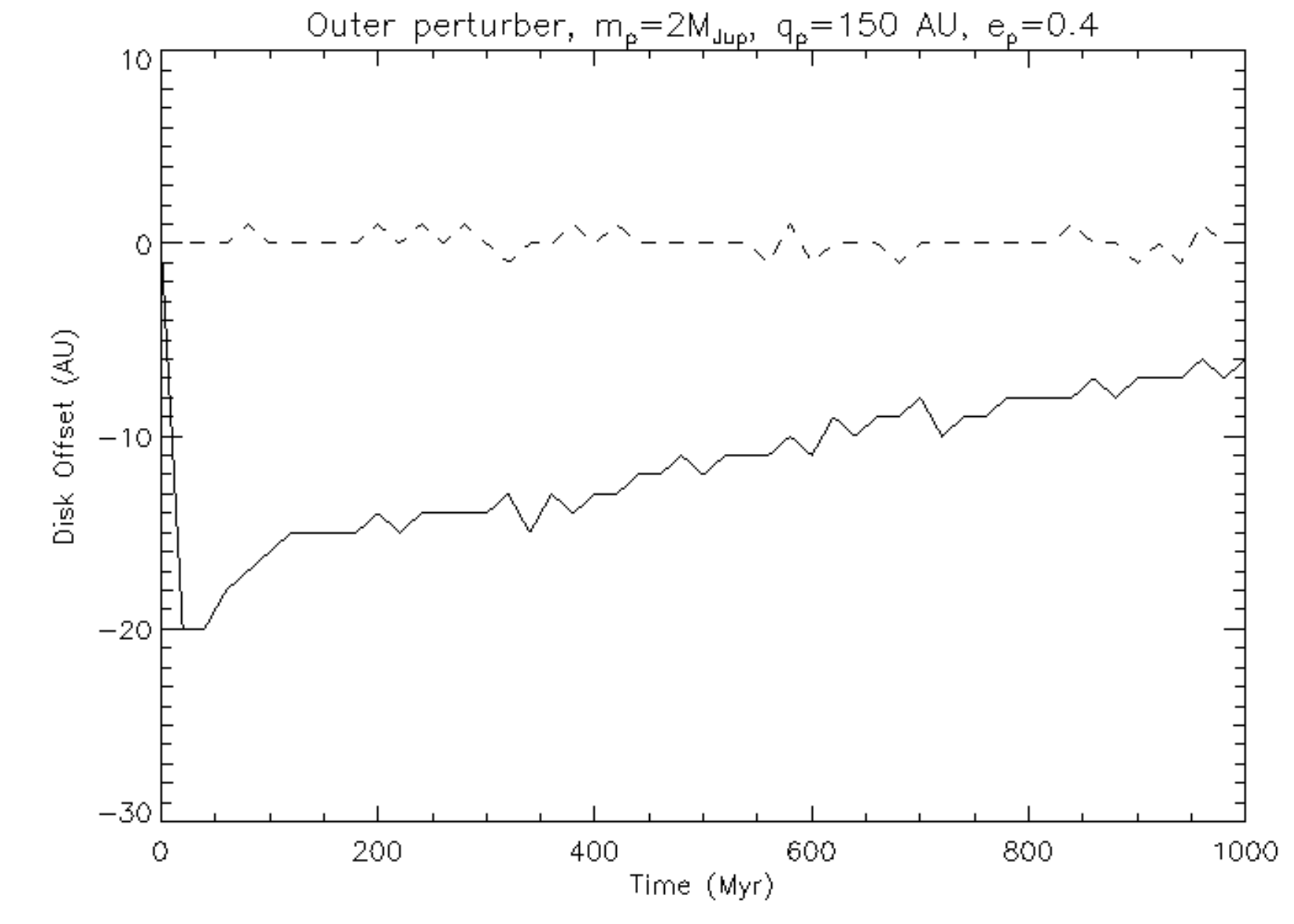}}
\caption[]{Example of the time evolution on 1 Gyr of the disk offset coordinates (X \emph{(solid line)} and Y \emph{(dashed line)}) for outer perturbers. \textbf{TOP} \emph{Left:} one of the best candidates (Case D). Since the orientation of the perturber orbit, i.e., its semi-major axis is along the X-axis with positive periastron, this corresponds, for a disk located at $a_\mathrm{c}\sim 100\,$AU, to a negative X-offset $\Delta X \sim - 30\,$AU and a zero Y-offset. \emph{Right:} the oscillations of the offset coordinates reveal the spiral-winding regime (Case E). \textbf{BOTTOM} The offset seems to relax. \emph{Left:} Case F.  \emph{Right:} Case G.}
\label{fig:outer_offsets}
\end{figure*}

Outer perturbers may lead to very significant scattering processes. We present such effects in Fig ~\ref{fig:outer_upperviews} (lower panels), where Case F and Case G are considered.

One can see that these processes are even more significant when the mass of the perturber increases. Indeed, very logically, more massive perturbers tend to scatter small bodies more efficiently. The distance to the disk plays a major part in this effect too, since close perturbers also tend to scatter more material. Additionally, when a perturber is on an eccentric orbit, it approaches even closer to the disk. In the most dramatic cases, the disk is completely destroyed.
Consequently, for a given distance to the disk, there is an upper limit to the mass of an outer companion. 

For the Case F, one might question the contribution of the scattered inner material to the emission of the disk, and whether it would be visible on resolved images. In this case, the potential visibility of material on real observations rely on the sensitivity and resolution of the instrument used for these observations, as well as on the distance of the object, the radiative properties of the material itself and the quantity of light it receives, i.e. on the host star properties. Therefore, only the production of synthetic images for direct comparison with observations can reveal whether this material is apparent or not, and refine constraints on the potential perturbers at work in the system.
Additionally, the evolution of the offset clearly suggests that the asymmetry relaxes asymptotically to a small but non-zero value, and indeed, the apparent ring structure appears to show little eccentricity.

More specifically concerning the $\zR$ system, our results allow us to place an upper mass limit of 2\,$\mjup$ on an outer perturber in a range of periastrons from 150 to 250 AU. However, one should note that this upper mass limit is expected to increase for periastrons greater than those explored, since scattering processes are expected to be less efficient for a given if the companion is further from the disk.

\subsection{Stellar binary companion}

We investigate here the influence of the binary companion
$\mathrm{\zeta^1 Reticuli}$ on the debris disk surrounding $\zR$. 
The aim is to determine whether it alone could generate 
the observed asymmetry. If this is to be the case, the companion must be on an eccentric orbit. 
The only observational constraint available so
far on the relative orbit of the binary system is the projected
distance between the two stars, namely 3713 AU
\citep{2001AJ....122.3466M}. Consequently, we cannot exclude the possibility that its
orbit is eccentric, with a present-day location near apastron.
We investigate whether a binary companion on an
eccentric orbit at such a distance could account alone for the observed
structure of the disk, i.e., an elliptic ring of minimum global
eccentricity 0.3, without any constraint on the binary eccentricity.
Both analytical and numerical methods are used.

We first examine the forced secular eccentricity applied 
to the planetesimals by an eccentric binary companion, which is assumed to be coplanar.
The debris disk surrounding $\zR$ is approximately centered on 
$a=100\,$AU, and the binary perturber is at 3713\,AU 
from $\zR$. This is only a projected distance, not a semi-major axis.
Consider that the binary companion is currently located at distance $r_\star$ from
$\zR$. The equation of its orbit around $\zR$ reads:
\begin{equation}
r_\star=\frac{a_\star(1-e_\star^2)}{1+e_\star\cos v_\star}\qquad,
\end{equation}
where $v_\star$, $a_\star$ and $e_\star$ are the binary companion current true anomaly, semi-major axis, and orbital eccentricity, respectively. The observed distance 
$d=3713\,$AU is related to $r_\star$ by $d=r_\star\cos\psi$, where $\psi$ is a 
projection angle. 
This gives: 
\begin{equation}\label{eq:astar}
a_\star=\frac{d(1+e_\star\cos v_\star)}{(1-e_\star^2)\cos\psi}\qquad.
\end{equation}
Now, from this result Eq.~\ref{eq:astar} and Eq.~\ref{eq:forced2}, with $a_\mathrm{p}$ and $e_\mathrm{p}$ being substituted by $a_\star$ and $e_\star$: 
\begin{equation}
2e_\mathrm{f} \simeq \frac{5}{2} \frac{a}{d}\frac{e_\star \cos\psi}{1+e_\star\cos v_\star}
\qquad.
\end{equation}
It is clear from this equation that the highest possible 
$e_\mathrm{f}$ values will be obtained for $\cos v_\star=-1$ 
(binary currently at apoastron) and
$\cos\psi=1$ (no projection factor). With these assumptions, one derives:
\begin{equation}\label{eq:ecc_bin}
2e_\mathrm{f,max} \simeq 0.068\frac{e_\star}{1-e_\star}\qquad.
\end{equation}
For $2e_\mathrm{f,max}$ to reach at least 0.3,
$e_\star\ge0.815$ is required. 

This seems highly eccentric and very unlikely at first sight. Yet, \citet[][see their Fig. 6.b]{1991A&A...248..485D} have shown that almost 25\% of binaries with orbital periods greater than $10^3\,$days have orbital eccentrities $e_\star = 0.825 \pm 0.075$. 
In the present case, $d=3713\,$AU and $e_\star = 0.815$ lead to $a_\star= 2046\,$AU and an orbital period $T_\star \sim 10^5\,$yrs. Thus, $\zeta^1\,$Reticuli being on an eccentric orbit, if not a highly eccentric one, is in fact possible. However, the derived orbit should also have an apoastron value of $q_\star=379~$AU and one might question the disk survival at $\sim 70-120\,$AU with a stellar-type perturber approaching so close to the system.

Eq.~1 of \citet{1999AJ....117..621H} gives the critical semi-major axis $a_\mathrm{crit}$ for orbital stability around a star perturbed by a binary. This is:

\begin{eqnarray}
a_\mathrm{crit} & = & [(0.464\pm 0.006)+(-0.380\pm 0.010)\mu \nonumber\\
& & +(-0.631\pm 0.034)e_\star + (0.586 \pm 0.061)\mu e_\star \nonumber\\
& & +(0.150\pm 0.041)e_\star^2+ (-0.198\pm 0.074)\mu e_\star^2]a_\star\qquad,
\end{eqnarray}
where $\mu = m_\star/(m_{\zeta^2\,\mathrm{Reticuli}} + m_\star)$ is the star mass ratio of value 1/2 if we assume here $m_\star=m_{\zeta^2\,\mathrm{Reticuli}}= 1 M_\odot$.
Material with $a\geq a_\mathrm{crit}$ will be on an unstable orbit, and most probably scattered out of the system.

Applying it to our case leads to $a_\mathrm{crit}=66_{-66}^{+236}$ AU. 
Uncertainties on $a_\mathrm{crit}$ are rather large, and this result shows that within uncertainties, the disk could exist at the observed distances, or, on the contrary, not exist at all. 
Therefore, this orbit is tested numerically.

We consider a ring of 150,000 massless planetesimals uniformly
distributed between 70 and 140 AU from their solar mass host star,
with proper initial eccentricities randomly distributed between 0 and
0.05, and initial inclinations between $\mathrm{\pm3^\circ}$. 
The test particles are perturbed
by another solar mass star in orbit around the primary with a
semi-major axis 2046 AU and eccentricity 0.815, coplanar to the disk.
We use the SWIFT-RMVS N-body symplectic code of \citet{1994Icar..108...18L}
to compute their orbital evolution for 2 Gyr. 

\begin{figure*}
\makebox[\textwidth]{\includegraphics[angle=-90,width=0.4\textwidth]{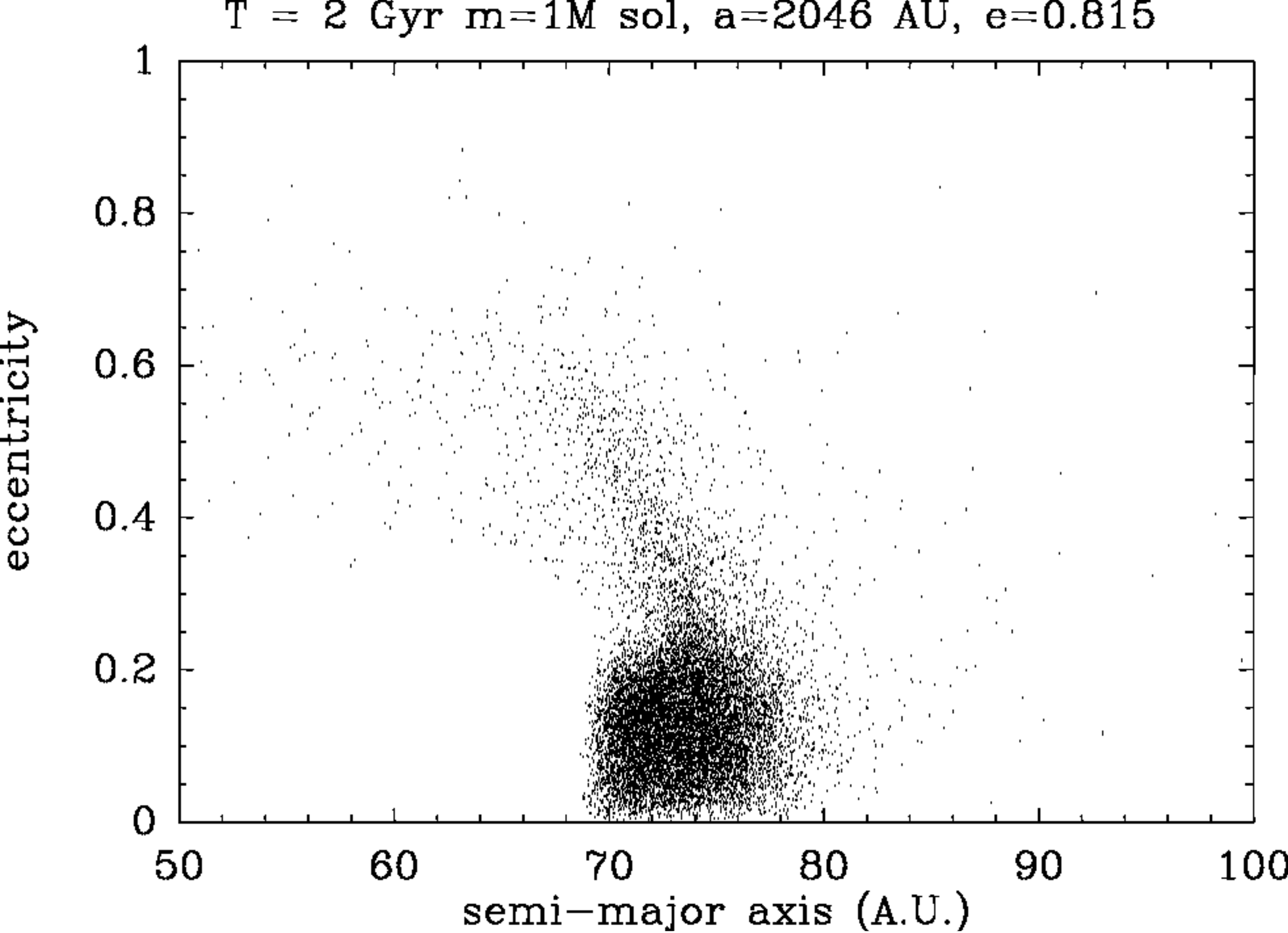}
\includegraphics[angle=-90,width=0.4\textwidth]{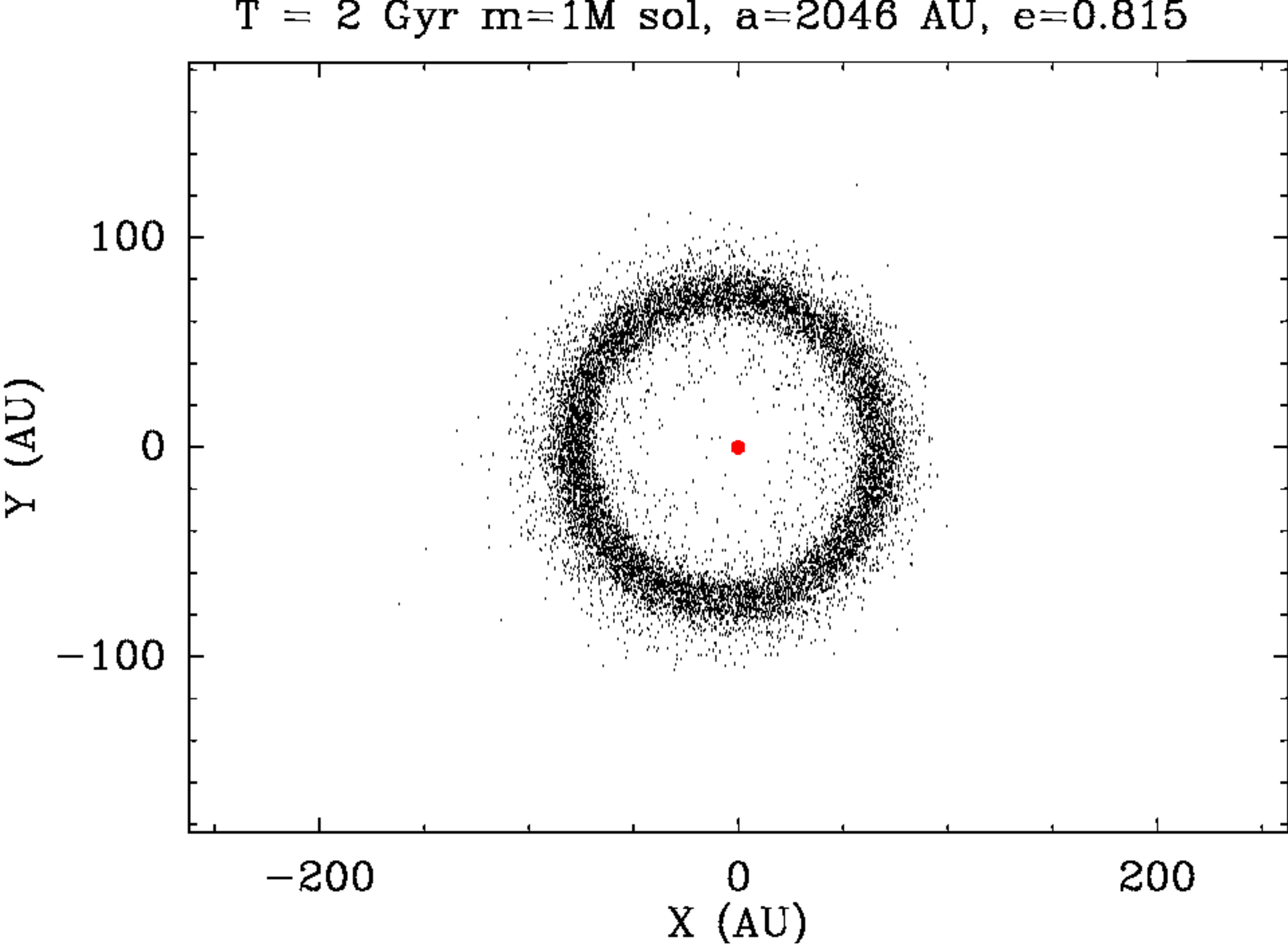}}
\caption[]{Semi-major axis vs eccentricity diagram (left) and pole-on projection(right) of a disk of planetesimals after 2 Gyr under the influence of a one solar mass star of semi-major axis $\mathrm{2046 AU}$ on an orbit of eccentricity $\mathrm{0.815}$. The disk is truncated at $\sim80~$AU and nearly circular. It has no clear offset: its center of symmetry is offset of $\sim6~$AU from the star along the perturber major axis. With a disk mean semi-major axis of $\sim70~$AU, this gives a global eccentricity of $\sim0.08$.
}
\label{fig:comp3}
\end{figure*}

Our results are consistent with the predictions of \citet{1999AJ....117..621H}: the disk is truncated at $\sim80~$AU (see Fig. ~\ref{fig:comp3}, left panel), which is incompatible with observational constraints of the debris disk of $\zR$, since it radially extends up to $\sim 120~$AU.

Moreover, at 2 Gyr, the disk has a rather symmetric shape, and no clear offset from the star (see Fig. ~\ref{fig:comp3}, right panel). These results suggest that the binary companion on an eccentric orbit can unlikely account for the disk structure, and that an unseen eccentric companion is more likely responsible for shaping the disk.

Another possible way for a stellar binary companion to generate high eccentricity orbiting bodies around a primary is the Kozai mechanism \citep{1962AJ.....67..591K}. This would happen if the disk and the binary orbit were mutually inclined by more than $\sim 39^{\circ}$. In that case, the orbit of a particle in the disk would suffer coupled modulations in inclination and eccentricity, i.e., a particle would periodically switch from a highly eccentric orbit, coplanar with the binary companion's orbit, to a circular orbit, very inclined with respect to the orbit of the binary companion. However, we can discard this mechanism from being a possible explanation for the eccentric global structure of the disk of $\zR$: indeed, the Kozai hamiltonian is invariant by rotation, i.e., if this mechanism is able to excite planetesimals eccentricities to high values, their longitudes of periastrons remain uniformely distributed, while they should be preferentially aligned for an eccentric and offset disk to be generated.

\section{Synthetic images}\label{sec:synthetic}
In this section, we will use the results of our best fit N-body simulations as ``seeds'' to produce a realistic dust population and synthetic images for comparison with \textit{Herschel}/PACS observations.
The procedure followed to create synthetic images is straightforward: a population of dust is created from the position of the parent planetesimals.

The main difference between dust particles and planetesimals 
is that the former are small enough to be affected by stellar radiation pressure. 
Radiation pressure is usually described for a particle via its constant ratio $\beta$ to stellar gravity.
The dust particles are supposed to be released by planetesimals which do not feel any radiation pressure. Hence the daugther particles assume an orbit that is very different from that of their parent bodies. It is well known that if the parent bodies move on circular orbits, the dust particles are unbound from the star as soon as $\beta\ge 0.5$. In our case, however, dust particles may be released by planetesimals orbiting on more or less eccentric orbits, which may change this threshold slightly. As planetesimal eccentricities are expected to be moderate on average, $\beta=0.5$ can nevertheless be considered as a reasonable approximation.

Small grains are released from seed planetesimal positions, at the planetesimal velocity, and are then spread along the orbits determined by these initial conditions and their $\beta$ value. We are aware that this simple procedure cannot accurately evaluate the spatial distribution of the smallest grains. To do so, complex models, such as the DyCoSS code of \citet{2012A&A...537A..65T}, the CGA of \citet{2009ApJ...707..543S}, or the LIDT-DD ode by \citet{2013A&A...558A.121K}, have to be used to evaluate the complex interplay between the rate at which grains at collisionally produced from parent planetesimals, the time they spend (because of their highly eccentric orbits) in empty collisionally inactive regions and the rate at which they can be affected or even ejected by close encounters with the perturbing planet \citep[see Sect. 4  of][]{2012A&A...547A..92T}, not to mention the Poynting-Robertson drag these small grains are subject to.
 
However, this caveat is acceptable for the present problem, because the role played by small micron-sized grains close to the blow-out size is very minor at wavelengths $>70~\mathrm{\mu m}$, so that our synthetic images will not be strongly affected by errors regarding their spatial distribution.

We set the dust grain sizes range from $0.5~\mathrm{\mu m}$ to 1 mm, with a classical \citet{1969JGR....74.2531D} power-law distribution\\ (index -3.5), which covers well the $\beta$ distribution from 0 to 0.5, since this parameter depends on grain size.

Their emission is then computed using the radiative transfer code $\textsc{GRaTeR}$ \citep[see, e.g.,][]{2012A&A...539A..17L}. To do this, the following parameters are required: distance of the star (12 pc), magnitude in band V ($V=5.24$), and total luminosity $0.96\,L_\odot$.

\begin{table}[htbp]
\begin{center}
\begin{tabular}{lcc}
\hline
\hline
Wavelength ($\mu \mathrm{m}$) & \hspace{0.25cm} Stellar Flux (mJy) \hspace{0.25cm}& Disk Flux (mJy) \hspace{0.25cm}\\
\hline
 70 & $24.9 \pm 0.8$ & $8.9 \pm 0.8$ \\
 100 & $13.4 \pm 1.0$ & $13.5 \pm 1.0$  \\
 160 & \multicolumn{2}{c}{$19.4 \pm 1.5$} \\
      & $\sim 4.7$ & $\sim 14.7$ \\
\hline
\end{tabular}
\end{center}
\caption{Stellar and disk fluxes at 70, 100 and 160 $\mu \mathrm{m}$ \citep{2010A&A...518L.131E}. The fluxes at 70 and 100 $\mu \mathrm{m}$, along with the total flux of the star-disk system at 160 $\mu \mathrm{m}$ are PACS measurements. The individual star and disk fluxes at 160 $\mu \mathrm{m}$ result from predictions.}
\label{tab:fluxes}
\end{table}

Because the disk is optically thin, its mass is linked linearly to the flux emission intensity and it can be easily scaled to fit with the intensity observed (see Table \ref{tab:fluxes}). 
The mass needed for the disk to produce a flux as observed on \textit{Herschel}/PACS will vary with the dust grain composition, and thus its density.
But since we do not have constraints on the dust composition, astrosilicate grains are used \citep{2003ARA&A..41..241D}, and the mass of the disk is simply scaled to obtain intensities compatible with observational constraints for a given wavelength.

Thermal emission images are produced with a resolution of $1^{\prime \prime}$/pixel at 70 and 100 $\mu \mathrm{m}$, and $2^{\prime \prime}$/pixel at 160 $\mu \mathrm{m}$.
Before convolving these images with the PSF, the star is added at the central pixel, with a flux intensity matching the predicted stellar photosphere flux density in each waveband.
The position angles of the disk observed with \textit{Herschel}/PACS, and of the disk in our synthetic images, but also the orientation of the telescope during the observations are taken into account (see Table ~\ref{tab:angles}). Our purpose is to match the observations.

\begin{table}[htbp]
\begin{center}
\begin{tabular}{lcc}
\hline
\hline
  & Wavelength ($\mu \mathrm{m}$)\hspace{0.25cm} & PA$(^\circ)$ \\
\hline
 Disk observed & - & 110 \\
 Disk simulated & - & 110  \\
 PSF & 70/160 & 127 \\
 PSF & 100/160 & 127 \\
  $\zR$    & 70/160 & 281 \\
  $\zR$    & 100/160 & 281 \\
\hline
\end{tabular}
\end{center}
\caption{Disk position angle observed with \textit{Herschel}/PACS, disk position angle on our synthetic images, and telescope orientation during \textit{Herschel}/PACS observations and PSF.}
\label{tab:angles}
\end{table}

We choose among our simulations one that lead to a clear and significantly eccentric disk at 1 Gyr, namely our Case A (see Table ~\ref{tab:results}, and upper left panels of Fig.~\ref{fig:outer_upperviews} and Fig.~\ref{fig:outer_offsets}), and aim to reproduce the \textit{Herschel}/PACS image at 100 $\mu \mathrm{m}$.
The mass of the disk is scaled so that a total flux of 13.5 mJy as observed with \textit{Herschel}/PACS at 100 $\mu \mathrm{m}$ is spread all over the disk.
The last parameter needed is the system inclination. Our best fit gives $65.5^{\circ}$ (see Appendix~\ref{sec:inclination}).

\begin{figure*}[htbp]
\makebox[\textwidth]{\includegraphics[width=0.5\textwidth]{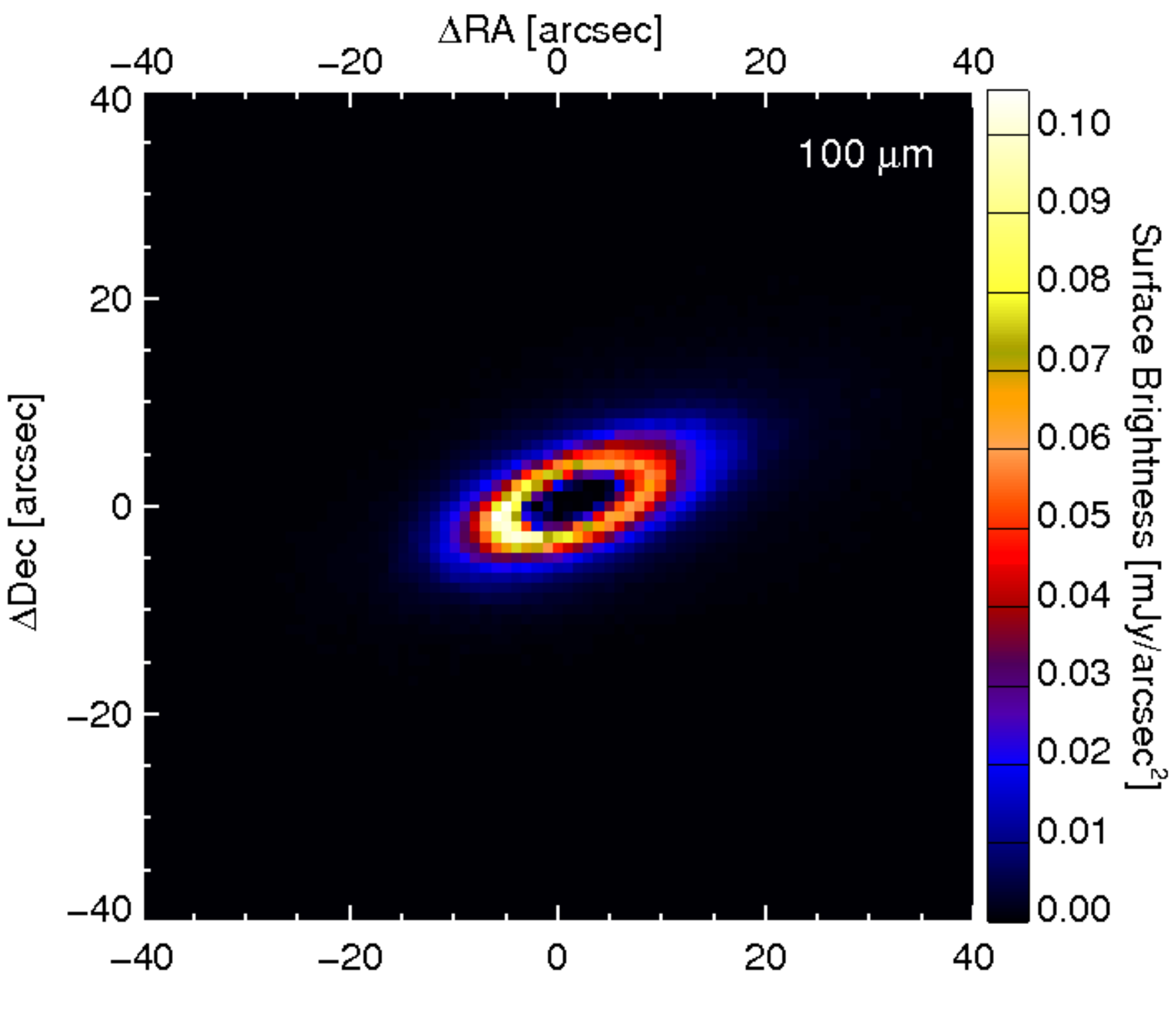}
\includegraphics[width=0.5\textwidth]{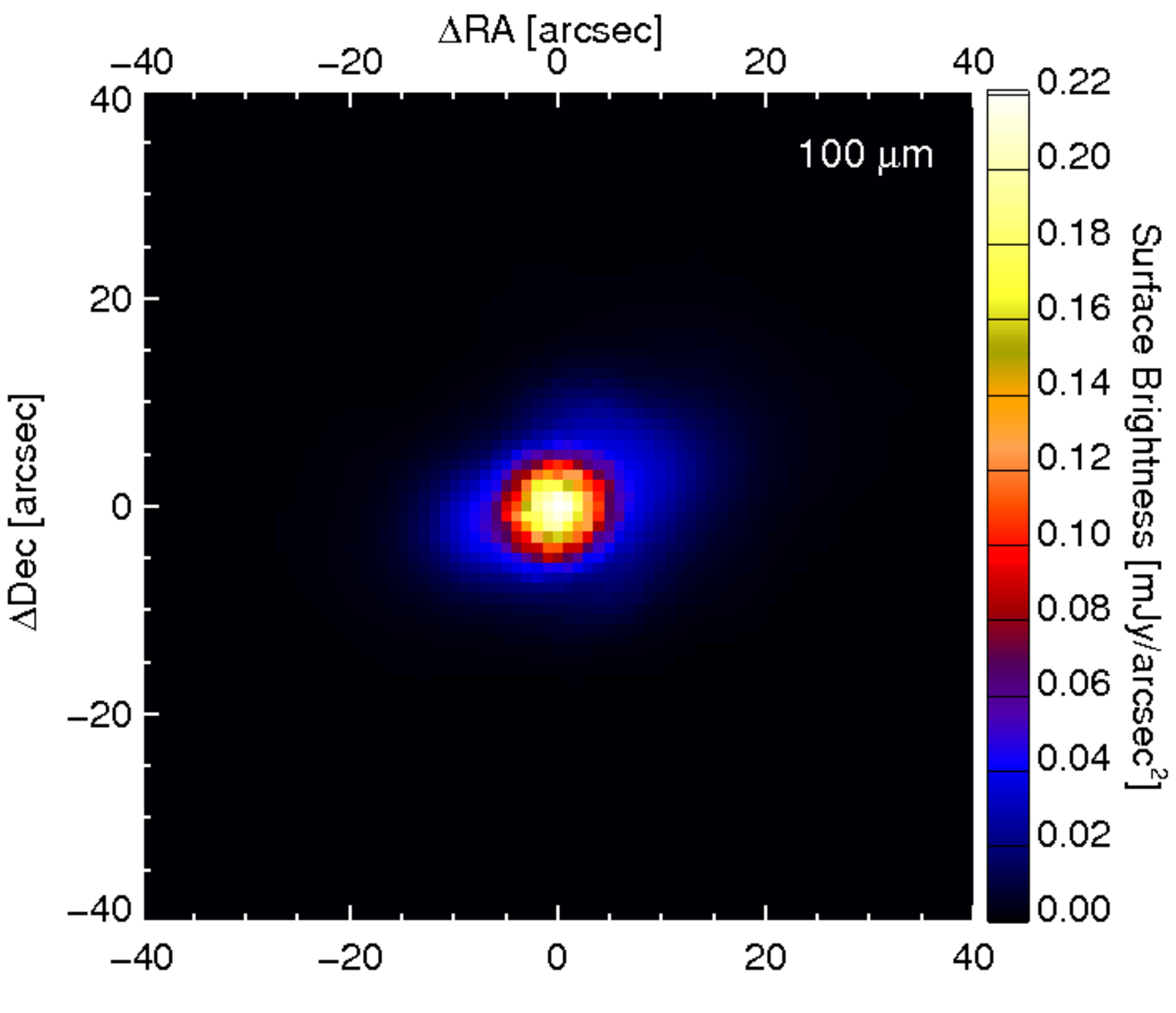}
}
\caption{Synthetic images at  $\lambda=100\,\mu \mathrm{m}$, unconvolved disk (no star) \emph{(left)} and convolved disk and star image with the PSF \emph{(right)}. Disk at 1 Gyr in Case A (see Table~\ref{tab:results}), seen with inclination 65$^\circ$. The star is at the centre of the image, and on the convolved image, the flux scale is set to match the one on \textit{Herschel}/PACS image.}
\label{init_unconvolved}
\end{figure*}

We present an unconvolved and convolved image of this disk on Fig.~\ref{init_unconvolved}.
The disk offset is clearly visible on unconvolved images. However, there seem to be no difference between the symmetric and asymmetric state once the images are convolved, because of the contrast between the star and the disk emission. Moreover, the disk flux per pixel is one order of magnitude smaller than the fluxes observed in \textit{Herschel}/PACS images. 

This is, however, not surprising: in order to estimate the total disk flux, the flux was measured in a small region of the disk before applying aperture correction. Even if the correct aperture correction for a point source was used, that would always be a lower limit, and the total disk flux will be underestimated. The parent ring also may be narrower, which would increase the flux per pixel.

Therefore, we investigate the impact of the width and of the total flux of the disk on the features visible with PACS, and produce convolved images of a dusty disk produced by an asymmetric eccentric parent ring of diverse total fluxes (1, 2 and 5 times the flux measured by \textit{Herschel}/PACS), and of diverse widths (semi major axis centered on 100 AU, widths 5, 10 and 20 AU). To do so, particles from a range of semi-major axis from our \textit{N}-body simulation output are being selected . An inclination angle of 65$^\circ$ is chosen, which is the most probable inclination derived for the system. 
 
\begin{figure*}[htbp]
\makebox[\textwidth]{\includegraphics[width=0.5\textwidth]{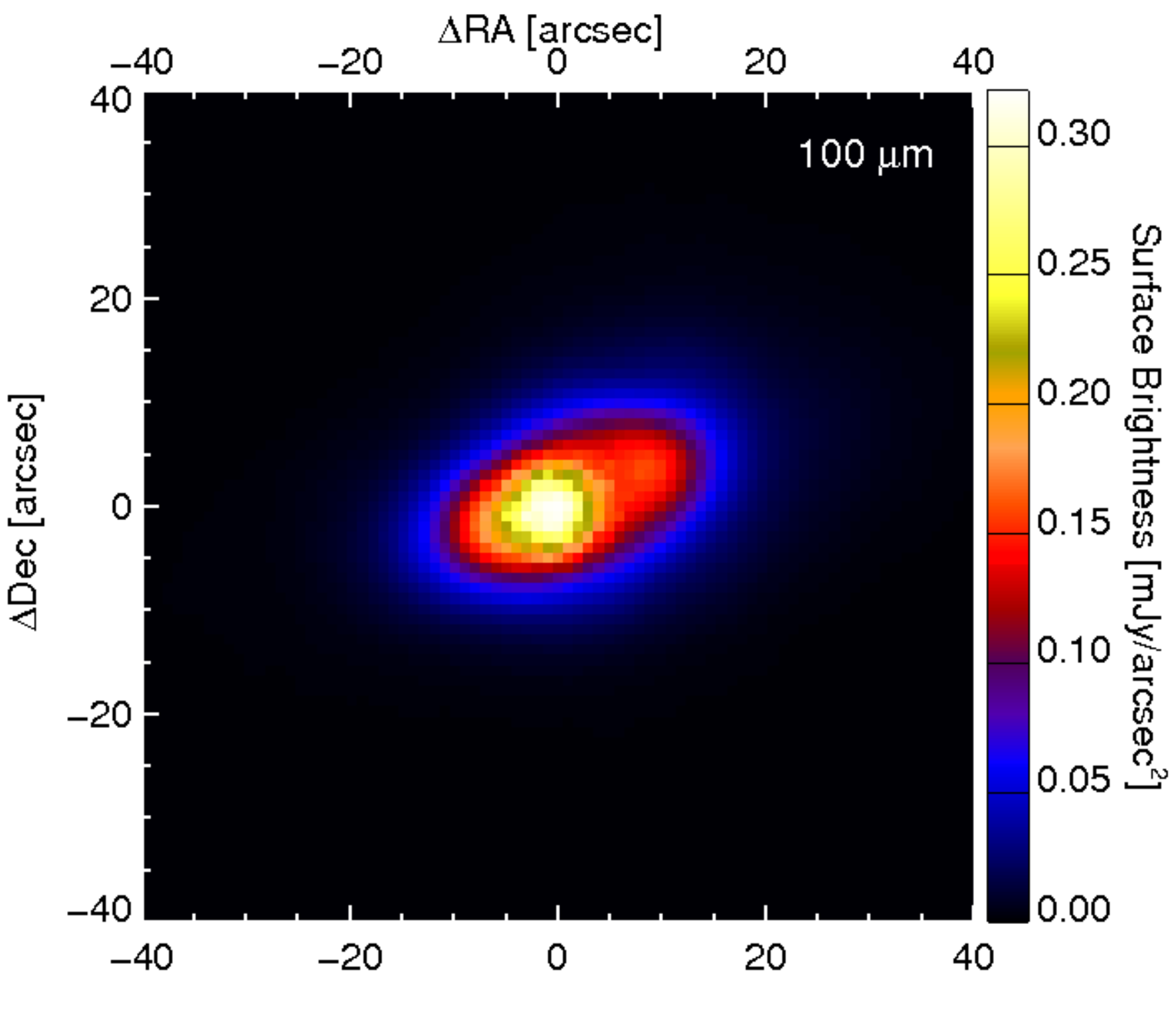}
\includegraphics[width=0.5\textwidth]{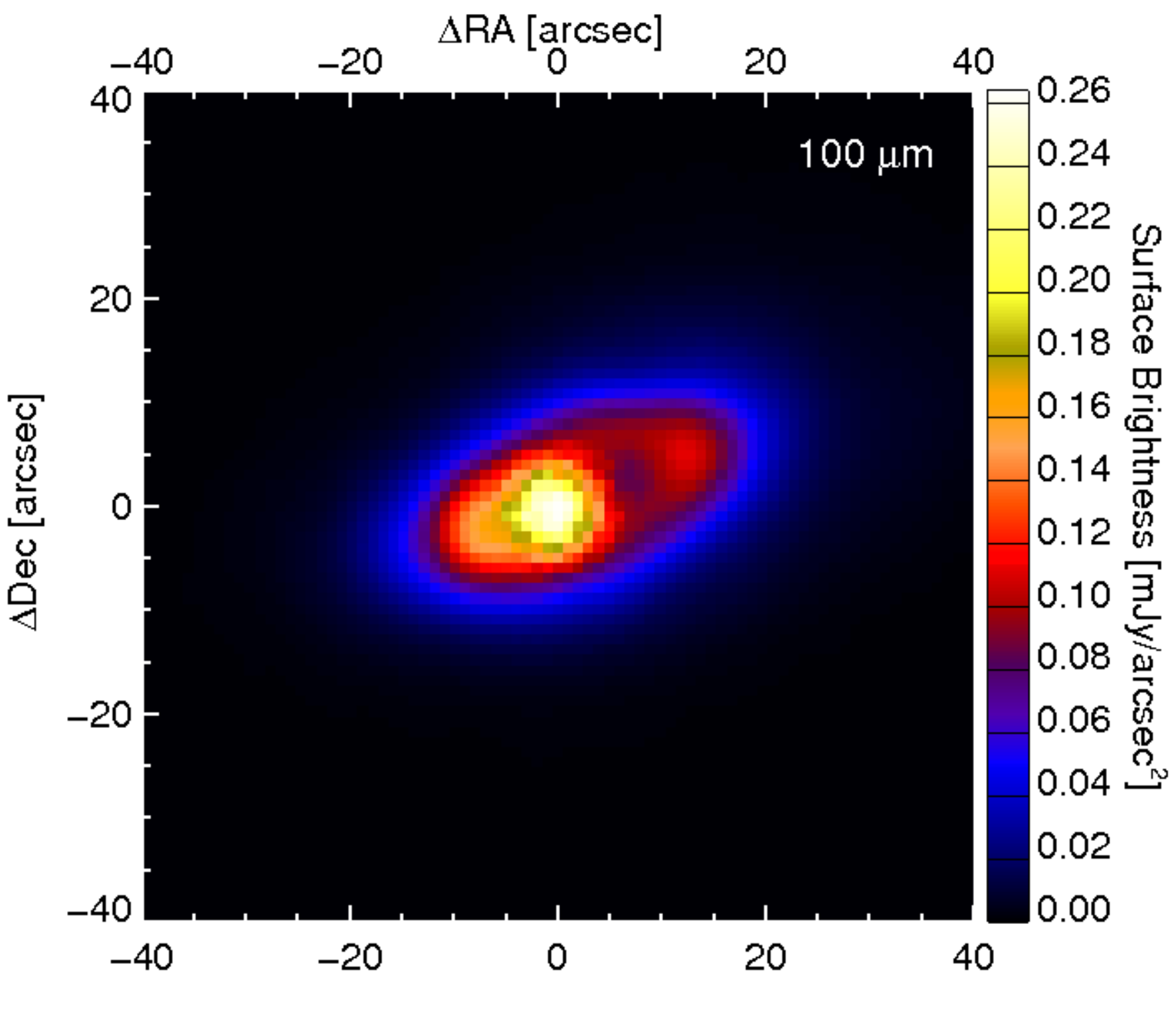}}
\caption{Synthetic images of Case A (see Table~\ref{tab:results}), at 100 $\mu \mathrm{m}$, and convolved with the PSF. The parent ring is centred at 100 AU (left) and 130 AU (right), has a width of 5 AU and is seen with an inclination of 65$^\circ$. The disk total flux is 5 times the flux measured by \textit{Herschel}/PACS. The star is at the centre of the image, and the flux scale is set to match that of \textit{Herschel}/PACS image.}
\label{extent}
\end{figure*}

The fluxes per pixel recovered on convolved images with a disk five times more massive than the mass initially derived from observations are more compatible with the observations. This provides a better constraint on the disk mass and total flux. We show this example in Fig. ~\ref{extent} (left panel): one can see that the asymmetric double lobe structure now appears clearly.

It is worth noting that the width of the parent ring has a limited influence on the flux per pixel compared to the mass of the disk. But it has an influence on the appearance of the disk: a narrow parent ring ($< 10\,$AU) leads to the apastron lobe to be resolved, which is more consistent with \textit{Herschel}/PACS images, although this lobe does not seem to be located as far from the star as it is on the \textit{Herschel}/PACS images.

Therefore, we also investigate the location of the disk, by producing  convolved images of a ring of dust produced by a narrow eccentric parent ring of width 5 AU and semi-major axis centered on 120, 130 and 140 AU . The disk total flux is set to be five times the flux derived from observations. The inclination angle is here again set to 65$^\circ$. As expected, the further the disk is located, the clearer the lobes appear (see Fig.~\ref{extent}).  

Finally, with all these insights, we conclude that the hypothesis of an eccentric dusty disk around $\zR$ is indeed compatible with \textit{Herschel}/PACS images, provided that the dust is produced by a narrow parent ring with width less than 10 AU and located slightly further away than derived by \citet{2010A&A...518L.131E}, i.e., it has a semi-major axis distribution centred between 120 and 140 AU.

This slightly changes the constraints derived on potential perturbers, but the forced eccentricity depends only on the ratio between the planet and planetesimals semi-major axis, and in a linear way at lowest order approximation. This means that in this approximation, the constraints can be completely scaled in a linear way, i.e., the potential semi-major axis for planets must also be increased by 20--40 \% and the disk should rather be centered at 120--140 AU, while the constraint on the perturber eccentricity ($e\gtrsim 0.3$), remains identical.

\section{Discussion \& Conclusion}\label{sec:conclusion}
In this paper, the case of $\zR$ is used as an example to discuss the shaping of a disk into an eccentric ring on Gyr timescales. 

We show that eccentric patterns in debris disks can be maintained on Gyr timescales, but also that eccentric perturbers can produce other patterns than eccentric rings. 

The general results of our simulations show that both inner and outer perturbers can generate extremely significant scattering processes. They can lead a disk to adopt structures that do not show any clear elliptic ring, namely the inner part of the disk is filled or the structure destroyed.
These scattering processes endanger the survival of an eccentric ring and their investigation with numerical experiments allows us to put constraints on potential perturbers. From these constraints, we derive an upper mass limit for outer perturbers in a certain range of periastrons, and a lower mass limit for inner perturbers.

Moreover, the timescale for spiral structures to vanish is longer with smaller mass perturbers, and thus, investigation of this timescale with numerical experiments permits to place a lower mass limit on perturbers, provided that the presence of spiral structures in a disk can definitely be ruled out from observations. The role of numerical experiments is crucial here, since the analytical timescale at the disk centre of distribution underestimates the effective timescale in a real extended disk.

The offset of the disk centre with respect to the star is mostly stable; however, we note that in rare cases, it seems to relax very slowly. While the former evolution is characteristic for pericentre glow dynamics, the latter one is surprising. The relaxation of the eccentric structure is not expected in the first order secular analysis described by \citet{1999ApJ...527..918W} and \citet{2005A&A...440..937W}. It may be the result of higher order terms that have been neglected in the analytical study, and more probably to erratic short-term variations of the planetesimals' semi-major axes due to moderately distant approaches to the planet. These effects that can lead to scattering of the planetesimals are eliminated in the analytical averaging process of the perturbations, and thus cannot be predicted analytically in the secular approximations used here. A more detailed study of this relaxation phenomenon of pericentre glow structures is nevertheless beyond the scope of the present paper, and will be the purpose of future work.

Transient spiral structures, filled inner holes, sparesly populated scattered disks, and resonant clumpy structures are all possible outcomes when an eccentric perturber acts on a debris disk. They can be put in evidence with numerical simulations, but more importantly, used to put constraints on a perturber in a system (mass, eccentricity).
Therefore, we provide a method to investigate and model eccentric ring structures, based on a complementary analytic and numerical approach, where one can derive potential orbits from analytics and put them to test numerically using N-body codes.
This method can be easily applied to other systems, and it is expected to be useful in a near future.

Indeed, \citet{2013Natur.493..381K} have pointed at the fact that wide binary star systems, i.e systems with separations greater than 1000 AU, can produce eccentric planets around a primary star on Gyr timescales. This is due to Galactic tides and passing stars perturbations, which are able, sooner or later, to put the secondary star on a highly eccentric orbit. The proportion of wide binary systems is by no mean negligible \citep[$\sim 50 \%$,][]{1991A&A...248..485D}, and although Gyr-old debris disks are faint and difficult to detect, this will be overcome with the unique capabilities of ALMA, JWST and SPICA.
Therefore, old eccentric patterns in debris disks are expected to be commonly observed in the future.

The $\zR$ disk is one such example of Gyr-old eccentric debris disk. Moreover, $\zR$ is part of a wide binary star system, which may provide an explanation for the presence of an eccentric perturber around $\zR$.
We show that the binary companion cannot be directly responsible for the eccentric ring structure, and show that the asymmetry is rather due to a closer companion, either interior or exterior to the disk. In all cases, the eccentric companion should have an eccentricity $e\gtrsim 0.3$ to produce such a pattern.

Investigation of the disk structure due to scattering processes provides an upper mass limit of $2~\mjup$ for an outer perturber located in a range of periastrons $150-250$AU, whereas a lower mass limit of $0.1~\mjup$ is associated with inner perturbers in the $\zR$ system.

By producing synthetic images, we show that the original interpretation of the double-lobed feature around $\zR$, i.e., the observed eccentric ring $e\gtrsim 0.3$, is clearly supported, although the disk is located slightly further (20-40\%) than originally derived.  Moreover, we find that the dusty disk should be created by a narrow parent ring (width $<$10 AU), which should have a slight inclination with the line of sight, compatible with the most probable inclination derived for the system, and it should also have a significantly greater flux than that estimated from the \textit{Herschel}/PACS measurements (at least five times).


\section*{Aknowledgements: }

We thank the referee, A. Mustill, for very useful comments that contributed to this paper.
Computations presented in this paper were performed at the Service 
Commun de Calcul Intensif de l'Observatoire de Grenoble (SCCI) on the 
super-computer funded by Agence Nationale pour la Recherche under 
contracts ANR-07-BLAN-0221, ANR-2010-JCJC-0504-01 and ANR-2010-JCJC-0501-01.
B. Montesinos, C. Eiroa and J.P. Marshall are supported by Spanish grant AYA 2011-26202.
A. Bonsor and S. Ertel acknowledge the support of the ANR-2010 BLAN-0505-
01 (EXOZODI).
The authors wish to thank the PNP/CNES for their financial support.
This work has also greatly benefited from the software resulting from Thomas Tintillier's training project.

\bibliography{biblio}

\begin{thebibliography}{79}
\expandafter\ifx\csname natexlab\endcsname\relax\def\natexlab#1{#1}\fi

\bibitem[{{Allard} {et~al.}(2011){Allard}, {Homeier}, \&
  {Freytag}}]{Allard2011}
{Allard}, F., {Homeier}, D., \& {Freytag}, B. 2011, in Astronomical Society of
  the Pacific Conference Series, Vol. 448, 16th Cambridge Workshop on Cool
  Stars, Stellar Systems, and the Sun, ed. C.~{Johns-Krull}, M.~K. {Browning},
  \& A.~A. {West}, 91

\bibitem[{{Augereau} {et~al.}(2001){Augereau}, {Nelson}, {Lagrange},
  {Papaloizou}, \& {Mouillet}}]{2001A&A...370..447A}
{Augereau}, J.~C., {Nelson}, R.~P., {Lagrange}, A.~M., {Papaloizou}, J.~C.~B.,
  \& {Mouillet}, D. 2001, \aap, 370, 447

\bibitem[{{Augereau} \& {Papaloizou}(2004)}]{2004A&A...414.1153A}
{Augereau}, J.~C. \& {Papaloizou}, J.~C.~B. 2004, \aap, 414, 1153

\bibitem[{{Aumann} {et~al.}(1984){Aumann}, {Beichman}, {Gillett}, {de Jong},
  {Houck}, {Low}, {Neugebauer}, {Walker}, \& {Wesselius}}]{1984ApJ...278L..23A}
{Aumann}, H.~H., {Beichman}, C.~A., {Gillett}, F.~C., {et~al.} 1984, \apjl,
  278, L23

\bibitem[{{Backman} \& {Paresce}(1993)}]{1993prpl.conf.1253B}
{Backman}, D.~E. \& {Paresce}, F. 1993, in Protostars and Planets III, ed.
  {E.~H.~Levy \& J.~I.~Lunine}, 1253--1304

\bibitem[{{Baraffe} {et~al.}(2003){Baraffe}, {Chabrier}, {Barman}, {Allard}, \&
  {Hauschildt}}]{Baraffe2003}
{Baraffe}, I., {Chabrier}, G., {Barman}, T.~S., {Allard}, F., \& {Hauschildt},
  P.~H. 2003, \aap, 402, 701

\bibitem[{{Beust} {et~al.}(2013){Beust}, {Augereau}, {Bonsor}, {Graham},
  {Kalas}, {Lebreton}, {Lagrange}, {Ertel}, {Faramaz}, \&
  {Th\'ebault}}]{beust2013}
{Beust}, H., {Augereau}, J.-C., {Bonsor}, A., {et~al.} 2013, in prep.

\bibitem[{{Beust} \& {Dutrey}(2006)}]{2006A&A...446..137B}
{Beust}, H. \& {Dutrey}, A. 2006, \aap, 446, 137

\bibitem[{{Boley} {et~al.}(2012){Boley}, {Payne}, {Corder}, {Dent}, {Ford}, \&
  {Shabram}}]{2012ApJ...750L..21B}
{Boley}, A.~C., {Payne}, M.~J., {Corder}, S., {et~al.} 2012, \apjl, 750, L21

\bibitem[{{Chiang} {et~al.}(2009){Chiang}, {Kite}, {Kalas}, {Graham}, \&
  {Clampin}}]{2009ApJ...693..734C}
{Chiang}, E., {Kite}, E., {Kalas}, P., {Graham}, J.~R., \& {Clampin}, M. 2009,
  \apj, 693, 734

\bibitem[{{Dohnanyi}(1969)}]{1969JGR....74.2531D}
{Dohnanyi}, J.~S. 1969, \jgr, 74, 2531

\bibitem[{{Draine}(2003)}]{2003ARA&A..41..241D}
{Draine}, B.~T. 2003, \araa, 41, 241

\bibitem[{{Duncan} {et~al.}(1989){Duncan}, {Quinn}, \&
  {Tremaine}}]{1989Icar...82..402D}
{Duncan}, M., {Quinn}, T., \& {Tremaine}, S. 1989, \icarus, 82, 402

\bibitem[{{Duquennoy} \& {Mayor}(1991)}]{1991A&A...248..485D}
{Duquennoy}, A. \& {Mayor}, M. 1991, \aap, 248, 485

\bibitem[{{Eiroa} {et~al.}(2010){Eiroa}, {Fedele}, {Maldonado},
  {Gonz{\'a}lez-Garc{\'{\i}}a}, {Rodmann}, {Heras}, {Pilbratt}, {Augereau},
  {Mora}, {Montesinos}, {Ardila}, {Bryden}, {Liseau}, {Stapelfeldt},
  {Launhardt}, {Solano}, {Bayo}, {Absil}, {Ar{\'e}valo}, {Barrado},
  {Beichmann}, {Danchi}, {Del Burgo}, {Ertel}, {Fridlund}, {Fukagawa},
  {Guti{\'e}rrez}, {Gr{\"u}n}, {Kamp}, {Krivov}, {Lebreton}, {L{\"o}hne},
  {Lorente}, {Marshall}, {Mart{\'{\i}}nez-Arn{\'a}iz}, {Meeus}, {Montes},
  {Morbidelli}, {M{\"u}ller}, {Mutschke}, {Nakagawa}, {Olofsson}, {Ribas},
  {Roberge}, {Sanz-Forcada}, {Th{\'e}bault}, {Walker}, {White}, \&
  {Wolf}}]{2010A&A...518L.131E}
{Eiroa}, C., {Fedele}, D., {Maldonado}, J., {et~al.} 2010, \aap, 518, L131

\bibitem[{{Eiroa} {et~al.}(2013){Eiroa}, {Marshall}, {Mora}, {Montesinos},
  {Absil}, {Augereau}, {Bayo}, {Bryden}, {Danchi}, {del Burgo}, {Ertel},
  {Fridlund}, {Heras}, {Krivov}, {Launhardt}, {Liseau}, {L{\"o}hne},
  {Maldonado}, {Pilbratt}, {Roberge}, {Rodmann}, {Sanz-Forcada}, {Solano},
  {Stapelfeldt}, {Th{\'e}bault}, {Wolf}, {Ardila}, {Ar{\'e}valo}, {Beichmann},
  {Faramaz}, {Gonz{\'a}lez-Garc{\'{\i}}a}, {Guti{\'e}rrez}, {Lebreton},
  {Mart{\'{\i}}nez-Arn{\'a}iz}, {Meeus}, {Montes}, {Olofsson}, {Su}, {White},
  {Barrado}, {Fukagawa}, {Gr{\"u}n}, {Kamp}, {Lorente}, {Morbidelli},
  {M{\"u}ller}, {Mutschke}, {Nakagawa}, {Ribas}, \&
  {Walker}}]{2013A&A...555A..11E}
{Eiroa}, C., {Marshall}, J.~P., {Mora}, A., {et~al.} 2013, \aap, 555, A11

\bibitem[{{Ertel} {et~al.}(2011){Ertel}, {Wolf}, {Eiroa}, {Augereau},
  {Metchev}, {Schneider}, {Silverstone}, \& {Rodmann}}]{2011epsc.conf..678E}
{Ertel}, S., {Wolf}, S., {Eiroa}, C., {et~al.} 2011, in EPSC-DPS Joint Meeting
  2011, 678

\bibitem[{{Ertel} {et~al.}(2012){Ertel}, {Wolf}, \&
  {Rodmann}}]{2012A&A...544A..61E}
{Ertel}, S., {Wolf}, S., \& {Rodmann}, J. 2012, \aap, 544, A61

\bibitem[{{Ford} {et~al.}(2000){Ford}, {Kozinsky}, \&
  {Rasio}}]{2000ApJ...535..385F}
{Ford}, E.~B., {Kozinsky}, B., \& {Rasio}, F.~A. 2000, \apj, 535, 385

\bibitem[{{Gray} {et~al.}(2006){Gray}, {Corbally}, {Garrison}, {McFadden},
  {Bubar}, {McGahee}, {O'Donoghue}, \& {Knox}}]{2006AJ....132..161G}
{Gray}, R.~O., {Corbally}, C.~J., {Garrison}, R.~F., {et~al.} 2006, \aj, 132,
  161

\bibitem[{{Guilloteau} {et~al.}(2011){Guilloteau}, {Dutrey}, {Pi{\'e}tu}, \&
  {Boehler}}]{2011A&A...529A.105G}
{Guilloteau}, S., {Dutrey}, A., {Pi{\'e}tu}, V., \& {Boehler}, Y. 2011, \aap,
  529, A105

\bibitem[{{Henry} {et~al.}(1996){Henry}, {Soderblom}, {Donahue}, \&
  {Baliunas}}]{1996AJ....111..439H}
{Henry}, T.~J., {Soderblom}, D.~R., {Donahue}, R.~A., \& {Baliunas}, S.~L.
  1996, \aj, 111, 439

\bibitem[{{Holman} \& {Wiegert}(1999)}]{1999AJ....117..621H}
{Holman}, M.~J. \& {Wiegert}, P.~A. 1999, \aj, 117, 621

\bibitem[{{Johnson} {et~al.}(1966){Johnson}, {Mitchell}, {Iriarte}, \&
  {Wisniewski}}]{1966CoLPL...4...99J}
{Johnson}, H.~L., {Mitchell}, R.~I., {Iriarte}, B., \& {Wisniewski}, W.~Z.
  1966, Communications of the Lunar and Planetary Laboratory, 4, 99

\bibitem[{{Kaib} {et~al.}(2013){Kaib}, {Raymond}, \&
  {Duncan}}]{2013Natur.493..381K}
{Kaib}, N.~A., {Raymond}, S.~N., \& {Duncan}, M. 2013, \nat, 493, 381

\bibitem[{{Kalas} {et~al.}(2008){Kalas}, {Graham}, {Chiang}, {Fitzgerald},
  {Clampin}, {Kite}, {Stapelfeldt}, {Marois}, \& {Krist}}]{2008Sci...322.1345K}
{Kalas}, P., {Graham}, J.~R., {Chiang}, E., {et~al.} 2008, Science, 322, 1345

\bibitem[{{Kalas} {et~al.}(2005){Kalas}, {Graham}, \&
  {Clampin}}]{2005Natur.435.1067K}
{Kalas}, P., {Graham}, J.~R., \& {Clampin}, M. 2005, \nat, 435, 1067

\bibitem[{{Kalas} {et~al.}(2013){Kalas}, {Graham}, {Fitzgerald}, \&
  {Clampin}}]{2013arXiv1305.2222K}
{Kalas}, P., {Graham}, J.~R., {Fitzgerald}, M.~P., \& {Clampin}, M. 2013, ArXiv
  e-prints

\bibitem[{{Kozai}(1962)}]{1962AJ.....67..591K}
{Kozai}, Y. 1962, \aj, 67, 591

\bibitem[{{Kral} {et~al.}(2013){Kral}, {Th{\'e}bault}, \&
  {Charnoz}}]{2013A&A...558A.121K}
{Kral}, Q., {Th{\'e}bault}, P., \& {Charnoz}, S. 2013, \aap, 558, A121

\bibitem[{{Krist} {et~al.}(2012){Krist}, {Stapelfeldt}, {Bryden}, \&
  {Plavchan}}]{2012AJ....144...45K}
{Krist}, J.~E., {Stapelfeldt}, K.~R., {Bryden}, G., \& {Plavchan}, P. 2012,
  \aj, 144, 45

\bibitem[{{Krivov}(2010)}]{2010RAA....10..383K}
{Krivov}, A.~V. 2010, Research in Astronomy and Astrophysics, 10, 383

\bibitem[{{Krymolowski} \& {Mazeh}(1999)}]{1999MNRAS.304..720K}
{Krymolowski}, Y. \& {Mazeh}, T. 1999, \mnras, 304, 720

\bibitem[{{Lebreton} {et~al.}(2012){Lebreton}, {Augereau}, {Thi}, {Roberge},
  {Donaldson}, {Schneider}, {Maddison}, {M{\'e}nard}, {Riviere-Marichalar},
  {Mathews}, {Kamp}, {Pinte}, {Dent}, {Barrado}, {Duch{\^e}ne}, {Gonzalez},
  {Grady}, {Meeus}, {Pantin}, {Williams}, \& {Woitke}}]{2012A&A...539A..17L}
{Lebreton}, J., {Augereau}, J.-C., {Thi}, W.-F., {et~al.} 2012, \aap, 539, A17

\bibitem[{{Lenzen} {et~al.}(2003){Lenzen}, {Hartung}, {Brandner}, {Finger},
  {Hubin}, {Lacombe}, {Lagrange}, {Lehnert}, {Moorwood}, \&
  {Mouillet}}]{2003SPIE.4841..944L}
{Lenzen}, R., {Hartung}, M., {Brandner}, W., {et~al.} 2003, in Society of
  Photo-Optical Instrumentation Engineers (SPIE) Conference Series, Vol. 4841,
  Society of Photo-Optical Instrumentation Engineers (SPIE) Conference Series,
  ed. M.~{Iye} \& A.~F.~M. {Moorwood}, 944--952

\bibitem[{{Levison} \& {Duncan}(1994)}]{1994Icar..108...18L}
{Levison}, H.~F. \& {Duncan}, M.~J. 1994, \icarus, 108, 18

\bibitem[{{L{\"o}hne} {et~al.}(2008){L{\"o}hne}, {Krivov}, \&
  {Rodmann}}]{2008ApJ...673.1123L}
{L{\"o}hne}, T., {Krivov}, A.~V., \& {Rodmann}, J. 2008, \apj, 673, 1123

\bibitem[{{Luhman} {et~al.}(2007){Luhman}, {Patten}, {Marengo}, {Schuster},
  {Hora}, {Ellis}, {Stauffer}, {Sonnett}, {Winston}, {Gutermuth}, {Megeath},
  {Backman}, {Henry}, {Werner}, \& {Fazio}}]{2007ApJ...654..570L}
{Luhman}, K.~L., {Patten}, B.~M., {Marengo}, M., {et~al.} 2007, \apj, 654, 570

\bibitem[{{Malhotra}(1993)}]{1993Natur.365..819M}
{Malhotra}, R. 1993, \nat, 365, 819

\bibitem[{{Malhotra}(1995)}]{1995AJ....110..420M}
{Malhotra}, R. 1995, \aj, 110, 420

\bibitem[{{Mamajek}(2012)}]{2012ApJ...754L..20M}
{Mamajek}, E.~E. 2012, \apjl, 754, L20

\bibitem[{{Mardling} \& {Lin}(2002)}]{2002ApJ...573..829M}
{Mardling}, R.~A. \& {Lin}, D.~N.~C. 2002, \apj, 573, 829

\bibitem[{{Marois} {et~al.}(2006){Marois}, {Lafreni{\`e}re}, {Doyon},
  {Macintosh}, \& {Nadeau}}]{Marois2006}
{Marois}, C., {Lafreni{\`e}re}, D., {Doyon}, R., {Macintosh}, B., \& {Nadeau},
  D. 2006, \apj, 641, 556

\bibitem[{{Marois} {et~al.}(2008){Marois}, {Macintosh}, {Barman}, {Zuckerman},
  {Song}, {Patience}, {Lafreni{\`e}re}, \& {Doyon}}]{2008Sci...322.1348M}
{Marois}, C., {Macintosh}, B., {Barman}, T., {et~al.} 2008, Science, 322, 1348

\bibitem[{{Marois} {et~al.}(2010){Marois}, {Zuckerman}, {Konopacky},
  {Macintosh}, \& {Barman}}]{2010Natur.468.1080M}
{Marois}, C., {Zuckerman}, B., {Konopacky}, Q.~M., {Macintosh}, B., \&
  {Barman}, T. 2010, \nat, 468, 1080

\bibitem[{{Masana} {et~al.}(2006){Masana}, {Jordi}, \&
  {Ribas}}]{2006A&A...450..735M}
{Masana}, E., {Jordi}, C., \& {Ribas}, I. 2006, \aap, 450, 735

\bibitem[{{Mason} {et~al.}(2001){Mason}, {Wycoff}, {Hartkopf}, {Douglass}, \&
  {Worley}}]{2001AJ....122.3466M}
{Mason}, B.~D., {Wycoff}, G.~L., {Hartkopf}, W.~I., {Douglass}, G.~G., \&
  {Worley}, C.~E. 2001, \aj, 122, 3466

\bibitem[{{Mayor} {et~al.}(2003){Mayor}, {Pepe}, {Queloz}, {Bouchy},
  {Rupprecht}, {Lo Curto}, {Avila}, {Benz}, {Bertaux}, {Bonfils}, {Dall},
  {Dekker}, {Delabre}, {Eckert}, {Fleury}, {Gilliotte}, {Gojak}, {Guzman},
  {Kohler}, {Lizon}, {Longinotti}, {Lovis}, {Megevand}, {Pasquini}, {Reyes},
  {Sivan}, {Sosnowska}, {Soto}, {Udry}, {van Kesteren}, {Weber}, \&
  {Weilenmann}}]{2003Msngr.114...20M}
{Mayor}, M., {Pepe}, F., {Queloz}, D., {et~al.} 2003, The Messenger, 114, 20

\bibitem[{{Mayor} \& {Queloz}(1995)}]{1995Natur.378..355M}
{Mayor}, M. \& {Queloz}, D. 1995, \nat, 378, 355

\bibitem[{{Moro-Mart{\'{\i}}n}(2012)}]{2012arXiv1203.0005M}
{Moro-Mart{\'{\i}}n}, A. 2012, ArXiv e-prints

\bibitem[{{Moro-Mart{\'{\i}}n} \& {Malhotra}(2002)}]{2002AJ....124.2305M}
{Moro-Mart{\'{\i}}n}, A. \& {Malhotra}, R. 2002, \aj, 124, 2305

\bibitem[{{Mouillet} {et~al.}(1997){Mouillet}, {Larwood}, {Papaloizou}, \&
  {Lagrange}}]{1997MNRAS.292..896M}
{Mouillet}, D., {Larwood}, J.~D., {Papaloizou}, J.~C.~B., \& {Lagrange}, A.~M.
  1997, \mnras, 292, 896

\bibitem[{{Mustill} \& {Wyatt}(2009)}]{2009MNRAS.399.1403M}
{Mustill}, A.~J. \& {Wyatt}, M.~C. 2009, \mnras, 399, 1403

\bibitem[{{Noyes} {et~al.}(1984){Noyes}, {Hartmann}, {Baliunas}, {Duncan}, \&
  {Vaughan}}]{1984ApJ...279..763N}
{Noyes}, R.~W., {Hartmann}, L.~W., {Baliunas}, S.~L., {Duncan}, D.~K., \&
  {Vaughan}, A.~H. 1984, \apj, 279, 763

\bibitem[{{Quillen}(2006)}]{2006MNRAS.372L..14Q}
{Quillen}, A.~C. 2006, \mnras, 372, L14

\bibitem[{{Reche} {et~al.}(2008){Reche}, {Beust}, {Augereau}, \&
  {Absil}}]{2008A&A...480..551R}
{Reche}, R., {Beust}, H., {Augereau}, J.-C., \& {Absil}, O. 2008, \aap, 480,
  551

\bibitem[{{Reiners} \& {Schmitt}(2003)}]{RS03}
{Reiners}, A. \& {Schmitt}, J.~H.~M.~M. 2003, \aap, 398, 647

\bibitem[{{Rousset} {et~al.}(2003){Rousset}, {Lacombe}, {Puget}, {Hubin},
  {Gendron}, {Fusco}, {Arsenault}, {Charton}, {Feautrier}, {Gigan}, {Kern},
  {Lagrange}, {Madec}, {Mouillet}, {Rabaud}, {Rabou}, {Stadler}, \&
  {Zins}}]{2003SPIE.4839..140R}
{Rousset}, G., {Lacombe}, F., {Puget}, P., {et~al.} 2003, in Society of
  Photo-Optical Instrumentation Engineers (SPIE) Conference Series, Vol. 4839,
  Society of Photo-Optical Instrumentation Engineers (SPIE) Conference Series,
  ed. P.~L. {Wizinowich} \& D.~{Bonaccini}, 140--149

\bibitem[{{Shaya} \& {Olling}(2011)}]{2011ApJS..192....2S}
{Shaya}, E.~J. \& {Olling}, R.~P. 2011, \apjs, 192, 2

\bibitem[{{Soummer} {et~al.}(2012){Soummer}, {Pueyo}, \&
  {Larkin}}]{Soummer2012}
{Soummer}, R., {Pueyo}, L., \& {Larkin}, J. 2012, \apjl, 755, L28

\bibitem[{{Stapelfeldt} {et~al.}(2004){Stapelfeldt}, {Holmes}, {Chen}, {Rieke},
  {Su}, {Hines}, {Werner}, {Beichman}, {Jura}, {Padgett}, {Stansberry},
  {Bendo}, {Cadien}, {Marengo}, {Thompson}, {Velusamy}, {Backus}, {Blaylock},
  {Egami}, {Engelbracht}, {Frayer}, {Gordon}, {Keene}, {Latter}, {Megeath},
  {Misselt}, {Morrison}, {Muzerolle}, {Noriega-Crespo}, {Van Cleve}, \&
  {Young}}]{2004stapelfeldt}
{Stapelfeldt}, K.~R., {Holmes}, E.~K., {Chen}, C., {et~al.} 2004, \apjs, 154,
  458

\bibitem[{{Stapelfeldt} {et~al.}(2012){Stapelfeldt}, {Krist}, {Bryden}, \&
  {Plavchan}}]{2012AAS...22050603S}
{Stapelfeldt}, K.~R., {Krist}, J.~E., {Bryden}, G.~C., \& {Plavchan}, P. 2012,
  in American Astronomical Society Meeting Abstracts, Vol. 220, American
  Astronomical Society Meeting Abstracts 220, 506.03

\bibitem[{{Stark} \& {Kuchner}(2008)}]{stark08}
{Stark}, C.~C. \& {Kuchner}, M.~J. 2008, \apj, 686, 637

\bibitem[{{Stark} \& {Kuchner}(2009)}]{2009ApJ...707..543S}
{Stark}, C.~C. \& {Kuchner}, M.~J. 2009, \apj, 707, 543

\bibitem[{{Th{\'e}bault}(2012)}]{2012A&A...537A..65T}
{Th{\'e}bault}, P. 2012, \aap, 537, A65

\bibitem[{{Th{\'e}bault} \& {Augereau}(2007)}]{2007A&A...472..169T}
{Th{\'e}bault}, P. \& {Augereau}, J.-C. 2007, \aap, 472, 169

\bibitem[{{Thebault} {et~al.}(2012){Thebault}, {Kral}, \&
  {Ertel}}]{2012A&A...547A..92T}
{Thebault}, P., {Kral}, Q., \& {Ertel}, S. 2012, \aap, 547, A92

\bibitem[{{Th{\'e}bault} {et~al.}(2012){Th{\'e}bault}, {Kral}, \&
  {Ertel}}]{2012arXiv1209.3969T}
{Th{\'e}bault}, P., {Kral}, Q., \& {Ertel}, S. 2012, ArXiv e-prints

\bibitem[{{Torres} {et~al.}(2006){Torres}, {Quast}, {da Silva}, {de La Reza},
  {Melo}, \& {Sterzik}}]{2006A&A...460..695T}
{Torres}, C.~A.~O., {Quast}, G.~R., {da Silva}, L., {et~al.} 2006, \aap, 460,
  695

\bibitem[{{Trilling} {et~al.}(2008){Trilling}, {Bryden}, {Beichman}, {Rieke},
  {Su}, {Stansberry}, {Blaylock}, {Stapelfeldt}, {Beeman}, \&
  {Haller}}]{2008ApJ...674.1086T}
{Trilling}, D.~E., {Bryden}, G., {Beichman}, C.~A., {et~al.} 2008, \apj, 674,
  1086

\bibitem[{{van Leeuwen}(2007)}]{2007A&A...474..653V}
{van Leeuwen}, F. 2007, A\&A, 474, 653

\bibitem[{{Watson} {et~al.}(2011){Watson}, {Littlefair}, {Diamond}, {Collier
  Cameron}, {Fitzsimmons}, {Simpson}, {Moulds}, \&
  {Pollacco}}]{2011MNRAS.413L..71W}
{Watson}, C.~A., {Littlefair}, S.~P., {Diamond}, C., {et~al.} 2011, \mnras,
  413, L71

\bibitem[{{Wisdom}(1980)}]{1980AJ.....85.1122W}
{Wisdom}, J. 1980, \aj, 85, 1122

\bibitem[{{Wyatt}(1999)}]{1999PhDT........12W}
{Wyatt}, M.~C. 1999, PhD thesis, Royal Observatory, Blackford Hill, Edinburgh
  EH9 3HJ, UK

\bibitem[{{Wyatt}(2004)}]{2004AIPC..713...93W}
{Wyatt}, M.~C. 2004, in American Institute of Physics Conference Series, Vol.
  713, The Search for Other Worlds, ed. S.~S. {Holt} \& D.~{Deming}, 93--102

\bibitem[{{Wyatt}(2005)}]{2005A&A...440..937W}
{Wyatt}, M.~C. 2005, \aap, 440, 937

\bibitem[{{Wyatt} \& {Dent}(2002)}]{2002MNRAS.334..589W}
{Wyatt}, M.~C. \& {Dent}, W.~R.~F. 2002, \mnras, 334, 589

\bibitem[{{Wyatt} {et~al.}(2000){Wyatt}, {Dermott}, \&
  {Telesco}}]{2000ASPC..219..289W}
{Wyatt}, M.~C., {Dermott}, S.~F., \& {Telesco}, C.~M. 2000, in Astronomical
  Society of the Pacific Conference Series, Vol. 219, Disks, Planetesimals, and
  Planets, ed. G.~{Garz{\'o}n}, C.~{Eiroa}, D.~{de Winter}, \& T.~J. {Mahoney},
  289

\bibitem[{{Wyatt} {et~al.}(1999){Wyatt}, {Dermott}, {Telesco}, {Fisher},
  {Grogan}, {Holmes}, \& {Pi{\~n}a}}]{1999ApJ...527..918W}
{Wyatt}, M.~C., {Dermott}, S.~F., {Telesco}, C.~M., {et~al.} 1999, \apj, 527,
  918

\end{thebibliography}

\appendix
\begin{onecolumn}

\section{SPIRE images of $\zR$} \label{sec:spire}
\begin{figure*}[h]
\makebox[\textwidth]{
\includegraphics[width=1\textwidth]{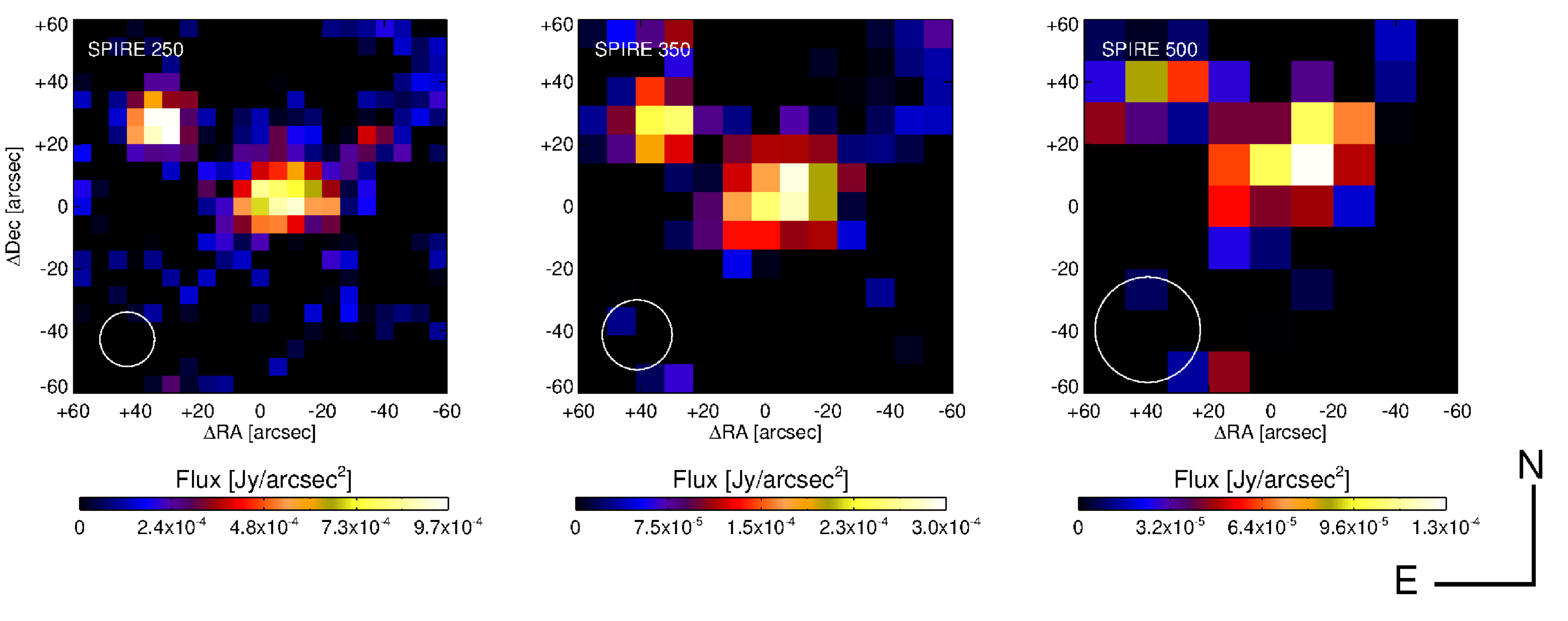}}
\caption[]{\textit{Herschel}/SPIRE images of $\zR$ at 250, 350 and 500~$\mu$m (left--right). The images were reduced using the standard reduction scripts ofn HIPE, version 8.2 and SPIRE CAL 8.1. Image orientation is North up, East left. The pixel scales are $6^{\prime \prime}$, $10^{\prime \prime}$ and $14^{\prime \prime}$~at 250, 350 and 500~$\mu$m, respectively. The SPIRE beam FWHM in each band is denoted by the white circle in the bottom left corner of each image.}  
\label{fig:SPIRE}
\end{figure*}

\begin{table}[h]
\begin{center}
\begin{tabular}{lcc}
\hline
\hline
Wavelength ($\mu \mathrm{m}$) & \hspace{0.25cm} Stellar Flux (mJy) \hspace{0.25cm}& Disk+Star Flux (mJy) \hspace{0.25cm}\\
\hline
 250 & $2.03 \pm 0.03$ & $59.72 \pm 6.70$ \\
 350 & $1.04 \pm 0.02$ & $24.68 \pm 6.89$ \\
 500 & $0.51 \pm 0.10$ & $20.29 \pm 7.66$ \\
\hline
\end{tabular}
\end{center}
\caption{Stellar predicted fluxes and SPIRE fluxes measurements of the star-disk system at 250, 350 and 500 $\mu \mathrm{m}$ \citep{2013A&A...555A..11E}.}
\label{tab:SPIREfluxes}
\end{table}

\section{Inclination of $\zR$}\label{sec:inclination}

Observations of the debris disk surrounding $\zR$ reveal a double-lobed asymmetric feature.
The inclination of this system relative to the line of sight is a key parameter to interprete correctly the observations. If the system is seen pole-on, one would rather expect the observed feature to be the signature of resonant clumps, whereas an eccentric ring signature would be more plausible if the system is observed edge-on.

In general, observations suggest that stellar and disk inclinations are aligned \citep{2011MNRAS.413L..71W,2011A&A...529A.105G}. Under this assumption, one can estimate the disk inclination from the observed stellar inclination.
Consequently, we aim here to measure the star's inclination $i$, i.e., the angle formed by its rotation axis with respect to the line of sight. With this convention, the system is seen pole-on if $i=0^\circ$ and edge-on if $i=90^\circ$.

The method used requires the knowledge of the following stellar properties : the colour index $(B-V)$, the radius $R_\star$, the projected rotational velocity, $v_{\mathrm{rot}} \sin(i)$, and finally $R'_{\mathrm{HK}}$, an activity indicator defined as $F'_{\mathrm{HK}}/\sigma T^4_\star$, where $F'_{\mathrm{HK}}$ is the chromospheric flux in the H and K lines of $\caii$, and $T_\star$ is the star effective temperature.
These properties for $\zR$ are summarized in Table ~\ref{tab:inclination}.

\begin{table}[htbp]
\caption{Stellar properties of $\zR$.}
\tablebib{(1) \citet{1966CoLPL...4...99J}; (2) this study; (3) \citet{2013A&A...555A..11E}; (4) \citet{RS03}; (5) \citet{1996AJ....111..439H}. }
\begin{center}
\begin{tabular}{lcc}
\hline
\hline
Stellar Property & Value & Reference \\
\hline
$(B-V)$ & $0.60$ & 1 \\
$R_\star$ ($\mathrm{R_{\odot}}$) & $\sim 0.965 \pm 0.05 $ & 2 \\
$L_\star$ ($L_\odot$) & $0.97$ & 3 \\
$T_\star$ (K) & $5851$ & 3 \\
$v_{\mathrm{rot}}\sin{i}$ (km/s) & $2.7 \pm 0.3$ & 4 \\
$\log(R'_{\mathrm{HK}})$ & $-4.79 \pm 0.03 $ & 5 \\
$\log(\mathrm{Ro})$ & $ \sim 0.185 \pm 0.085 $ & 2 \\
$\tau_\mathrm{c}$ (days) & $\sim 9.10 $ & 2 \\
$P_{\mathrm{rot}}$ (days) & $\sim 14.20 \pm 2.75 $ & 2 \\
$v_{\mathrm{rot}}$ (km/s) & $ \sim 3.42 \pm 0.66 $ & 2 \\
$i (^\circ)$ & $ \sim 65.5_{-31.5}^{+24.5} $ & 2 \\
\hline  
\end{tabular}
\end{center}
\label{tab:inclination}
\end{table}

We first use the activity/rotation diagram built by \citet{1984ApJ...279..763N}, which plots $\mathrm{\log(R'_{HK})}$ versus $\mathrm{\log(Ro)}$ and shows a relationship between these two quantities for late-type stars. $\mathrm{Ro}=P_{\mathrm{rot}}/\tau_\mathrm{c}$ is the Rossby number, $P_{\mathrm{rot}}$ being the rotational period of the star and $\mathrm{\tau_c}$ a model-dependent, typical, convective time, called the `turnover time'. Using Fig. 6(b) of \citet{1984ApJ...279..763N} and the observed value of $\log(R'_{\mathrm{HK}})=-4.79$ found by \citet{1996AJ....111..439H} for $\zR$ allows us to estimate $\log(P_{\mathrm{rot}}/\tau_\mathrm{c})\sim 0.185 \pm 0.085 $. 

Then, using equation (4) of \citet{1984ApJ...279..763N} where $x$ is defined with the star colour index $(B-V)$ by $x=1-(B-V)$, one can estimate $\tau_\mathrm{c}$:

\begin{center}
\begin{equation}
\log \tau_\mathrm{c} = 
\left\{
\begin{array}{rl}
1.362-0.166x+0.025x^2-5.323x^3 & \mbox{ , }x > 0 \\
1.362-0.14x & \mbox{ , }x <0 \\
\end{array}
\right.\qquad.
\end{equation}
\end{center}

$\zR$ has a colour index $(B-V)=0.60$ \citep{1966CoLPL...4...99J}, which gives $\tau_\mathrm{c}=9.10~$days. The corresponding range of possible rotation periods is $P_{\mathrm{rot}} \sim 14.20 \pm 2.75~$days.

Since the equatorial rotation velocity is defined as $v_{\mathrm{rot}}=2\pi R_\star / P_{\mathrm{rot}}$, knowing the stellar radius $R_\star$ allows us to obtain a range of possible values for $v_{\mathrm{rot}}$.
Using $T_\star$ and $L_{\mathrm{bol}}$ and corrections prescribed by \citet{2006A&A...450..735M}, we find that for $\zR$, $R_\star=0.965R_{\odot} \pm 0.05$.
The corresponding value of equatorial velocity is $ \sim 3.42 \pm 0.66~$km/s.

We confront this with an independent measurement of $v_{\mathrm{rot}}\sin{i}$ by \citet{RS03}. They find $v_{\mathrm{rot}} \sin(i)=2.7 \pm 0.3~$km/s, which combined with the calculated rotational velocity, allows us to estimate the stellar inclination. We find that the inclination can range from 34$^\circ$ to 90$^\circ$ (see Fig ~\ref{fig:act_rot}). This is rather consistent with an inclined disk.

\begin{figure*}[ht]
\makebox[\textwidth]{\includegraphics[width=0.3\textwidth,angle=-90]{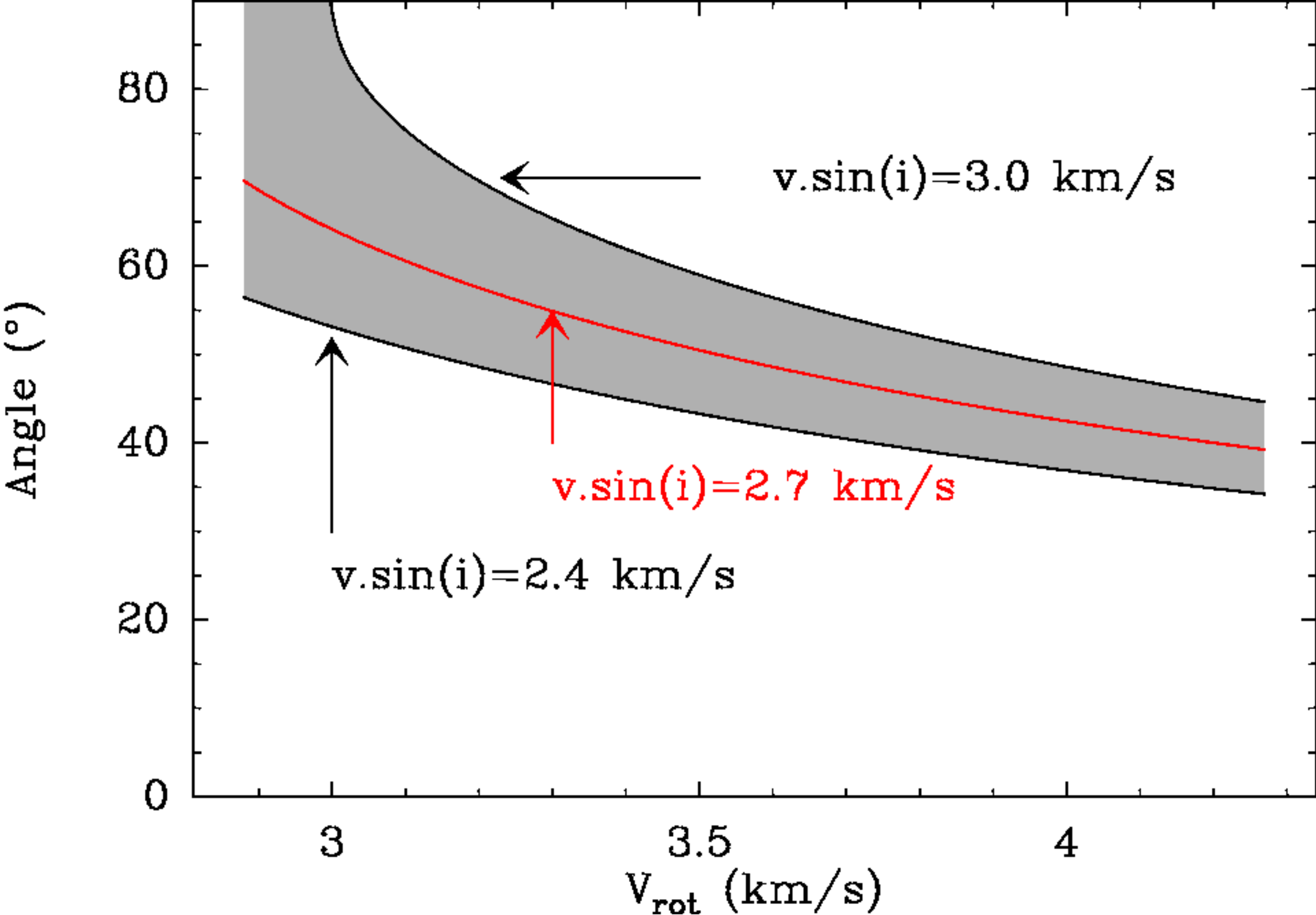}}
\caption[]{Possible inclination angles (taken from the pole) for $\zR$ as a function of $v_{\mathrm{rot}}$ ranging between $\sim 2.76$ and $4.08~$km/s, and $i$ is computed using the values of $v_{\mathrm{rot}} \sin(i)=2.7 \pm 0.3~$km/s \citep{RS03}. The acceptable zone for the inclination is grey-shaded.}
\label{fig:act_rot}
\end{figure*}

However, two angles are required to fully constraint the stellar rotation and disk axis. Therefore a degree of freedom remains and different orientations may lead to the same inclination $i$. Namely, the range of possible orientations leading to a same inclination is the set of axes describing a solid angle $2\pi \sin i$ about the line of sight.
But the number of axes which leads to the same inclination increases with i, since it follows a $\sin i$ distribution. This means that the inclinations in a range  $[34^\circ ;90^\circ]$ are not equiprobable. 
The probability for a given inclination to be between $i$ and $i+di$, provided it is in the range $[34^\circ ;90^\circ]$, can thus be written :

\begin{equation}
dP(i) = \frac{\sin i di}{\int^{90^\circ}_{34^\circ} \sin i di}\qquad,
\end{equation}
and the probability to find inclinations between $i_1$ and $i_2$ in the range $[34^\circ ; 90^\circ]$ is :
\begin{center}
\begin{equation}
P(i \in [i_1;i_2]) = \frac{\int^{i_2}_{i_1} \sin i di}{\int^{90^\circ}_{34^\circ} \sin i di}\qquad.
\end{equation}
\end{center}

Applying this to the case of $\zR$, among the possible range of $[34^\circ ; 90^\circ]$, we have $\sim 50 \%$ chance that the observed inclination is in the range $[65.5^\circ ; 90^\circ]$. Thus the system inclination is $i=65.5^{\circ +24.5}_{-31.5}$.
Consequently, the disk is more probably seen almost edge-on, with a pure edge-on configuration not having been ruled out. 

Because of the large uncertainties, this constraint does not really allow us to say with absolute confidence whether the disk exhibits an eccentric ring or resonant clumps. However, resonant structures are in general thin structures which tend to be hidden by non-resonant bodies and are difficult to detect, even in case of pole-on observations \citep{2008A&A...480..551R}. This argument clearly tends to support the interpretation of \citet{2010A&A...518L.131E}, i.e., an eccentric ring structure with $e \gtrsim 0.3$ seen edge-on, and extending from $\sim 70$ to $\sim 120~$AU, is observed.

\section{Constraints on  $\zeta^2$ Ret set by direct imaging}
\label{App_det_lim} 

VLT/NaCo \citep{2003SPIE.4841..944L,2003SPIE.4839..140R} Ks-band data were retrieved from the ESO archive (ID 086.C-0732(A); PI: L\"ohne,71574). Two epochs were available in August 2010 and November 2010, the former missing photometric calibration so only the latter could be used to set detection limits on the presence of bound companions. Nevertheless, both data sets were reduced and no companion is detected. The data from November, 11th 2010 were obtained in field stabilized-mode with five manual offsets of the derotator to simulate field rotation, with the S27 camera providing a pixel scale of 27 mas/pixel. Twenty image cubes with a DITxNDIT of 1.5s x 42 were obtained, for a total observing time on target of 21min. The semi-transparent mask C\_0.7\_sep\_10 with a diameter of $0.7^{\prime \prime}$ and a central transmission  of $3.5\times10^{-3}$ was used. 
Each individual image was bad pixel-corrected and flat- fielded. Background subtraction was made for each cube using the closest sky images. Recentering of the images was done using a Gaussian fit of the attenuated central star. Data selection was made within each data cube using criteria based on the attenuated central star flux and the encircled energy between $0.4^{\prime \prime}$ and $0.55^{\prime \prime}$. The images were then binned every 6s, and derotated into a reference frame where the pupil is stabilized in order to simulate Angular Differential Imaging, ADI \citep{Marois2006}. In this reference frame, the total field rotation provided by the manual offsets plus the natural pupi/field rotation is  $17^\circ$. This data cube is then reduced using Principal Components Analysis \citep{Soummer2012}, retaining 4 components out of 105.  

The noise in the final reduced image was estimated using a sliding 9 pixel wide box to get a preliminary map of detection limits in magnitude. We corrected this map by computing the flux losses due to the PCA reductionThey were estimated by injecting fake planets in the data cube at a $10-\sigma$ level and processing again the data. Last, these detection limits in magnitude were converted into detection limits in masses, using the COND models \citep{Baraffe2003} or BT-settl models \citep{Allard2011}, assuming an age of 2 Gyr. The 2D-detection limits given the COND evolutionary models is presented in Fig. \ref{Fig_det_map}

  \begin{figure}
   \centering
   \makebox[\textwidth]{
   \includegraphics[width=\columnwidth]{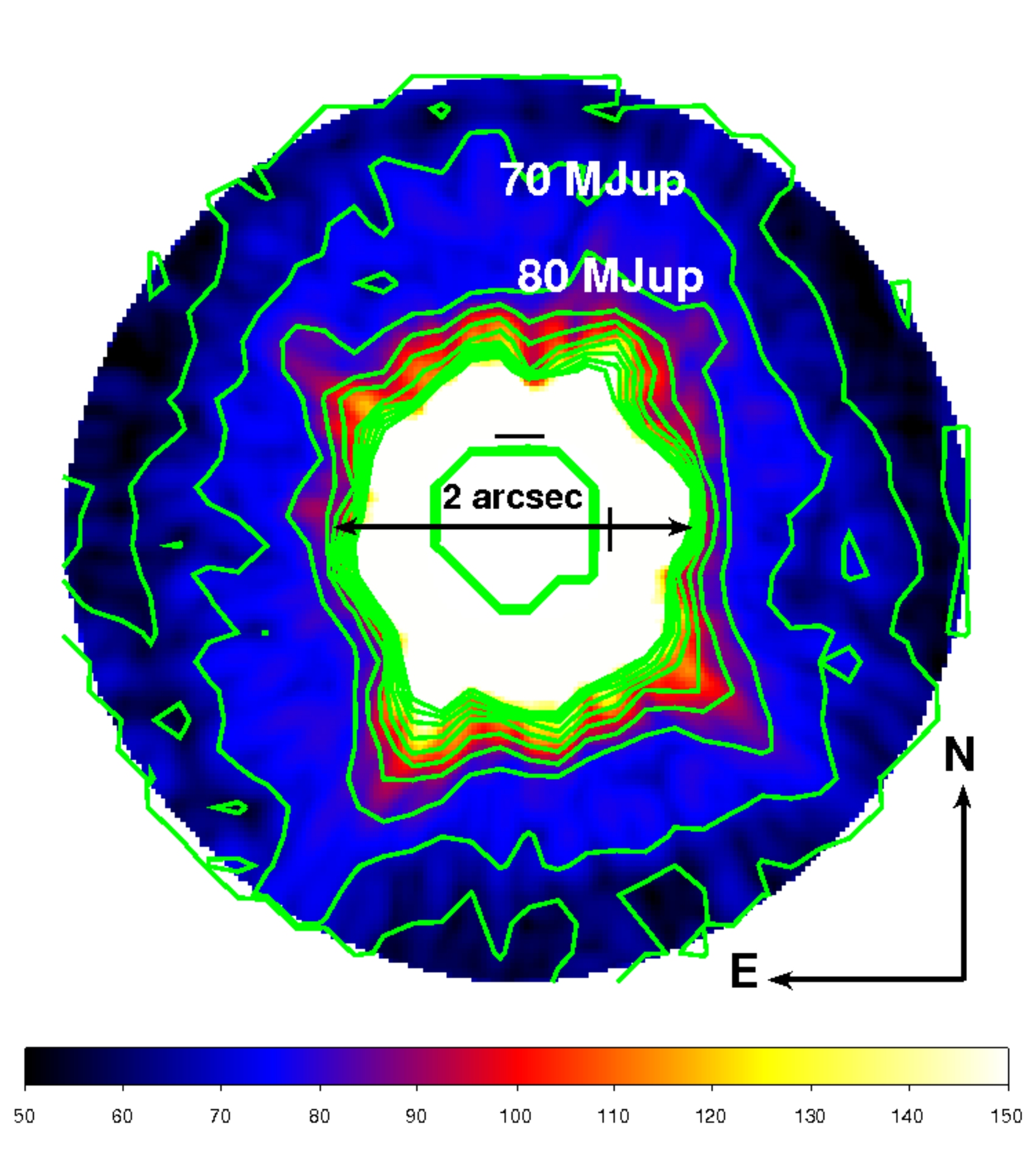}}
   \caption{Map of the detection limits in Jupiter masses set by the COND evolutionary models. The contours range from  60 to 150 $M_\mathrm{Jup}$ with a step of 10 $M_\mathrm{Jup}$.}
    \label{Fig_det_map}%
    \end{figure}
    
\end{onecolumn}
\end{document}